\documentclass[11pt]{article}
\pdfoutput=1
\usepackage[utf8x]{inputenc}	
\usepackage[english]{babel}
\usepackage{amssymb}
\usepackage{amsthm}
\usepackage{MnSymbol}
\usepackage{mathtools}		
\usepackage{dsfont}		
\usepackage{stmaryrd}		
\usepackage{cite}		
\usepackage[svgnames]{xcolor}
\usepackage{enumerate}
\usepackage{setspace}
\usepackage{booktabs}
\usepackage{longtable}
\usepackage{tikz}
\usetikzlibrary{shapes}
\usetikzlibrary{positioning}
\usetikzlibrary{chains}
\usetikzlibrary{arrows,fit,decorations.pathreplacing, shapes.misc}
\tikzstyle{every picture}+=[remember picture]
\tikzstyle{na} = [baseline=-.5ex]
\usepackage{todonotes}
\usepackage{paralist}
\usepackage[font={small,it},labelfont=bf]{caption} 
\usepackage{subcaption}
\usepackage[pdftex,breaklinks,colorlinks=true,urlcolor=RoyalBlue,linkcolor=blue,
citecolor=blue]{hyperref}		
\usepackage[normalem]{ulem}
\usepackage{geometry}
\usepackage{graphicx}
\usepackage{tikz}
\usetikzlibrary{shapes}

\usepackage{multirow}

\catcode`@=11
\renewcommand{\section}{\@startsection{section}{1}{0pt}{\medskipamount}
{\medskipamount}{\Large\bf}}
\numberwithin{equation}{section}
\catcode`@=12

\def\dim{\mathrm{dim}}

\newcommand{\diff}{\mathrm{d}}

\newcommand{\C}{\mathbb{C}}
\newcommand{\R}{\mathbb{R}}

\newcommand{\HH}{\mathbb{H}}
\newcommand{\NN}{\mathbb{N}}  
\newcommand{\Z}{\mathbb{Z}}
\newcommand{\D}{\mathbb{D}}

\newcommand{\Coulomb}{\mathcal{C}}

\newcommand{\Higgs}{\mathcal{H}}

\newcommand{\clorbit}[1]{\overline{\mathcal{O}}_{#1}}
\newcommand{\height}[1]{\text{ht}(#1)}
\newcommand{\slice}[1]{\mathcal{S}_{#1}}

\newcommand{\Ncal}{\mathcal{N}}

\newcommand{\uo}{{ \mathrm{U}(1)}}

\newcommand{\surm}{{{\rm SU}}}

\newcommand{\sorm}{{{\rm SO}}}
\newcommand{\orm}{{{\rm O}}}

\newcommand{\sprm}{{{\rm Sp}}}

\newcommand{\usprm}{{{\rm USp}}}

\newcommand{\Pcal}{\mathcal{P}}

\newtheorem{myConj}{Conjecture}

\newcommand{\NS}{\text{NS5}}
\newcommand{\Done}{\text{D1}}
\newcommand{\Dtwo}{\text{D2}}
\newcommand{\Dthree}{\text{D3}}
\newcommand{\Ot}{\text{O3}}
\newcommand{\Otm}{\text{O3${}^{-}$}}
\newcommand{\Otmt}{\text{$\widetilde{\text{O3}}^{-}$}}
\newcommand{\Otp}{\text{O3${}^{+}$}}
\newcommand{\Otpt}{\text{$\widetilde{\text{O3}}^{+}$}}
\newcommand{\Df}{\text{D4}}
\newcommand{\Dfive}{\text{D5}}
\newcommand{\Ds}{\text{D6}}
\newcommand{\Os}{\text{O6}}
\newcommand{\Osm}{\text{O6${}^{-}$}}
\newcommand{\Osmt}{\text{$\widetilde{\text{O6}}^{-}$}}
\newcommand{\Osp}{\text{O6${}^{+}$}}
\newcommand{\Ospt}{\text{$\widetilde{\text{O6}}^{+}$}}

\newcommand{\De}{\text{D8}}

\newcommand{\Mf}{\text{M5}}
\newcommand{\Mn}{\text{M9}}
\newcommand{\magQuiv}{\mathsf{Q}}
\newcommand{\magQuivText}{\substack{\mathrm{magnetic} \\ \mathrm{quiver}}}
\newcommand{\elecTheory}{\substack{\mathrm{electric} \\ \mathrm{theory}}}
\newcommand{\fin}{\mathrm{fin}}
\newcommand{\dalg}{\mathfrak{d}}
\newcommand{\balg}{\mathfrak{b}}
\newcommand{\calg}{\mathfrak{c}}
\newcommand{\cc}[1]{$\scriptstyle{\calg_{#1}}$}
\newcommand{\bb}[1]{$\scriptstyle{\balg_{#1}}$}
\newcommand{\dd}[1]{$\scriptstyle{\dalg_{#1}}$}
\usepackage{xargs}

\newlength{\myline}
\newcommandx*{\doublearrow}[4][1=0, 2=1]{
  \draw[line width=\myline,double distance=3\myline,#3] #4;
}

\newcommandx*{\triplearrow}[4][1=0, 2=1]{
  \draw[line width=\myline,double distance=5\myline,#3] #4;
  \draw[line width=\myline,shorten <=#1\myline,shorten >=#2\myline,#3] #4;
}

\newcommandx*{\quadarrow}[4][1=0, 2=2.5]{
  \draw[line width=\myline,double distance=5\myline,#3] #4;
  \draw[line width=\myline,double distance=\myline,shorten <=#1\myline,shorten 
>=#2\myline,#3] #4;
}

\def\ns#1{
	\node[circle, draw, fill=white] at (#1){};
	\node[cross out, draw] at (#1){};
}

\def\DeightMany#1#2#3{
   \foreach \i in {1,...,#1} {
	\draw({#2+#3*0.\i-0.1*#3},-1)--({#2+#3*0.\i-0.1*#3},+1);
      }
   }

\def\o#1#2{
	\draw[dotted](#1)--(#2);
}

\def\oAux#1#2{
	\draw[dashed](#1)--(#2);
}

\def\OPlus#1#2{
	\draw[dotted](#1)--(#2);
}

\def\OPlusTilde#1#2{
	\draw[dashed](#1)--(#2);
}

\def\Dbrane#1#2{
	\draw(#1)--(#2);
}

\def\DsixEvenFree#1#2{
	\draw(#2,0.1)--(#2+1,0.1);
	\draw(#2,-0.1)--(#2+1,-0.1);
	\node[label=above:{{\footnotesize{#1}}}] at (#2+0.65,0.2) {};
}

\def\DsixOPlus#1#2{
	\draw(#2,0.1)--(#2+1,0.1);
	\draw(#2,-0.1)--(#2+1,-0.1);
	\draw[dotted] (#2,0)--(#2+1,0);
	\node[label=above:{{\scriptsize{#1}}}] at (#2+0.65,0) {};
}

\def\DsixOPlusTilde#1#2{
	\draw(#2,0.1)--(#2+1,0.1);
	\draw(#2,-0.1)--(#2+1,-0.1);
	\draw[dashed] (#2,0)--(#2+1,0);
	\node[label=above:{{\scriptsize{#1}}}] at (#2+0.65,0) {};
}

\def\DsixOMinus#1#2{
	\draw(#2,0.1)--(#2+1,0.1);
	\draw(#2,-0.1)--(#2+1,-0.1);
	\node[label=above:{{\scriptsize{#1}}}] at (#2+0.65,0) {};
}

\def\DsixOMinusTilde#1#2{
	\draw(#2,0.1)--(#2+1,0.1);
	\draw(#2,-0.1)--(#2+1,-0.1);
	\draw (#2,0)--(#2+1,0);
	\node[label=above:{{\scriptsize{#1}}}] at (#2+0.65,0) {};
}
\def\DsixOPlusFree#1#2#3{
	\draw(#2,0.3+0.1*#3)--(#2+1,0.3+0.1*#3);
	\draw(#2,-0.3-0.1*#3)--(#2+1,-0.3-0.1*#3);
	\draw[dotted] (#2,0)--(#2+1,0);
	\node[label=above:{{\scriptsize{#1}}}] at (#2+0.65,0.1+0.1*#3) {};
}

\def\DsixOPlusTildeFree#1#2#3{
	\draw(#2,0.3+0.1*#3)--(#2+1,0.3+0.1*#3);
	\draw(#2,-0.3-0.1*#3)--(#2+1,-0.3-0.1*#3);
	\draw[dashed] (#2,0)--(#2+1,0);
	\node[label=above:{{\scriptsize{#1}}}] at (#2+0.65,0.1+0.1*#3) {};
}

\def\DsixOMinusFree#1#2#3{
	\draw(#2,0.3+0.1*#3)--(#2+1,0.3+0.1*#3);
	\draw(#2,-0.3-0.1*#3)--(#2+1,-0.3-0.1*#3);
	\node[label=above:{{\scriptsize{#1}}}] at (#2+0.65,0.1+0.1*#3) {};
}

\def\DsixOMinusTildeFree#1#2#3{
	\draw(#2,0.3+0.1*#3)--(#2+1,0.3+0.1*#3);
	\draw(#2,-0.3-0.1*#3)--(#2+1,-0.3-0.1*#3);
	\draw (#2,0)--(#2+1,0);
	\node[label=above:{{\scriptsize{#1}}}] at (#2+0.65,0.1+0.1*#3) {};
}

\newcommand{\Sp}[1]{{\tiny{$Sp(#1)$}}}

\newcommand{\SO}[1]{{\tiny{$SO(#1)$}}}

\def\ArrowMagnetic#1#2{
	\draw (#1,#2)--(#1+1,#2);
	\draw (#1+0.8,#2-.2)--(#1+1,#2);
	\draw (#1+0.8,#2+.2)--(#1+1,#2);
	\node (label) at (#1+0.5,#2+0.7)[]{magnetic quiver};
}

\allowdisplaybreaks

\begin{document}
\begin{titlepage}
\setcounter{page}{0}
%
%

\begin{flushright}
Imperial/TP/19/AH/05
\end{flushright}

\vskip 2cm

\begin{center}

{\Large\bf 
Magnetic Quivers, Higgs Branches, and $6$d $\Ncal=(1,0)$ Theories \\[5mm]
---\\[5mm] Orthogonal and Symplectic Gauge Groups
}

\vspace{15mm}

{\large Santiago Cabrera${}^{1}$},\ {\large Amihay Hanany${}^{1}$}, and \ 
{\large 
Marcus Sperling${}^{2}$} 
\\[5mm]
\noindent ${}^1${\em Theoretical Physics Group, Imperial College London\\
Prince Consort Road, London, SW7 2AZ, UK}\\
{Email: {\tt santiago.cabrera13@imperial.ac.uk}, \\ {\tt 
a.hanany@imperial.ac.uk}}
\\[5mm]
\noindent ${}^{2}${\em Yau Mathematical Sciences Center, Tsinghua University}\\
{\em Haidian District, Beijing, 100084, China}\\
Email: {\tt marcus.sperling@univie.ac.at}
\\[5mm]

\vspace{15mm}

\begin{abstract}
\Mf\ branes on a $D$-type ALE singularity display various phenomena that 
introduce additional massless degrees of freedom. The \Mf\ branes are known to 
fractionate on a $D$-type singularity. Whenever two fractional \Mf\ branes 
coincide, tensionless strings arise. Therefore, these systems do not admit a 
low-energy Lagrangian description.
Focusing on the $6$-dimensional $\Ncal=(1,0)$ world-volume theories on the \Mf\ 
branes, the vacuum moduli space has two branches were either the scalar fields 
in the tensor multiplet or the scalars in the hypermultiplets acquire a 
non-trivial vacuum expectation value.
As suggested in previous work, the Higgs branch may change drastically whenever 
a BPS-string becomes tensionless. Recently, \emph{magnetic quivers} have been 
introduced with the aim to capture all Higgs branches over any point of the 
tensor branch.
In this paper, the formalism is extended to Type IIA brane configurations 
involving \Os\ planes. Since the $6$d $\Ncal=(1,0)$ theories are composed 
of orthosymplectic gauge groups, the derivation rules for the 
magnetic quiver in the presence of \Os\ planes have to be conjectured.
This is achieved by analysing the $6$d theories for a single \Mf\ brane on a 
$D$-type singularity and deriving the magnetic quivers for the finite and 
infinite gauge coupling Higgs branch from a brane configuration. The 
validity of the proposed derivation rules is underpinned by deriving the 
associated Hasse diagram.
For multiple \Mf\ branes, the approach of this paper provides magnetic 
quivers for all Higgs branches over any point of the tensor branch. In 
particular, an interesting infinite gauge coupling transition is found that is 
related to the $\sorm(8)$ non-Higgsable cluster.

\end{abstract}

\end{center}

\end{titlepage}

{\baselineskip=12pt
{\footnotesize
\tableofcontents
}
}
  \section{Introduction}
\label{sec:intro}
Starting from the $6$-dimensional $\Ncal=(2,0)$ world-volume theories living on 
a 
stack of \Mf\ branes \cite{Witten:1995zh,Strominger:1995ac}, the $6$-dimensional 
theories derived from \Mf\ branes in various settings have been studied 
intensively, but many aspects still remain mysterious. 
One of the simplest classes of $6$-dimensional $\Ncal=(1,0)$ theories is 
obtained from multiple \Mf\ branes transverse to $\R \times \C^2\slash \Gamma$
with $\Gamma = \Z_k$ or $\D_{k-2}$, i.e.\ the A or D-type singularities. The 
main advantage of this class is the existence of a dual Type IIA construction 
via 
\Ds-\De-\NS\ brane configurations with or without \Os\ orientifolds 
\cite{Hanany:1997gh,Brunner:1997gk,Brunner:1997gf,Hanany:1999sj}. These brane 
constructions pointed towards the existence of non-trivial conformal 
fixed-points at the origin of the tensor branch, where all \NS\ branes become 
coincident. A classification for more general $6$d $\Ncal=(1,0)$ superconformal 
theories obtainable from F-theory compactifications has been proposed in 
\cite{Heckman:2013pva,Heckman:2015bfa}.

A $6$-dimensional $\Ncal=(1,0)$ supersymmetric theory has massless 
degrees of freedom encoded in three types of supermultiplets --- tensor 
multiplet, 
vector multiplet, and hypermultiplet --- as well as other degrees of freedom 
which arise 
from tensionless strings \cite{Witten:1995zh}. For consistence, the 
gravitational anomaly cancellation \cite{Green:1984bx} for a $6$d 
$\Ncal=(1,0)$ theory requires 
\cite{RandjbarDaemi:1985wc,Dabholkar:1996zi} 
\begin{align}
 n_h + 29 n_t-n_v=\text{constant} \; ,
\label{eq:anomaly_cancellation}
\end{align}
where $n_t$, $n_v$, $n_h$ denote the numbers of tensor, vector, and 
hypermultiplets, respectively. In general, anomalies in $6$-dimensional 
$\Ncal=(1,0)$ theories have been studied in works like 
\cite{Sagnotti:1992qw,Danielsson:1997kt,Bershadsky:1997sb}.

In contrast to lower dimensional theories, the gauge coupling in $6$ 
dimensions is not a mere 
parameter, but inversely proportional to the vacuum expectation value of the  
scalar field in the tensor 
multiplet. Moreover, the inverse gauge coupling serves as tension for 
BPS-strings, and is given by the distance of \NS\ branes in the Type IIA 
realisation. On a generic point of the tensor branch, i.e.\ a point in which 
all gauge couplings are finite, the Higgs branch moduli spaces is a 
hyper-K\"ahler quotient 
realised by the vanishing locus of the F and D-terms modulo gauge 
equivalence \cite{Hitchin:1986ea}. 
Whenever one gauge coupling approaches infinity, i.e.\ at a singular locus of 
the tensor branch, certain BPS-strings become tensionless and new massless 
degrees of freedom contribute to the Higgs branch.
Due to the amount of supersymmetry, the Higgs branches over tensor branch 
singularities are still hyper-K\"ahler, but generically not hyper-K\"ahler 
quotients anymore. 
For instance, the jump in the dimension between the Higgs branch over a 
generic point and the  Higgs branch at the origin of the tensor branch has been 
computed in \cite{Mekareeya:2017sqh}. This indicates a non-trivial change in 
the Higgs branch along the tensor branch.
Therefore, alternative descriptions are desirable to capture the changes of 
the Higgs branch geometry. Fortunately, Coulomb branches of $3$-dimensional 
$\Ncal=4$ gauge theories are a most suitable class of hyper-K\"ahler moduli 
spaces, as detailed extensively in \cite{Cabrera:2019izd}.
More generally, Higgs branches of theories with $8$ supercharges can be 
enlarged at the UV fixed point due to massless BPS-objects, and the classical 
hyper-K\"ahler 
description breaks down. 
For instance, Coulomb branches have already been employed successfully to 
describe Higgs 
branches of Argyres-Douglas theories \cite{DelZotto:2014kka}, as 
well as infinite coupling limits of $5$-dimensional 
theories \cite{Cremonesi:2015lsa,Ferlito:2017xdq,Cabrera:2018jxt} and of
$6$-dimensional gauge theories 
\cite{Mekareeya:2017jgc,Hanany:2018uhm,Hanany:2018vph,Cabrera:2019izd}. In 
fact, by interpreting these moduli spaces as symplectic 
singularities\cite{beauville2000symplectic}, the \emph{magnetic quiver} 
techniques allowed to derive the Hasse diagrams for the various Higgs branches 
\cite{Bourget:2019aer}.  

In this paper, the focus is placed on a class of $6$d $\Ncal=(1,0)$ 
supersymmetric gauge theories that originate from multiple \Mf\ branes on a 
$D$-type ALE singularity. As shown in \cite{DelZotto:2014hpa}, new massless 
tensor 
multiplets appear once the \Mf\ branes reach the fixed point of the ALE space 
$\C^2 \slash \D_{k-2}$. In other words, an \Mf\ brane fractionates into two 
parts on the singularity; a phenomenon, known as \NS\ branes splitting into two 
half \NS s on an \Os\ orientifold plane \cite{Evans:1997hk}. The associated 
class of $6$d 
$\Ncal=(1,0)$ theories has been studied extensively 
\cite{Intriligator:1997kq,Blum:1997mm,Hanany:1997gh,Brunner:1997gk, 
Intriligator:1997dh,Brunner:1997gf,Ferrara:1998vf,Heckman:2013pva,
DelZotto:2014hpa}; interestingly, the Higgs branches at the origin of the 
tensor 
branch have only been addressed in 
\cite{Mekareeya:2016yal,Mekareeya:2017sqh,Hanany:2018uhm} recently. For a 
single \Mf\ brane on $\C^2 \slash \D_{k-2}$, the Higgs branch dimension jumps 
by $29$ quaternionic units between a 
generic point and the origin of the tensor branch \cite{Mekareeya:2017sqh}.  In 
\cite{Hanany:2018uhm}
this phenomenon has been identified with the \emph{small $E_8$ instanton 
transition} 
\cite{Ganor:1996mu}, see also 
\cite{Seiberg:1996vs,Intriligator:1997kq,Blum:1997mm,Hanany:1997gh}. For $n$ 
\Mf\ branes on $\C^2\slash \D_{k-2}$, the Higgs branch dimension jumps by 
$n{+}\dim\ \sorm(8)$ 
quaternionic units between a generic point and the origin 
\cite{Mekareeya:2017sqh}. In \cite{Hanany:2018uhm} a description for the Higgs 
branch at the CFT point has been conjectured, but a more detailed analysis is 
still missing.

The common reason behind the, perhaps surprising, feature that the Higgs 
branches change discontinuously over tensor branch lies in BPS-strings becoming 
tensionless. As put forward in \cite{Cabrera:2019izd} (see also 
\cite{Mekareeya:2017sqh}), the different singular loci of 
the tensor branch can be associated with different subsets of order parameters 
being zero. Here, the inverse gauge couplings $\frac{1}{g_i^2}$ serve as 
suitable order parameters for the Higgs branch phases $\Pcal_i$ of a given 
$6$d $\Ncal=(1,0)$ theory. A unified analysis of the Higgs branch phases is 
possible by changing the phase of the Type IIA \Ds-\De-\NS\ brane configuration 
to the 
phase where all (as many as possible) \Ds\ branes are suspended between \De\ 
branes instead of \NS\ branes. This is quite intuitive, because Higgs branch 
degrees of freedom are 
read off from \Ds\ branes suspended between \De\ branes. This brane system 
phase enables one to systematically read off an associated 
\emph{magnetic quiver} $\magQuiv(\Pcal_i)$ such that its data considered as 
defining a $3$d $\Ncal=4$ Coulomb branch correctly describes the $6$d 
$\Ncal=(1,0)$ Higgs branch of the point (phase) $\Pcal_i$ of the tensor branch, 
i.e.\
\begin{align}
 \Higgs^{6\diff} \left( \text{phase }\Pcal_i \right)
 =
 \Coulomb^{3\diff} \left( \magQuivText \ \magQuiv(\Pcal_i) \right)
 \label{eq:objective}
\end{align}
holds as \emph{equality of moduli spaces}.

The key technique\cite{Cabrera:2019izd} for achieving \eqref{eq:objective} is 
to 
generalise the notion of \emph{electric} and \emph{magnetic theory} from the 
Type IIB construction \cite{Hanany:1996ie} of $3$d $\Ncal=4$ world-volume 
theories from \Dthree-\Dfive-\NS\ branes. Since the Type IIA system of 
\Ds-\De-\NS\ branes (with or without \Os\ planes) is T-dual to the Type IIB 
configuration of \Dthree-\Dfive-\NS\ branes (with or without \Ot\ planes), the 
magnetic quiver is derived from the possible ways virtual \Df\ branes can be 
suspend between \Ds , \De, and \NS\ branes. Again, this is in complete analogy 
to 
D-string in the Type IIB \Dthree-\Dfive-\NS\ systems. The main purpose of this 
paper is to develop the formalism of magnetic quivers for $6$d $\Ncal=(1,0)$ 
theories with orthosymplectic gauge nodes. Therefore, the inclusion of \Os\ 
orientifold planes is of central importance and one needs to 
suitably generalise \Ot\ plane arguments of \cite{Feng:2000eq}.

The proposed formalism, as extension of \cite{Cabrera:2019izd}, is heavily based 
on 
various $3$d $\Ncal=4$ Coulomb branch techniques 
developed after the Coulomb branch realisation as \emph{space of dressed 
monopole operators} \cite{Cremonesi:2013lqa}. Relevant techniques include
Kraft-Procesi transitions and transverse slices 
\cite{Cabrera:2016vvv,Cabrera:2017njm,Hanany:2018uhm},
quiver subtraction \cite{Cabrera:2018ann}, and
discrete quotients \cite{Hanany:2018vph,Hanany:2018cgo,Hanany:2018dvd}.

The outline of the paper is as follows: after introducing the set-up in 
Section \ref{sec:setup}, the 
concept of a \emph{magnetic quiver} is detailed in Section 
\ref{sec:el-mag_quiver}. Thereafter, the cases of one \Mf\ and multiple \Mf s 
transverse to a $\R\times \C^2\slash \D_{k-2}$ are focused on in Sections 
\ref{sec:single_M5} and \ref{sec:multiple_M5}. In particular, 
the derivations of the \emph{magnetic quivers} and the geometry of the 
transitions of the different Higgs branches are elaborated. 
In Section \ref{sec:Hasse} the geometry of the finite and infinite coupling 
Higgs 
branch of a single \Mf\ is explored via Kraft-Procesi transitions; moreover, 
the corresponding Hasse diagrams are derived. 
Lastly, Section \ref{sec:conclusion} provides a conclusion and outlook.
Appendix \ref{app:orientifolds} summarises details of \Os\ planes, and Appendix 
\ref{app:global_sym} reviews Coulomb branch symmetries of $3$d $\Ncal=4$ 
orthosymplectic quiver gauge theories.

  \section{Magnetic Quiver}
\label{sec:magnetic_quiver}
\subsection{Set-up}
\label{sec:setup}
Consider \Mf\ branes and a $D_{k}$ ALE singularity $\C^2 \slash \D_{k-2}$
stretching the space-time dimensions as indicated in Table 
\ref{tab:directions}. Here $\D_{k-2}$ denotes the \emph{binary dihedral group} 
of order $4k-8$ such that the crepant 
resolution of $\C^2\slash \D_{k-2}$ has associated Dynkin diagram 
$\widehat{D}_k$. 
The singularity at the origin of $\C^2 \slash \D_{k-2}$ is localized in 
directions 
$x^{7}$, $x^8$, $x^9$, and $x^{10}$, and spans directions $x^0,x^1, 
\ldots, x^6$. Therefore, it is represented as a horizontal line.
\begin{table}[t]
\centering
\begin{tabular}{c|ccccccccccc}
\toprule
 M-theory & $x^0$ &  $x^1$ & $x^2$ & $x^3$ & $x^4$ & $x^5$ & $x^6$ & $x^7$ & 
$x^8$ & $x^9$  & $x^{10}$ \\ \midrule 
\Mf\ & $\times$ & $\times$ & $\times$ & $\times$ & $\times$ & $\times$ & & & & 
& \\
$\C^2 \slash \D_{k-2}$ & $\times$ & $\times$ & $\times$ & $\times$ & $\times$ & 
$\times$ & $\times$ & & & 
& \\ \midrule
Type IIA & $x^0$ &  $x^1$ & $x^2$ & $x^3$ & $x^4$ & $x^5$ & $x^6$ & $x^7$ & 
$x^8$ & $x^9$  & \\ \midrule 
\NS & $\times$ & $\times$ & $\times$ & $\times$ & $\times$ & $\times$ & & & & 
& \\
 \De & $\times$ & $\times$ & $\times$ & $\times$ & $\times$ & $\times$ & 
&$\times$ & $\times$ &$\times$ &   \\
\Ds, \Os & $\times$  & $\times$ & $\times$ & $\times$ & $\times$ & $\times$ & 
$\times$ &  &  & &   \\
\midrule
F1 & $\times$  &  &  & $\times$  &  &  &  &  &  &    \\
\Df & $\times$  &  &  & $\times$ & $\times$ & $\times$ &  & $\times$ &  &    
\\
\bottomrule
\end{tabular}
\caption{Upper part: Occupation of space-time directions by \Mf , and 
$D_{k}$ 
singularity in M-theory. Lower part: Occupation of space-time directions by 
\NS, \De, \Ds, and \Os\ in Type IIA. The fundamental string F1 and the \Df\ 
branes are virtual objects which are used to read off the electric and magnetic 
quivers.}
\label{tab:directions}
\end{table}
The M-theory picture can be presented as 
\begin{align}\label{branes:11d}
\raisebox{-.5\height}{
 \begin{tikzpicture}
  \draw(0,0)--(6,0);
  \draw (0,0.2) node {$D_{k}$};
  \draw (1,0) node {$\times$};
  \draw (2,1) node {$\times$};  
  \draw (3,0) node {$\times$};
  \draw (4,0) node {$\times$};
  \draw (5,1) node {$\times$};
  \draw (2.3,0.8) node {$\Mf$};
  \draw[thick,->] (8,0)--(9,0);
  \draw (9.2,0.2) node {$x^6$};
  \draw[thick,->] (8,0)--(8,1);
  \draw (8.2,1.2) node {$x^{7,8,9,10}$};
 \end{tikzpicture}
 }
\end{align}
The 
corresponding description in Type IIA is obtained by an identification as 
follows: the \NS\ originates from the \Mf\ which is point-like in the $x^{10}$ 
direction. The 
$D_{k}$ ALE space $\C^2 \slash \D_{k-2}$ in M-theory provides a local 
description 
of $k$ coincident \Ds\ branes on an \Osm\ orientifold in Type IIA on flat 
space. 
In particular, the 
directions $x^7, x^8, \ldots, x^{10}$, in which the singular origin of the
ALE singularity is localised in, become the three directions transverse to 
the \Ds s and the direction of the M-theory circle.

An important phenomenon is that \Mf\ branes fractionate on 
ALE-singularities \cite{DelZotto:2014hpa}. While for A-type singularities the 
number of fractions is just one, the \Mf\ splits into two fractions on D-type 
orbifolds.  
Hence, $n$ \Mf\ on the D-type orbifold correspond to $n$ pairs of two half \NS\ 
branes in the dual Type IIA description. (The splitting of a full \NS\ brane 
into two half \NS\ branes along an \Os\ plane in Type IIA had already been 
observed earlier in \cite{Evans:1997hk}.)
The corresponding Type IIA diagram for \eqref{branes:11d} is:
\begin{align}\label{branes:IIA}
\raisebox{-.5\height}{
 \begin{tikzpicture}
 \DsixOMinus{$k$}{0}
 \DsixOPlus{$k{-}4$}{1}
 \DsixOMinus{$k$}{2}
 \Dbrane{3,0.1}{4,0.1}
 \Dbrane{3,-0.1}{4,-0.1}
 \DsixOPlus{$k{-}4$}{4}
 \DsixOMinus{$k$}{5}
 \DsixOPlus{$k{-}4$}{6}
 \DsixOMinus{$k$}{7}
 \Dbrane{8,0.1}{9,0.1}
 \Dbrane{8,-0.1}{9,-0.1}
  \ns{1,0};
  \ns{2,0};
  \ns{3,1}; 
  \ns{3,-1}; 
  \ns{4,0}; 
  \ns{5,0};
  \ns{6,0}; 
  \ns{7,0};
  \ns{8,1};
  \ns{8,-1}; 
  \draw (3.85,1) node {$\tfrac{1}{2}$\NS };
  \draw (8.85,0.45) node {$\tfrac{1}{2}$\Ds };
  \draw[thick,->] (11,0)--(12,0);
  \draw (12.2,0.2) node {$x^6$};
  \draw[thick,->] (11,0)--(11,1);
  \draw (11.2,1.2) node {$x^{7,8,9}$};
 \end{tikzpicture}
 }
\end{align}
and the numbers displayed count full \Ds\ branes.
Note that the \Os\ orientifolds change whenever they cross a half \NS\ or half 
\De\ brane as summarised in Appendix \ref{app:orientifolds}. Moverover, the 
different numbers of \Ds\ branes follow from the charges of the orientifolds 
and the charge conservation.
%
%
\subsection{Electric and magnetic quiver}
\label{sec:el-mag_quiver}
In the study of the Higgs branches of 6d $\Ncal=(1,0)$ theories resulting from 
\Mf\ branes on an A-type singularity $\C^2\slash \Z_k$, the concept of magnetic 
quivers has been introduced in \cite{Cabrera:2019izd}.
In this section, this concept is reviewed and extended for the 
application of D-type singularities.

To begin with, recall the \Dthree-\Dfive-\NS\ brane configurations of 
\cite{Hanany:1996ie} supplemented by orientifold 3-planes \cite{Feng:2000eq}, 
which yield $3$d $\Ncal=4$ world-volume theories with alternating orthogonal 
and symplectic gauge groups. Table \ref{tab:directions_3d} provides an overview 
of the set-up. 
In this scenario, there exists a natural notion of \emph{electric} and 
\emph{magnetic} gauge theory.
\Dthree\ branes suspended between \NS\ branes give rise to the electric gauge 
theory on their world-volume and the low-energy degrees of freedom are deduced 
from suspended fundamental strings.  Adding \Dfive\ branes introduces electric 
hypermultiplets. 
Conversely, \Dthree\ branes in between \Dfive\ branes lead to a magnetic gauge 
theory on the \Dthree\ world-volume and it is the D-string that induces the 
relevant degrees of freedom. Consequently, \NS\ branes are responsible for 
magnetic hypermultiplets.

The effect of \Ot\ planes lies in a projection that reduces unitary gauge and 
flavour symmetries to orthogonal and symplectic symmetries, see Table 
\ref{tab:orientifold_3d} for an overview. The characteristic sign of the 
low-energy effective theories is a quiver gauge theory with alternating 
orthogonal and symplectic gauge nodes.

By virtue of S-duality or $3$d mirror symmetry \cite{Intriligator:1996ex}, the 
maximal branches of the moduli spaces of electric and magnetic theory are 
related via 
\begin{align}
 \Higgs^{3\diff} \left( \text{electric theory} \right) = 
  \Coulomb^{3\diff} \left( \text{magnetic theory} \right) \,.
  \label{eq:3d_miror}
\end{align}
Nevertheless, $3$d mirror symmetry is a full-fledged IR-duality between the 
electric and magnetic theory, but for the 
purposes of this paper relations of the type \eqref{eq:3d_miror} are the 
central objective.

\begin{table}[t]
\centering
\begin{tabular}{c|cccccccccc}
\toprule
 Type IIB & $x^0$ &  $x^1$ & $x^2$ & $x^3$ & $x^4$ & $x^5$ & $x^6$ & $x^7$ & 
$x^8$ & $x^9$   \\ \midrule 
\NS & $\times$ & $\times$ & $\times$ & $\times$ & $\times$ & $\times$ & & & &\\
\Dfive & $\times$ & $\times$ & $\times$ & & & & &$\times$ & $\times$ &$\times$  
\\
\Dthree , \Ot & $\times$  & $\times$ & $\times$ &  &  &  & $\times$ &  &  &    
\\ 
\midrule
F1 & $\times$  &  &  &  $\times$ &  &  &  &  &  &    \\
\Done & $\times$  &  &  &  &  &  &  & $\times$ &  &    \\
\bottomrule
\end{tabular}
\caption{Occupation of space-time directions by \NS, \Dfive, \Dthree, and \Ot\ 
in Type IIB. The fundamental string F1 induces the electric theory, while the 
D-string \Done\ induces the magnetic theory.}
\label{tab:directions_3d}
\end{table}

\begin{table}
 \centering
 \begin{tabular}{c|c|c||c}
 \toprule
orientifold & gauge group & flavour group & S-dual  
\\\midrule
\Otm & $\orm(2n)$ & $\usprm(2k)$ & \Otm  \\
\Otmt & $\orm(2n+1)$ & $\usprm(2k)$ & \Otp  \\
\Otp & $\usprm(2n)$ & $\orm(2k)$ or $\orm(2k+1)$ & \Otmt \\
\Otpt & $\usprm'(2n)$ & $\orm(2k)$ or $\orm(2k+1)$ & \Otpt \\
\bottomrule
 \end{tabular}
 \caption{Effect of \Ot\ orientifolds on the low-energy effective theories. 
A stack of $n$ full \Dthree\ branes and an \Ot\ plane suspended 
between two \NS\ branes gives rise to the electric gauge group. Moreover, 
if there are $2k$ half \Dfive\ branes, or in the case of \Otp\ and \Otpt\ there 
may also be $2k{+}1$ half \Dfive s, intersecting the \Dthree -\Ot\ stack 
then 
an electric flavour group arises. Lastly, S-duality transforms the \Ot\ planes 
among each other and the resulting magnetic gauge groups are the GNO duals 
\cite{Goddard:1976qe} of 
the electric gauge groups.}
\label{tab:orientifold_3d}
\end{table}

Returning to the \Ds-\De-\NS\ brane configurations 
\cite{Hanany:1997gh,Brunner:1997gk} supplemented by orientifold 6-planes, 
a central point in the argument of \cite{Cabrera:2019izd} is that the system is 
T-dual 
to the \Dthree-\Dfive-\NS\ system upon three T-dualities along $x^3$, $x^4$, 
$x^5$. 
The conventional quiver gauge theory on a generic point 
of the tensor branch of the $6$d $\Ncal=(1,0)$ theory is read off from the 
phase of the Type IIA brane configuration in which all \NS\ branes are well 
separated along the orientifold. The effect of the \Os\ orientifold planes  
is analogous to the $3$d setting and is summarised in the left-hand-side of 
Table 
\ref{tab:orientifold} for convenience. The condition for an anomaly-free $6$d 
theory is 
equivalent to charge conservation in the Type IIA brane configuration 
\cite{Hanany:1997gh,Brunner:1997gk}, see also Appendix \ref{app:orientifolds} 
and \cite{Danielsson:1997kt} for $6$d anomaly-free theories. This type of 
quiver gauge theory is denoted as \emph{electric theory} in the remainder of 
this paper.

In $p$-dimensional world-volume theories (with $8$ supercharges) originating 
from 
D$p$-D$(p{+}2)$-\NS\ brane configurations, the Higgs branch degrees of freedom 
are associated with freely moving D$p$ branes suspended between D$(p{+}2)$ 
branes. As such, the proposal of \cite{Cabrera:2019izd} is to employ this phase 
of 
the brane configuration to read off a \emph{magnetic quiver}, such that
\begin{align}
 \Higgs^{p\diff} \left( \text{electric theory} \right) = 
 \Coulomb^{3\diff} \left( \text{magnetic theory} \right)
\end{align}
holds as an \emph{equality of moduli spaces}.

Inspired from $3$d mirror symmetry, the magnetic degrees of freedom are 
associated to the suspension pattern of \Df\ branes  in the \Ds-\De-\NS\ 
configuration, because following the three T-dualities of the \Done\ branes in 
Type IIB precisely lead to \Df\ branes in Type IIA. A major difference in the 
magnetic quiver of the 
\Ds-\De-\NS\ system compared to the \Dthree-\Dfive-\NS\ system is the role 
played by the \NS\ branes. Since the \NS s and the \Ds s suspended between 
\De\ branes both share a $6$-dimensional world-volume, the \NS\ branes 
contribute as magnetic gauge degrees of freedom as opposed to flavour degrees 
of freedom.

The inclusion of \Os\ planes presents a major conceptual challenge in the 
derivation of the associated magnetic quivers. That is because 
\eqref{eq:3d_miror} for systems with \Ot\ planes involves S-duality of the 
orientifold $3$-planes, and there is no S-duality in Type IIA. 
To overcome this obstacle, one recalls the logic of \cite{Cabrera:2019izd} for 
A-type  
singularities (see also \cite{Hanany:2018uhm} for D-type): the magnetic quiver 
associated to the conventional electric quiver gauge theory in the finite 
coupling phase is essentially given by $3$d mirror symmetry, up to taking care 
of anomalous $\uo$ gauge nodes in transition $6$d to $3$d and back.
The point of \cite{Cabrera:2019izd} is to promote the Higgs branch phase or 
magnetic 
phase, i.e.\ D$p$ branes suspended between D$(p{+}2)$, and the associated 
quiver theories as valid moduli space description at \emph{any} value of the 
electric gauge coupling.
Therefore, inspired from $3$d mirror symmetry of orientifolds 
\cite{Feng:2000eq}, the proposed prescription to read off the magnetic quiver 
is as follows:
\begin{compactenum}[(i)]
    \item Change the brane system to the phase where as many \Ds\ branes are 
suspended between \De\ branes as possible.
    \item Change the physical (electric) orientifolds to virtual \emph{magnetic 
orientifolds}, which follow the logic of GNO or Langlands duality. These are 
summarised in Table \ref{tab:orientifold}.
\end{compactenum}

\begin{table}[t]
\centering
\begin{tabular}{c|c||c|c}
\toprule
\multirow{2}{*}{orientifold} & electric & magnetic & magnetic \\
  & group & orientifold & algebra \\ \midrule
 $O6^-$ & $\sorm(2n)$ & $O6^-$ &   $\dalg_n$ \\     
 $\widetilde{O6}^-$ & $\sorm(2n+1)$ & $O6^+$ &  $\calg_n$ \\
 $O6^+$ & $\usprm(2n)$ & $\widetilde{O6}^-$ & $\balg_n$\\
 $\widetilde{O6}^+$ & $\usprm'(2n)$ & $\widetilde{O6}^+$ & $\calg_n$
 \\ \bottomrule
 \end{tabular}
 \caption{The two left columns display low-energy gauge group of a stack 
of $n$ physical \Ds\ branes on top of an \Os\ plane, all suspended in between 
\NS\ branes. The two columns on the right-hand-side display the proposed 
magnetic orientifold to read off the magnetic gauge algebra of a stack of 
$n$ physical \Ds\ on top of an magnetic orientifold, all suspended 
between \De\ branes.}
 \label{tab:orientifold}
\end{table}

The main point of this paper is to extend the techniques of 
\cite{Cabrera:2019izd} to 
the study of $6$d $\Ncal=(1,0)$ Higgs branches originating from $n$ \Mf\ branes 
on a $D_k$ singularity $\C^2 \slash \D_{k-2}$.
\paragraph{Notation.}
In the remainder, the notation is adjusted to differentiate electric and 
magnetic 
quivers, as well as to accommodate for known subtleties. The gauge nodes in the 
relevant electric theories are denoted by $\sorm(2k)$ and $\sprm(k)$. For the 
magnetic quiver, only the gauge algebra are detailed, i.e.\ 
$\balg_k$, $\calg_k$, or $\dalg_k$. This is partly due to 
known issues about magnetic theories with orthogonal gauge groups. For example, 
in $T^\rho(G)$ theories \cite{Gaiotto:2008ak,Benini:2010uu,Cremonesi:2014uva} 
with $G$ of type $B$, $C$, or $D$, there exists several possible quivers for a 
single partition $\rho$. The corresponding Coulomb branches differ by 
projections of certain discrete groups and the correct identification of the 
required quotient is subtle \cite{Cabrera:2017ucb}.
%
%
\subsection{Single M5 on a D-type singularity}
\label{sec:single_M5}
Consider a single \Mf\ brane transverse to $\R\times \C^2 \slash \D_{k-2}$ for 
$k\geq 4$. For the dual Type IIA 
description, see Table \ref{tab:directions}, one recalls
\begin{align}
 \raisebox{-.5\height}{
\begin{tikzpicture}
		\DsixOMinus{$k$}{0}
        \DsixOPlus{$k{-}4$}{1}
		\DsixOMinus{$k$}{2}
		\o{1,0}{2,0}
		\ns{1,0}
		\ns{2,0}
	\end{tikzpicture}
	}
	\label{eq:1M5_electric_brane_system}
\end{align}
and the conventions on \Os\ planes are summarised in Appendix 
\ref{app:orientifolds}. 
The low-energy effective 6d $\Ncal=(1,0)$ 
theory\cite{Intriligator:1997kq,Blum:1997mm,Hanany:1997gh,Brunner:1997gk, 
Intriligator:1997dh,Brunner:1997gf,Ferrara:1998vf,DelZotto:2014hpa} contains a 
single tensor 
multiplet as well as hyper and vector multiplets encoded in the following 
electric quiver  
\begin{align}
 \raisebox{-.5\height}{
 	\begin{tikzpicture}
	\tikzstyle{gauge} = [circle, draw];
	\tikzstyle{flavour} = [regular polygon,regular polygon sides=4, draw];
	\node (g1) [gauge,label=below:{\Sp{k-4}}] {};
	\node (f1) [flavour,above of=g1,label=above:{\SO{4k}}] {};
	\draw (g1)--(f1);
	\end{tikzpicture}
	} 
	\label{eq:1M5_electric_quiver}
\end{align}
and the interest is placed on the moduli space of vacua.
For completeness, there exists one decoupled tensor multiplet, which can be 
neglected for the purposes of this paper.
Since there exists 
only one non-decoupled tensor multiplet, the interesting part of the tensor 
branch is effectively $\R_{\geq0}$, which 
exhibits a singularity at the origin. Therefore, the objective is to study two 
spaces:
\begin{compactenum}[(i)]
 \item The Higgs branch $\Higgs^{6\diff}_\fin$ of the theory over a generic 
point of the tensor branch, i.e.\ one tensor multiplet together with the 
gauge theory \eqref{eq:1M5_electric_quiver} at finite gauge coupling. 
 \item The Higgs branch $\Higgs^{6\diff}_\infty$ over the origin of the tensor 
branch, i.e.\ no tensor multiplets, but the quiver theory 
\eqref{eq:1M5_electric_quiver} at infinite coupling.
\end{compactenum}
Physically, whenever a gauge coupling diverges, certain BPS-strings become 
tensionless and contribute to the massless degrees of freedom. As, for 
instance, detailed in \cite{Hanany:2018uhm}, these originate from \Dtwo\ branes 
stretched between the half \NS\ branes in the brane configuration 
\eqref{eq:1M5_electric_brane_system}. Since the \Dtwo s are codimension 4 
objects for the \Ds\ branes, they are gauge instantons with corresponding 
zero-modes. The quantised zero-modes have been argued to finitely generate all 
massless degrees of freedom stemming from the tensionless BPS-strings.  
Consequently, there exists a natural inclusion of moduli spaces
\begin{align}
 \Higgs^{6\diff}_\fin \subset \Higgs^{6\diff}_\infty
 \label{eq:Higgs_Hasse-type_1M5}
\end{align}
because $\Higgs^{6\diff}_\infty$ is generated by all classical Higgs branch 
generators of $\Higgs^{6\diff}_\fin$ plus the additional generators for the 
massless string modes.

In this section, the transition between $\Higgs^{6\diff}_\fin$ and 
$\Higgs^{6\diff}_\infty$ as well as their geometry are derived from a brane 
construction.
\subsubsection{Minimal case \texorpdfstring{$k=4$}{k=4}}
\label{sec:1M5_k=4}
For $k=4$, the electric gauge theory \eqref{eq:1M5_electric_quiver} is 
trivial as well as the Higgs branch at finite gauge
coupling. As a warm up, one begins by studying how the trivial finite coupling 
phase manifests itself in the magnetic phase.
\paragraph{Finite coupling.}
From the view point of the Type IIA  brane system, the magnetic phase 
for \eqref{eq:1M5_electric_brane_system} is realised by, 
firstly, pulling in $8$ half \De\ from $x^6= \pm \infty$ 
\begin{align}
\begin{tikzpicture}
        \DeightMany{1}{0}{1}
        \Dbrane{0,0}{1,0}
		\DeightMany{1}{1}{1}
		\DsixOMinus{1}{1}
		\DeightMany{1}{2}{1}
        \DsixOMinusTilde{1}{2}
        \DeightMany{1}{3}{1}
        \DsixOMinus{2}{3}
        \DeightMany{1}{4}{1}
        \DsixOMinusTilde{2}{4}        
        \DeightMany{1}{5}{1}
        \DsixOMinus{3}{5}
        \DeightMany{1}{6}{1}
        \DsixOMinusTilde{3}{6}
        \DeightMany{1}{7}{1}
        \DsixOMinus{4}{7}
        \OPlus{8,0}{9,0}
        \DsixOMinus{4}{9}
        \ns{8,0}
        \ns{9,0}
        \DeightMany{1}{10}{1}
        \DsixOMinusTilde{3}{10}
        \DeightMany{1}{11}{1}
        \DsixOMinus{3}{11}        
        \DeightMany{1}{12}{1}
        \DsixOMinusTilde{2}{12}
        \DeightMany{1}{13}{1}
        \DsixOMinus{2}{13}
        \DeightMany{1}{14}{1}
        \DsixOMinusTilde{1}{14}
        \DeightMany{1}{15}{1}
		\DsixOMinus{1}{15}
		\DeightMany{1}{16}{1}
        \Dbrane{16,0}{17,0}
        \DeightMany{1}{17}{1}
\draw[decoration={brace,mirror,raise=10pt},decorate,thick]
  (-0.2,-1) -- node[below=10pt] {$8$ half \De } (7.2,-1);
\draw[decoration={brace,mirror,raise=10pt},decorate,thick]
  (10-0.2,-1) -- node[below=10pt] {$8$ half \De } (17.2,-1);
	\end{tikzpicture}
	\label{eq:k=4_D8_pulled_in}
\end{align}
such that the configuration obeys the S-rule.
The only \Ds\ branes present are the ones suspended between \De\ and \NS\ 
branes; hence, they are \emph{frozen} in the sense that the boundary conditions 
do not allow any degrees of freedom. One can eliminate these frozen branes by 
moving the half \NS\ branes through the half \De\ branes, taking care of brane 
anniliation, see \eqref{eq:brane_creation_all}. Consequently, the second step 
to reach the magnetic phase becomes
\begin{align}
\begin{tikzpicture}
        \DeightMany{1}{0}{1}
        \Dbrane{0,0}{0.5,0}
        \OPlusTilde{0.5,0}{1,0}
        \ns{0.5,0}
		\DeightMany{1}{1}{1}
        \OPlus{1,0}{2,0}
		\DeightMany{1}{2}{1}
        \OPlusTilde{2,0}{3,0}
        \DeightMany{1}{3}{1}
        \OPlus{3,0}{4,0}
        \DeightMany{1}{4}{1}
        \OPlusTilde{4,0}{5,0}
        \DeightMany{1}{5}{1}
        \OPlus{5,0}{6,0}
        \DeightMany{1}{6}{1}
        \OPlusTilde{6,0}{7,0}
        \DeightMany{1}{7}{1}
        \OPlus{7,0}{10,0}
        \DeightMany{1}{10}{1}
        \OPlusTilde{10,0}{11,0}
        \DeightMany{1}{11}{1}
        \OPlus{11,0}{12,0}
        \DeightMany{1}{12}{1}
        \OPlusTilde{12,0}{13,0}
        \DeightMany{1}{13}{1}
        \OPlus{13,0}{14,0}
        \DeightMany{1}{14}{1}
        \OPlusTilde{14,0}{15,0}
        \DeightMany{1}{15}{1}
        \OPlus{15,0}{16,0}
		\DeightMany{1}{16}{1}
        \OPlusTilde{16,0}{16.5,0}
        \Dbrane{16.5,0}{17,0}
        \ns{16.5,0}
        \DeightMany{1}{17}{1}
\draw[decoration={brace,mirror,raise=10pt},decorate,thick]
  (-0.2,-1) -- node[below=10pt] {$8$ half \De } (7.2,-1);
\draw[decoration={brace,mirror,raise=10pt},decorate,thick]
  (10-0.2,-1) -- node[below=10pt] {$8$ half \De } (17.2,-1);
	\end{tikzpicture}
	\label{eq:k=4_finite_coupling}
\end{align}
and there are no \Ds\ branes left. This simply reflects that the Higgs branch 
is trivial. Nonetheless, being very explicit, the magnetic quiver is read off 
by converting the physical \Os\ planes into \emph{magnetic orientifolds}, 
and then assigning a magnetic gauge node for $n$ \Ds\ branes on top of a \Os\ 
plane suspended between half \De\ branes, as in Table \ref{tab:orientifold}. 
The logic is as in \cite{Cabrera:2019izd}, the motion of the \Ds\ in transverse 
$x^{7,8,9}$ direction is identified with magnetic vector multiplet 
contributions due to \Df\ branes suspended between the \Ds\ branes.
Here, the gauge part of the magnetic quiver becomes trivial as there are 
no \Ds s to begin with, but one can write
\begin{align}
 \raisebox{-.5\height}{
 	\begin{tikzpicture}
 	\tikzset{node distance = 0.5cm};
	\tikzstyle{gauge} = [circle, draw,inner sep=2.5pt];
	\tikzstyle{flavour} = [regular polygon,regular polygon sides=4,inner 
sep=2.5pt, draw];
	\node (g1) [gauge,label={[rotate=-45]below right:{\cc{0}}}] {};
	\node (g2) [gauge,right of=g1,label={[rotate=-45]below right:{\bb{0}}}] 
{};
	\node (g3) [gauge,right of=g2,label={[rotate=-45]below right:{\cc{0}}}] {};
	\node (g4) [gauge,right of=g3,label={[rotate=-45]below 
right:{\bb{0}}}] {};
	\node (g5) [gauge,right of=g4,label={[rotate=-45]below right:{\cc{0}}}] {};
    \node (g6) [gauge,right of=g5,label={[rotate=-45]below 
right:{\bb{0}}}] {};
	\node (g7) [gauge,right of=g6,label={[rotate=-45]below right:{\cc{0}}}] {};
    \node (g8) [gauge,right of=g7,label={[rotate=-45]below right:{\bb{0}}}] {};
	\node (g9) [gauge,right of=g8,label={[rotate=-45]below right:{\cc{0}}}] {};
    \node (g10) [gauge,right of=g9,label={[rotate=-45]below right:{\bb{0}}}] {};
	\node (g11) [gauge,right of=g10,label={[rotate=-45]below right:{\cc{0}}}] 
{};
    \node (g12) [gauge,right of=g11,label={[rotate=-45]below right:{\bb{0}}}] 
{};
	\node (g13) [gauge,right of=g12,label={[rotate=-45]below right:{\cc{0}}}] 
{};
	\node (g14) [gauge,right of=g13,label={[rotate=-45]below right:{\bb{0}}}] 
{};
	\node (g15) [gauge,right of=g14,label={[rotate=-45]below right:{\cc{0}}}] 
{};
    \node (f1) [flavour,above of=g1, label=left:{\bb{0}}] {};
    \node (f15) [flavour,above of=g15, label=right:{\bb{0}}] {};
	\draw (g1)--(g2) (g2)--(g3) (g3)--(g4) (g4)--(g5) (g5)--(g6) (g6)--(g7) 
(g7)--(g8) (g8)--(g9) (g9)--(g10) (g10)--(g11) (g11)--(g12) (g12)--(g13)
(g13)--(g14) (g14)--(g15) (g1)--(f1) (g15)--(f15);
	\end{tikzpicture}
	} 
	\,.
	\label{eq:magQuiver_k=4_finite}
\end{align}
Besides the trivial gauge nodes, the quiver \eqref{eq:magQuiver_k=4_finite} 
displays two flavour nodes. These originate from the two half \NS\ branes which 
are stuck on the orientifold plane. Since these are not free to move in 
transverse $x^{7,8,9}$ direction, one does not associate any magnetic gauge 
degrees of freedom with them. 
As a consequence, the Coulomb branch of the magnetic quiver 
\eqref{eq:magQuiver_k=4_finite} is trivial
\begin{align}
 \Coulomb^{3\diff}\left(\substack{\text{magnetic} \\ \text{quiver}} 
\eqref{eq:magQuiver_k=4_finite} \right) 
= \{0\} =
\Higgs^{6\diff}_{\fin} \left( \elecTheory \eqref{eq:1M5_electric_quiver}|_{k=4} 
\right)
\;,
\end{align}
which agrees with the Higgs branch of \eqref{eq:1M5_electric_quiver} for $k=4$ 
at finite coupling. 
\paragraph{Infinite coupling.}
Proceeding to infinite gauge coupling, two effects are expected 
\cite{Hanany:2018uhm} to happen in order to fit the result of 
\cite{Ganor:1996mu}. These are the following:
\begin{compactenum}[(i)]
 \item The $\sorm(16)$ flavour symmetry enhances to $E_8$.
 \item The Higgs branch $\Higgs_\infty^{6\diff}$ is the minimal (non-trivial) 
hyper-K\"{a}hler cone with an $E_8$ symmetry, i.e.\ the closure of the minimal 
nilpotent orbit of $E_8$.
\end{compactenum}
Now, how to reconcile these features from the magnetic quiver approach?
Tuning the gauge coupling of \eqref{eq:1M5_electric_quiver} to infinity means 
that the two half \NS\ branes in \eqref{eq:k=4_D8_pulled_in} need to become 
coincident along the $x^6$ direction. Since the number of \Ds\ branes on the 
left-hand-side of the left \NS\ coincides with the number of \Ds\ branes on the 
right-hand-side of the right \NS\, due to charge conservation, the pair of \NS\ 
branes can leave the orientifold plane while the \Ds\ brane reconnect 
simultaneously. As the \Ds s are solely suspended between \De\ branes, they 
become free to move in transverse $x^{7,8,9}$ direction.
Hence, the magnetic brane configuration for the infinite gauge coupling 
phase is
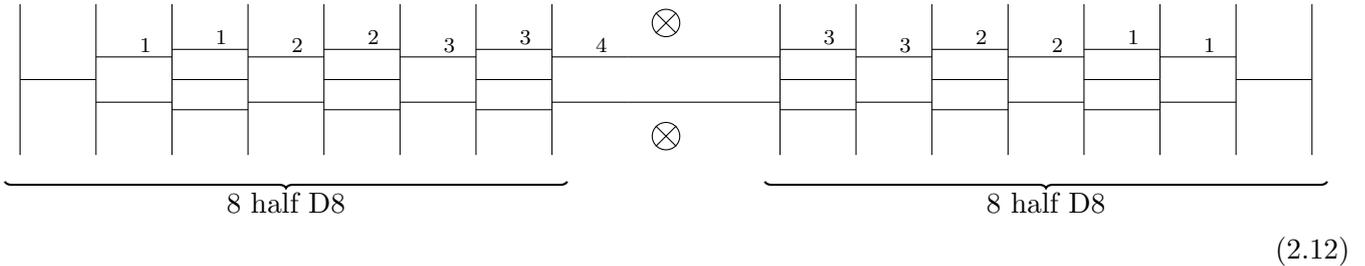
\begin{align}
\begin{tikzpicture}
        \DeightMany{1}{0}{1}
        \Dbrane{0,0}{1,0}
		\DeightMany{1}{1}{1}
		\DsixOMinusFree{1}{1}{0}
		\DeightMany{1}{2}{1}
        \DsixOMinusTildeFree{1}{2}{1}
        \DeightMany{1}{3}{1}
        \DsixOMinusFree{2}{3}{0}
        \DeightMany{1}{4}{1}
        \DsixOMinusTildeFree{2}{4}{1}        
        \DeightMany{1}{5}{1}
        \DsixOMinusFree{3}{5}{0}
        \DeightMany{1}{6}{1}
        \DsixOMinusTildeFree{3}{6}{1}
        \DeightMany{1}{7}{1}
        \DsixOMinusFree{4}{7}{0}
        \Dbrane{8,0.3}{10,0.3}
        \Dbrane{8,-0.3}{10,-0.3}
        \ns{8.5,0.75}
        \ns{8.5,-0.75}
        \DeightMany{1}{10}{1}
        \DsixOMinusTildeFree{3}{10}{1}
        \DeightMany{1}{11}{1}
        \DsixOMinusFree{3}{11}{0}        
        \DeightMany{1}{12}{1}
        \DsixOMinusTildeFree{2}{12}{1}
        \DeightMany{1}{13}{1}
        \DsixOMinusFree{2}{13}{0}
        \DeightMany{1}{14}{1}
        \DsixOMinusTildeFree{1}{14}{1}
        \DeightMany{1}{15}{1}
		\DsixOMinusFree{1}{15}{0}
		\DeightMany{1}{16}{1}
        \Dbrane{16,0}{17,0}
        \DeightMany{1}{17}{1}
\draw[decoration={brace,mirror,raise=10pt},decorate,thick]
  (-0.2,-1) -- node[below=10pt] {$8$ half \De } (7.2,-1);
\draw[decoration={brace,mirror,raise=10pt},decorate,thick]
  (10-0.2,-1) -- node[below=10pt] {$8$ half \De } (17.2,-1);
	\end{tikzpicture}
	\label{eq:k=4_infinite_coupling}
\end{align}
where there are clearly non-frozen \Ds\ branes. This already indicates that the 
Higgs branch of this phase is non-trivial.

To read off the magnetic quiver, one associates to a stack of \Ds\ branes in 
between \De\ branes a magnetic vector multiplet depending on which type of 
orientifold is present, see Table \ref{tab:orientifold}. These magnetic 
degrees of freedom are associated with the way virtual \Df\ branes are 
suspended. The action of the \emph{magnetic orientifold} projects out certain 
states, much like it does for \Done\ or F1 suspended between \Dthree\ on top of 
\Ot\ planes. 
Since the half \NS\ branes have been lifted from the orientifold 
they are now free to move in the transverse $x^{7,8,9} $ direction. To this 
motion one associates a magnetic vector multiplet, much like the reason for the 
magnetic vector multiplet coming from the transverse motion of the \Ds\ branes. 
To see the character of the gauge group one can suspend virtual \Df\ branes 
between a half \NS\ and its mirror image. Then the magnetic orientifold of an 
\Osm\ is again an \Osm\ such that the corresponding gauge group is of 
symplectic nature.   
In addition, one can suspend \Df\ branes between the pair of half \NS\ branes 
and the \Ds\ on top of the orientifold. This yields magnetic half 
hypermultiplets between the symplectic gauge group from the \NS\ branes and the 
orthogonal gauge group from the \Ds\ on top of the \Os\ plane.
Consequently, the resulting magnetic quiver is
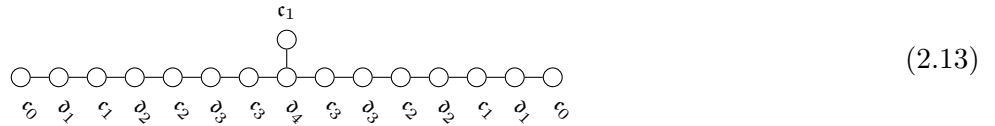
\begin{align}
 \raisebox{-.5\height}{
 	\begin{tikzpicture}
 	\tikzset{node distance = 0.5cm};
	\tikzstyle{gauge} = [circle, draw,inner sep=2.5pt];
	\tikzstyle{flavour} = [regular polygon,regular polygon sides=4,inner 
sep=2.5pt, draw];
	\node (g1) [gauge,label={[rotate=-45]below right:{\cc{0}}}] {};
	\node (g2) [gauge,right of=g1,label={[rotate=-45]below right:{\dd{1}}}] 
{};
	\node (g3) [gauge,right of=g2,label={[rotate=-45]below right:{\cc{1}}}] {};
	\node (g4) [gauge,right of=g3,label={[rotate=-45]below 
right:{\dd{2}}}] {};
	\node (g5) [gauge,right of=g4,label={[rotate=-45]below right:{\cc{2}}}] {};
    \node (g6) [gauge,right of=g5,label={[rotate=-45]below 
right:{\dd{3}}}] {};
	\node (g7) [gauge,right of=g6,label={[rotate=-45]below right:{\cc{3}}}] {};
    \node (g8) [gauge,right of=g7,label={[rotate=-45]below right:{\dd{4}}}] {};
	\node (g9) [gauge,right of=g8,label={[rotate=-45]below right:{\cc{3}}}] {};
    \node (g10) [gauge,right of=g9,label={[rotate=-45]below right:{\dd{3}}}] {};
	\node (g11) [gauge,right of=g10,label={[rotate=-45]below right:{\cc{2}}}] 
{};
    \node (g12) [gauge,right of=g11,label={[rotate=-45]below right:{\dd{2}}}] 
{};
	\node (g13) [gauge,right of=g12,label={[rotate=-45]below right:{\cc{1}}}] 
{};
	\node (g14) [gauge,right of=g13,label={[rotate=-45]below right:{\dd{1}}}] 
{};
	\node (g15) [gauge,right of=g14,label={[rotate=-45]below right:{\cc{0}}}] 
{};
	\node (g0) [gauge,above of=g8,label=above:{\cc{1}}] {};
	\draw (g1)--(g2) (g2)--(g3) (g3)--(g4) (g4)--(g5) (g5)--(g6) (g6)--(g7) 
(g7)--(g8) (g8)--(g9) (g9)--(g10) (g10)--(g11) (g11)--(g12) (g12)--(g13)
(g13)--(g14) (g14)--(g15) (g8)--(g0) ;
	\end{tikzpicture}
	}
\end{align}
and dropping the empty gauge nodes yields
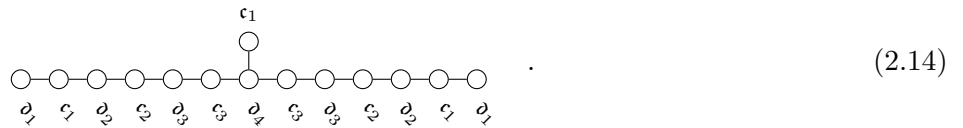
\begin{align}
 \raisebox{-.5\height}{
 	\begin{tikzpicture}
 	\tikzset{node distance = 0.5cm};
	\tikzstyle{gauge} = [circle, draw,inner sep=2.5pt];
	\tikzstyle{flavour} = [regular polygon,regular polygon sides=4,inner 
sep=2.5pt, draw];
	\node (g2) [gauge,label={[rotate=-45]below right:{\dd{1}}}] 
{};
	\node (g3) [gauge,right of=g2,label={[rotate=-45]below right:{\cc{1}}}] {};
	\node (g4) [gauge,right of=g3,label={[rotate=-45]below right:{\dd{2}}}] {};
	\node (g5) [gauge,right of=g4,label={[rotate=-45]below right:{\cc{2}}}] {};
    \node (g6) [gauge,right of=g5,label={[rotate=-45]below right:{\dd{3}}}] {};
	\node (g7) [gauge,right of=g6,label={[rotate=-45]below right:{\cc{3}}}] {};
    \node (g8) [gauge,right of=g7,label={[rotate=-45]below right:{\dd{4}}}] {};
	\node (g9) [gauge,right of=g8,label={[rotate=-45]below right:{\cc{3}}}] {};
    \node (g10) [gauge,right of=g9,label={[rotate=-45]below right:{\dd{3}}}] {};
	\node (g11) [gauge,right of=g10,label={[rotate=-45]below right:{\cc{2}}}] 
{};
    \node (g12) [gauge,right of=g11,label={[rotate=-45]below right:{\dd{2}}}] 
{};
	\node (g13) [gauge,right of=g12,label={[rotate=-45]below right:{\cc{1}}}] 
{};
	\node (g14) [gauge,right of=g13,label={[rotate=-45]below right:{\dd{1}}}] 
{};
	\node (g0) [gauge,above of=g8,label=above:{\cc{1}}] {};
	\draw  (g2)--(g3) (g3)--(g4) (g4)--(g5) (g5)--(g6) (g6)--(g7) 
(g7)--(g8) (g8)--(g9) (g9)--(g10) (g10)--(g11) (g11)--(g12) (g12)--(g13)
(g13)--(g14)  (g8)--(g0) ;
	\end{tikzpicture}
	} 
	\,.
	\label{eq:magQuiver_k=4_infinite}
\end{align}
The Coulomb branch dimension and the (naive) symmetry (see Appendix 
\ref{app:global_sym}) can be computed to be
\begin{align}
 \dim_\HH\ \Coulomb^{3d} \left(\magQuivText \eqref{eq:magQuiver_k=4_infinite}   
\right) &=2 \cdot \sum_{i=1}^{3} \left( \dim\ \dalg_i +\dim\ \calg_i 
\right) + \dim\ \dalg_4 + \dim\ \calg_1 = 29 \,, \\
G_J &= \sorm(16) \,.
\end{align}
In fact, more is true because \eqref{eq:magQuiver_k=4_infinite} is a 
star-shaped quiver constructed by gluing $T_{(1^8)}[\sorm(8)]$, 
$T_{(1^8)}[\sorm(8)]$, and $T_{(5,3)}[\sorm(8)]$ along the common flavour node. 
As such it is the mirror of the $S^1$ compactification of an class $\mathcal{S}$ 
theory of type $\sorm(8)$ with punctures $(1^8)$, $(1^8)$, $(5,3)$, which is 
known to be a rank-1 $E_8$ SCFT \cite[Sec.\ 3.2.2]{Chacaltana:2011ze}. 
Therefore, as concluded in \cite[Eq.\ (2.43)]{Hanany:2018uhm} and 
\cite{Zhenghao} the Coulomb 
branch 
of \eqref{eq:magQuiver_k=4_infinite} is the closure of the minimal nilpotent 
orbit of $E_8$, i.e.
\begin{align}
 \Coulomb^{3\diff} \left( \magQuivText \eqref{eq:magQuiver_k=4_infinite}  
\right) 
 = 
\clorbit{\mathrm{min}}^{E_8} 
= 
\Higgs_{\infty}^{6\diff} \left( \elecTheory 
\eqref{eq:1M5_electric_quiver}|_{k=4} \right)
\,.
\end{align}
The novel point here is that the brane construction 
\eqref{eq:k=4_infinite_coupling} allows to derive 
the correct magnetic quiver that describes $\Higgs_{\infty}^{6\diff}$.

The change in dimension of the 
Higgs branch from finite to infinite coupling follows straightforwardly from 
the anomaly cancellation condition \eqref{eq:anomaly_cancellation}, 
as discussed in \cite{Hanany:2018uhm}. At finite 
coupling, there are no hyper and vector multiplets, but only one tensor 
multiplet (ignoring the decoupled tensor multiplet). At 
infinite coupling, the tensor multiplet is lost and needs to be compensated by 
$29$ (additional) hypermultiplets, since there are no new gauge degrees of 
freedom.

In terms of geometry, the transition from \eqref{eq:magQuiver_k=4_finite} to 
\eqref{eq:magQuiver_k=4_infinite} is a simple case of a transverse slice 
for \eqref{eq:Higgs_Hasse-type_1M5}, in the sense that locally one may write
\begin{align}
 \Higgs_\infty^{6\diff} = \Higgs_{\fin}^{6\diff} \times \slice{}  = 
 \{0\} \times \slice{} \cong \slice{} = \clorbit{\mathrm{min}}^{E_8} \,.
 \label{eq:transverse_slice_k=4}
\end{align}
From the associated magnetic quivers \eqref{eq:magQuiver_k=4_finite} and 
\eqref{eq:magQuiver_k=4_infinite}, this statement can be deduced by 
\emph{quiver subtraction} as detailed in \cite{Cabrera:2018ann}, see also 
\eqref{eq:quiver_subtraction_1M5} below. 
For $k>4$, the relation \eqref{eq:transverse_slice_k=4} becomes more 
complicated, as $\Higgs_{\fin}^{6\diff}$ is non-trivial.
%
%
\subsubsection{Generic case \texorpdfstring{$k>4$}{k4}}
\label{sec:1M5_generic_k}
For $k>4$, the electric theory \eqref{eq:1M5_electric_quiver} as well as the 
Higgs branch at finite coupling are non-trivial. Hence, the first task is to 
derive the magnetic quiver for the finite gauge coupling phase.
\paragraph{Finite coupling.}
Starting from \eqref{eq:1M5_electric_brane_system} one moves to the phase of 
the brane system where all \Ds\ are suspended 
between \De ; therefore, from $x^6 = \pm \infty$ one pulls in 
$2k$ half \De\ each.
\begin{align}
 \raisebox{-.5\height}{
\begin{tikzpicture}
        \DeightMany{1}{0}{1}
        \Dbrane{0,0}{1,0}
		\DeightMany{1}{1}{1}
		\DsixOMinus{$1$}{1}
		\DeightMany{1}{2}{1}
        \DsixOMinusTilde{$1$}{2}
        \DeightMany{1}{3}{1}
        \draw (3.5,0) node {$\cdots$};
        \DeightMany{1}{4}{1}
        \DsixOMinus{$k{-}1$}{4}        
        \DeightMany{1}{5}{1}
        \DsixOMinusTilde{$k{-}1$}{5}
        \DeightMany{1}{6}{1}
        \DsixOMinus{$k$}{6}
        \DsixOPlus{$k{-}4$}{7}
        \DsixOMinus{$k$}{8}
        \o{7,0}{8,0}
        \ns{7,0}
        \ns{8,0}
        \DeightMany{1}{9}{1}
        \DsixOMinusTilde{$k{-}1$}{9}
        \DeightMany{1}{10}{1}
        \DsixOMinus{$k{-}1$}{10}        
        \DeightMany{1}{11}{1}
        \draw (11.5,0) node {$\cdots$};
        \DeightMany{1}{12}{1}
        \DsixOMinusTilde{$1$}{12}
        \DeightMany{1}{13}{1}
		\DsixOMinus{$1$}{13}
		\DeightMany{1}{14}{1}
        \Dbrane{14,0}{15,0}
        \DeightMany{1}{15}{1}
\draw[decoration={brace,mirror,raise=10pt},decorate,thick]
  (-0.2,-1) -- node[below=10pt] {$2k$ half \De } (6.2,-1);
\draw[decoration={brace,mirror,raise=10pt},decorate,thick]
  (9-0.2,-1) -- node[below=10pt] {$2k$ half \De } (15.2,-1);
\draw[very thick,dotted,red] (7.5,0.25)--(7.5,-1.25);
\draw (7.5,-1.5) node {\tiny{symmetry axis} };
	\end{tikzpicture}
	}
	\label{eq:branes_1M5_on_Dk_pulled_D8}
\end{align}
To reach the phase of finite gauge coupling, the \Ds\ branes have to be 
suspended 
purely between \De\ branes without the half \NS s leaving the orientifold plane.
Since the set-up is symmetric along the $x^6$ direction (see red dotted 
symmetry axis), one can focus on the 
left-hand-side without loss of generality.
Inspecting the left-hand-side of \eqref{eq:branes_1M5_on_Dk_pulled_D8}, there 
are $k$ full \Ds\ on the left and $(k{-}4)$ full \Ds\ on the right of the 
left-most half \NS\ brane. Therefore, one may consider $(k{-}4)$ \Ds\ branes 
going through the half \NS\ branes such that these \Ds s are really suspended 
between half \De\ branes. The remaining $4$ full \Ds\ on the left-hand-side are 
considered as frozen between \De\ branes and the half \NS\ brane. These frozen 
\Ds\ can be eliminated, without changing the physics, by passing the half \NS\ 
through seven half \De\ branes and taking care of brane anniliation 
\eqref{eq:brane_creation_all}.
One may wonder whether to cross an additional eighth \De\ brane, but it turns 
out that after conversion to magnetic orientifolds the correct brane 
configuration for the purpose of reading the magnetic quiver becomes 
\begin{align}
 \raisebox{-.5\height}{
\begin{tikzpicture}
        \DeightMany{1}{0}{1}
        \Dbrane{0,0}{1,0}
		\DeightMany{1}{1}{1}
		\DsixOMinusFree{$1$}{1}{0}
		\DeightMany{1}{2}{1}
        \DsixOMinusTildeFree{$1$}{2}{1}
        \DeightMany{1}{3}{1}
        \draw (3.5,0) node {$\cdots$};
        \DeightMany{1}{4}{1}
        \DsixOMinusFree{$k{-}4$}{4}{0}
        \DeightMany{1}{5}{1}
        \DsixOMinusFree{$k{-}4$}{5}{1} 
        \Dbrane{5.0,0}{5.5,0}
        \oAux{5.5,0}{6,0}
        \DeightMany{1}{6}{1}
        \DsixOPlusFree{$k{-}4$}{6}{0}
        \DeightMany{1}{7}{1}
        \DsixOPlusTildeFree{$k{-}4$}{7}{1}
        \DeightMany{1}{8}{1}
        \DsixOPlusFree{$k{-}4$}{8}{0}
        \DeightMany{1}{9}{1}
        \DsixOPlusTildeFree{$k{-}4$}{9}{1}
        \DeightMany{1}{10}{1}
        \DsixOPlusFree{$k{-}4$}{10}{0}
        \DeightMany{1}{11}{1}
        \DsixOPlusTildeFree{$k{-}4$}{11}{1}
        \DeightMany{1}{12}{1}
        \DsixOPlusFree{$k{-}4$}{12}{0}
        \draw (13.5,0) node {$\cdots$};
        \ns{5.5,0}
        \draw[very thick,dotted,red] (13,0.75)--(13,-0.75);
	\end{tikzpicture}
	}
	\label{eq:moving_NS}
\end{align}
and all the \Ds\ branes are suspended between \De\ branes.
The half \NS\ has no \Ds\ branes ending on it and is, moreover, stuck on the 
orientifold. Employing the same arguments as in the previous section --- 
reading off gauge nodes from stacks of \Ds\ branes on \emph{magnetic 
orientifolds} of Table \ref{tab:orientifold} and so forth --- one deduces the 
following magnetic quiver:
\begin{align}
 \raisebox{-.5\height}{
 	\begin{tikzpicture}
 	\tikzset{node distance = 0.5cm};
	\tikzstyle{gauge} = [circle, draw,inner sep=2.5pt];
	\tikzstyle{flavour} = [regular polygon,regular polygon sides=4,inner 
sep=2.5pt, draw];
	\node (g2) [gauge,label={[rotate=-45]below right:{\dd{1}}}] 
{};
	\node (g3) [gauge,right of=g2,label={[rotate=-45]below right:{\cc{1}}}] {};
	\node (g4) [right of=g3] {$\scriptstyle{\ldots}$};
	\node (g5) [gauge,right of=g4,label={[rotate=-45]below 
right:{\dd{k{-}5}}}] {};
	\node (g6) [gauge,right of=g5,label={[rotate=-45]below right:{\cc{k{-}5}}}] 
{};
    \node (g7) [gauge,right of=g6,label={[rotate=-45]below 
right:{\dd{k{-}4}}}] {};
	\node (g8) [gauge,right of=g7,label={[rotate=-45]below right:{\cc{k{-}4}}}] 
{};
    \node (g9) [gauge,right of=g8,label={[rotate=-45]below 
right:{\bb{k{-}4}}}] {};
	\node (g10) [gauge,right of=g9,label={[rotate=-45]below 
right:{\cc{k{-}4}}}] 
{};
    \node (g11) [right of=g10] {$\scriptstyle{\ldots}$};
    \node (g12) [gauge,right of=g11,label={[rotate=-45]below 
right:{\bb{k{-}4}}}] {};
	\node (g13) [gauge,right of=g12,label={[rotate=-45]below 
right:{\cc{k{-}4}}}] 
{};
    \node (g14) [gauge,right of=g13,label={[rotate=-45]below 
right:{\dd{k{-}4}}}] {};
	\node (g15) [gauge,right of=g14,label={[rotate=-45]below 
right:{\cc{k{-}5}}}] {};
	\node (g16) [gauge,right of=g15,label={[rotate=-45]below 
right:{\dd{k{-}5}}}] {};
	\node (g17) [right of=g16] {$\scriptstyle{\ldots}$};
	\node (g18) [gauge,right of=g17,label={[rotate=-45]below right:{\cc{1}}}] 
{};
	\node (g19) [gauge,right of=g18,label={[rotate=-45]below right:{\dd{1}}}] 
{};
	\node (f8) [flavour,above of=g8,label=above:{\bb{0}}] {};
	\node (f13) [flavour,above of=g13,label=above:{\bb{0}}] {};
	\draw (g2)--(g3) (g3)--(g4) (g4)--(g5) (g5)--(g6) (g6)--(g7) 
(g7)--(g8) (g8)--(g9) (g9)--(g10) (g10)--(g11) (g11)--(g12) (g12)--(g13)
(g13)--(g14) (g14)--(g15) (g15)--(g16) (g16)--(g17) (g17)--(g18) (g18)--(g19) 
(g8)--(f8) (g13)--(f13);
\draw[decoration={brace,mirror,raise=10pt},decorate,thick]
  (3.5-0.4,-0.6) -- node[below=10pt] {\tiny{$7$} \bb{k{-4}} \tiny{\& $8$} 
\cc{k{-}4} } (6,-0.6);
	\end{tikzpicture}
	} 
	\,,
	\label{eq:k>4_magQuiver_finite}
\end{align}
and potentially empty gauge nodes from \De\ intervals with no \Ds\ branes have 
been omitted. 
The dimension and symmetry of the Coulomb branch are computed to be
\begin{subequations}
\begin{align}
 \dim_\HH\ \Coulomb^{3\diff} \left( \magQuivText \eqref{eq:k>4_magQuiver_finite} 
\right) &= 
2\cdot \sum_{i=1}^{k-4} \left( \dim\ \dalg_i + \dim\ \calg_i \right) + 
7\cdot \dim\ \balg_{k-4} + 6 \cdot \dim\ \calg_{k-4} \notag \\
&= \dim\ \sorm(2k) -\dim\ \sorm(8) \,,
\\
G_J&= \sorm(4k) \,,
\end{align}
\end{subequations}
due to a chain of $(4k{-}3)$ balanced nodes with $\dalg_1$ nodes at each 
end, 
see Appendix \ref{app:global_sym}. These 
properties match the classical Higgs branch of 
\eqref{eq:1M5_electric_quiver}. Hence, the significance of 
\eqref{eq:k>4_magQuiver_infinite} lies in
\begin{align}
 \Coulomb^{3\diff}\left( \magQuivText \eqref{eq:k>4_magQuiver_finite} \right)
 =
 \Higgs^{6\diff}_{\fin} \left(\elecTheory \eqref{eq:1M5_electric_quiver} \right)
 \,,
\end{align}
which can also be derived by taking \eqref{eq:1M5_electric_quiver} as a $3$d 
$\Ncal=4$ system and computing the $3$d mirror, as shown in \cite[Fig.\ 
13]{Feng:2000eq}. While the quiver \eqref{eq:k>4_magQuiver_finite} has been 
conjectured in \cite{Hanany:2018uhm}; here, the magnetic quiver has been 
\emph{derived} from a \Ds-\De-\NS\ brane construction with \Os\ planes.
\paragraph{Infinite coupling.}
Having established the magnetic phase for the finite coupling regime, one can 
proceed to infinite gauge coupling. Physically, infinite gauge coupling means 
that the two half \NS\ in \eqref{eq:1M5_electric_brane_system} or 
\eqref{eq:branes_1M5_on_Dk_pulled_D8} have vanishing distance along $x^6$. 
However, when two half \NS\ branes are coincident on an \Os\ plane, they can 
leave the orientifold in transverse  $x^{7,8,9}$ direction as mirror pair of 
half \NS\ 
branes. The $(k{-}4)$ full \Ds\ branes that had originally been suspended 
between the two half \NS\ branes disappeared, and there are no \Ds\ branes 
attached 
between the pair of half \NS\ outside the \Os\ plane. However, the $k$ 
full \Ds\ branes\footnote{Due to charge conservation, the numbers of \Ds\ 
branes on the left and right of a pair of half \NS\ branes have to be equal.} 
that were attached from the left and right side of the 
pair of \NS\ branes can reconnect while the half \NS s leave the orientifold.  
Therefore, the brane configuration describing the infinite gauge coupling 
phase is reached by reuniting the two half \NS s such that they can leave the 
orientifold as pair of half \NS , i.e.\
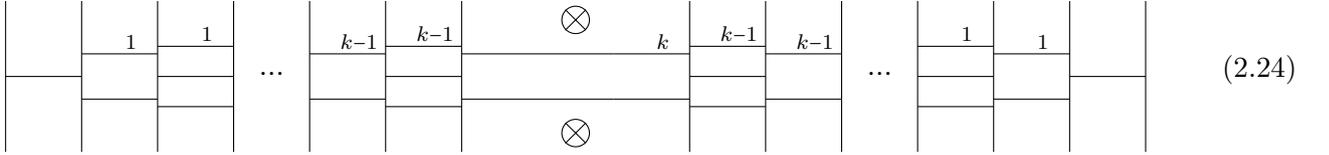
\begin{align}
 \raisebox{-.5\height}{
\begin{tikzpicture}
        \DeightMany{1}{0}{1}
        \Dbrane{0,0}{1,0}
		\DeightMany{1}{1}{1}
		\DsixOMinusFree{$1$}{1}{0}
		\DeightMany{1}{2}{1}
        \DsixOMinusTildeFree{$1$}{2}{1}
        \DeightMany{1}{3}{1}
        \draw (3.5,0) node {$\cdots$};
        \DeightMany{1}{4}{1}
        \DsixOMinusFree{$k{-}1$}{4}{0}        
        \DeightMany{1}{5}{1}
        \DsixOMinusTildeFree{$k{-}1$}{5}{1}
        \DeightMany{1}{6}{1}
        \Dbrane{6,0.3}{8,0.3}
        \Dbrane{6,-0.3}{8,-0.3}
        \DsixOMinusFree{$k$}{8}{0}
        \ns{7.5,0.75}
        \ns{7.5,-0.75}
        \DeightMany{1}{9}{1}
        \DsixOMinusTildeFree{$k{-}1$}{9}{1}
        \DeightMany{1}{10}{1}
        \DsixOMinusFree{$k{-}1$}{10}{0}        
        \DeightMany{1}{11}{1}
        \draw (11.5,0) node {$\cdots$};
        \DeightMany{1}{12}{1}
        \DsixOMinusTildeFree{$1$}{12}{1}
        \DeightMany{1}{13}{1}
		\DsixOMinusFree{$1$}{13}{0}
		\DeightMany{1}{14}{1}
        \Dbrane{14,0}{15,0}
        \DeightMany{1}{15}{1}
	\end{tikzpicture}
	}
	\label{eq:branes_1M5_on_Dk_infinite}
\end{align}
and the magnetic quiver is read off by using the orientifold conversion to 
\emph{magnetic orientifolds}, cf.\ Table \ref{tab:orientifold}, to be
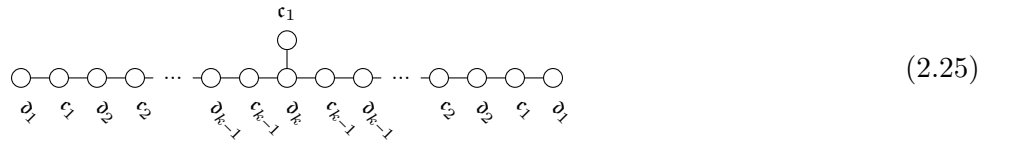
\begin{align}
 \raisebox{-.5\height}{
 	\begin{tikzpicture}
 	\tikzset{node distance = 0.5cm};
	\tikzstyle{gauge} = [circle, draw,inner sep=2.5pt];
	\tikzstyle{flavour} = [regular polygon,regular polygon sides=4,inner 
sep=2.5pt, draw];
	\node (g2) [gauge,label={[rotate=-45]below right:{\dd{1}}}] 
{};
	\node (g3) [gauge,right of=g2,label={[rotate=-45]below right:{\cc{1}}}] {};
	\node (g4) [gauge,right of=g3,label={[rotate=-45]below 
right:{\dd{2}}}] {};
	\node (g5) [gauge,right of=g4,label={[rotate=-45]below right:{\cc{2}}}] 
{};
	\node (g6) [right of=g5] {$\scriptstyle{\ldots}$};
    \node (g7) [gauge,right of=g6,label={[rotate=-45]below 
right:{\dd{k{-}1}}}] {};
	\node (g8) [gauge,right of=g7,label={[rotate=-45]below right:{\cc{k{-}1}}}] 
{};
    \node (g9) [gauge,right of=g8,label={[rotate=-45]below 
right:{\dd{k}}}] {};
	\node (g10) [gauge,right of=g9,label={[rotate=-45]below 
right:{\cc{k{-}1}}}] {};
    \node (g11) [gauge,right of=g10,label={[rotate=-45]below 
right:{\dd{k{-}1}}}] {};
    \node (g12) [right of=g11] {$\scriptstyle{\ldots}$};
	\node (g13) [gauge,right of=g12,label={[rotate=-45]below right:{\cc{2}}}] 
{};
    \node (g14) [gauge,right of=g13,label={[rotate=-45]below right:{\dd{2}}}] 
{};
	\node (g15) [gauge,right of=g14,label={[rotate=-45]below right:{\cc{1}}}] 
{};
	\node (g16) [gauge,right of=g15,label={[rotate=-45]below right:{\dd{1}}}] 
{};
	\node (g0) [gauge,above of=g9,label=above:{\cc{1}}] {};
	\draw  (g2)--(g3) (g3)--(g4) (g4)--(g5) (g5)--(g6) (g6)--(g7) 
(g7)--(g8) (g8)--(g9) (g9)--(g10) (g10)--(g11) (g11)--(g12) (g12)--(g13)
(g13)--(g14) (g14)--(g15) (g15)--(g16)  (g9)--(g0);
	\end{tikzpicture}
	} 
		\label{eq:k>4_magQuiver_infinite}
\end{align}
and the Coulomb branch dimension and symmetry is computed to be
\begin{subequations}
\label{eq:data_k>4_Coulomb_infinite}
\begin{align}
 \dim_\HH\ \Coulomb^{3\diff} \left( \magQuivText 
\eqref{eq:k>4_magQuiver_infinite} 
\right) &= 
2\cdot \sum_{i=1}^{k-1} \left( \dim\ \dalg_i + \dim\ \calg_i \right) + 
\dim\ \dalg_k + \dim\ \calg_1 \notag \\
&= \dim\ \sorm(2k) +1  \\
&=  \dim_\HH\ \Coulomb^{3\diff} \left( \magQuivText 
\eqref{eq:k>4_magQuiver_finite} 
\right) + 29 \,,\notag \\
G_J&= \sorm(4k) \,.
\end{align}
\end{subequations}
Note that the global symmetry matches the global $\sorm(4k)$ symmetry of 
\eqref{eq:1M5_electric_quiver}, which is expected to remain the symmetry in the 
infinite coupling case for $k>4$. Moreover, the Higgs branch dimension of 
\eqref{eq:1M5_electric_quiver} at infinite coupling has been computed in 
\cite[Eq.\ (1.2)]{Mekareeya:2017sqh}
and agrees with \eqref{eq:data_k>4_Coulomb_infinite}. Consequently, the 
significance of this result lies in
\begin{align}
 \Coulomb^{3\diff}\left( \magQuivText \eqref{eq:k>4_magQuiver_infinite} \right)
 =
 \Higgs^{6\diff}_\infty \left( \elecTheory \eqref{eq:1M5_electric_quiver} 
\right) 
\,.
\end{align}
Again, the magnetic quiver \eqref{eq:k>4_magQuiver_infinite} had been 
conjectured in \cite{Hanany:2018uhm}, but the formalism presented here allows to 
\emph{derive} it from a brane configuration.

The geometric relationship between finite and 
infinite gauge coupling phase is the subject of Section \ref{sec:Hasse}.
%
%
%
\subsection{Multiple M5s on D-type singularity}
\label{sec:multiple_M5}
Having discussed a single \Mf\ brane on a D-type singularity, it is time to 
include multiple \Mf\ branes. To be precise, consider $n$ M5 
branes on $\C^2 \slash \D_{k-2}$ for $k\geq 4$, then the dual Type IIA 
description 
yields
\begin{align}
 \raisebox{-.5\height}{
\begin{tikzpicture}
		\DsixOMinus{k}{0}
        \DsixOPlus{k{-}4}{1}
		\DsixOMinus{k}{2}
		\DsixOPlus{k{-}4}{3}
		\draw (4.5,0) node {$\cdots$};
        \DsixOPlus{k{-}4}{5}
		\DsixOMinus{k}{6}
		\DsixOPlus{k{-}4}{7}
		\DsixOMinus{k}{8}
		\ns{1,0}
		\ns{2,0}
        \ns{3,0}
		\ns{6,0}
		\ns{7,0}
        \ns{8,0}
    \draw[decoration={brace,mirror,raise=20pt},decorate,thick]
  (1-0.2,0) -- node[below=20pt] {\tiny{$2n$ half \NS }} (8.2,0);
	\end{tikzpicture}
	}
	\label{eq:branes_n_M5_on_Dk}
\end{align}
such that the resulting 6d $\Ncal =(1,0)$ theory consists of $(2n-1)$ tensor 
multiplets together with hyper and $(2n-1)$ vector multiplets encoded in the 
electric quiver gauge theory 
\cite{Intriligator:1997kq,Blum:1997mm,Hanany:1997gh,Brunner:1997gk, 
Intriligator:1997dh,Brunner:1997gf,Ferrara:1998vf,DelZotto:2014hpa}
\begin{align}
 \raisebox{-.5\height}{
 	\begin{tikzpicture}
 	\tikzset{node distance = 0.5cm};
	\tikzstyle{gauge} = [circle, draw,inner sep=2.5pt];
	\tikzstyle{flavour} = [regular polygon,regular polygon sides=4,inner 
sep=2.5pt, draw];
	\node (g1) [gauge,label={[rotate=-45]below right:{\Sp{k-4}}}] {};
	\node (g2) [gauge,right of=g1,label={[rotate=-45]below right:{\SO{2k}}}] 
{};
	\node (g3) [gauge,right of=g2,label={[rotate=-45]below right:{\Sp{k{-}4}}}] 
{};
	\node (g4) [right of=g3] {$\ldots$};
	\node (g5) [gauge,right of=g4,label={[rotate=-45]below right:{\Sp{k{-}4}}}] 
{};
    \node (g6) [gauge,right of=g5,label={[rotate=-45]below 
right:{\SO{2k}}}] {};
	\node (g7) [gauge,right of=g6,label={[rotate=-45]below right:{\Sp{k{-}4}}}] 
{};
	\node (f1) [flavour,left of=g1,label=above:{\SO{2k}}] {};
	\node (f7) [flavour,right of=g7,label=above:{\SO{2k}}] {};
	\draw (g1)--(g2) (g2)--(g3) (g3)--(g4) (g4)--(g5) (g5)--(g6) (g6)--(g7) 
(g1)--(f1) (g7)--(f7);
\draw[decoration={brace,mirror,raise=20pt},decorate,thick]
  (0,-0.45) -- node[below=20pt] {\tiny{$2n-1$ }} (3.8,-0.45);
	\end{tikzpicture}
	}
	\label{eq:electric_quiver_n_M5_Dk}
\end{align}
and one decoupled tensor multiplet.
The vacuum moduli spaces structure is more sophisticated than in the single 
\Mf\ brane case, simply because there are $(2n-1)$ 
non-decoupled tensor multiplets, or, equivalently, $(2n-1)$ independent 
gauge couplings in \eqref{eq:electric_quiver_n_M5_Dk}.
Again, there are various singular 
loci where BPS-strings become tensionless and the Higgs branches of the 
theories over these singularities have to be investigated carefully.
\begin{compactenum}[(i)]
 \item The Higgs branch $\Higgs^{6\diff}_\fin$ over a generic point of the 
tensor branch, i.e.\ the theory has $(2n-1)$ tensor multiplets and all 
couplings in the gauge theory \eqref{eq:electric_quiver_n_M5_Dk} are finite.
\item The Higgs branch $\Higgs^{6\diff}_{j,\gamma\in\sigma(j)}$ over a singular 
point of 
order 
$j$ ($1\leq j < 2n-1$) of the tensor branch, i.e.\ the theory has lost $j$ 
out 
of the $(2n-1)$ tensor multiplets. Note that there are multiple singular loci 
of the same order, meaning that there are $\sigma(j)$ different possibilities 
to 
take $j$ out of the $(2n-1)$ gauge couplings to infinity.
\item The Higgs branch $\Higgs^{6\diff}_\infty$ over the origin of the tensor 
branch, i.e.\ no tensor multiplets and all couplings in 
\eqref{eq:electric_quiver_n_M5_Dk} are infinite.
\end{compactenum}
The Higgs branches of the different phases as well as the transition between 
them are \emph{derived} from a brane configuration in this section. 
\paragraph{Generic point on tensor branch.}
The first step is to derive the 
magnetic quiver description for the finite coupling regime of 
\eqref{eq:electric_quiver_n_M5_Dk}. To achieve this, one pulls in $2k$ half 
\De\ branes from both $x^6 = \pm \infty$ and obtains 
\begin{align}
 \raisebox{-.5\height}{
\begin{tikzpicture}
        \DeightMany{1}{0}{1}
        \Dbrane{0,0}{1,0}
		\DeightMany{1}{1}{1}
		\DsixOMinus{$1$}{1}
		\DeightMany{1}{2}{1}
        \draw (2.5,0) node {$\cdots$};
        \DeightMany{1}{3}{1}
        \DsixOMinusTilde{$k{-}1$}{3}
        \DeightMany{1}{4}{1}
        \DsixOMinus{$k$}{4}
        \DsixOPlus{$k{-}4$}{5}
        \DsixOMinus{$k$}{6}
        \ns{5,0}
        \ns{6,0}
        \draw (7.5,0) node {$\cdots$};
        \DsixOMinus{$k$}{8}
        \DsixOPlus{$k{-}4$}{9}
        \DsixOMinus{$k$}{10}
        \ns{9,0}
        \ns{10,0}
        \DeightMany{1}{11}{1}
        \DsixOMinusTilde{$k{-}1$}{11}
        \DeightMany{1}{12}{1}
        \draw (12.5,0) node {$\cdots$};
        \DeightMany{1}{13}{1}
		\DsixOMinus{$1$}{13}
		\DeightMany{1}{14}{1}
        \Dbrane{14,0}{15,0}
        \DeightMany{1}{15}{1}
\draw[decoration={brace,mirror,raise=10pt},decorate,thick]
  (-0.2,-1) -- node[below=10pt] {$2k$ half \De } (4.2,-1);
\draw[decoration={brace,mirror,raise=10pt},decorate,thick]
  (11-0.2,-1) -- node[below=10pt] {$2k$ half \De } (15.2,-1);
\draw[decoration={brace,mirror,raise=10pt},decorate,thick]
  (5-0.2,-1) -- node[below=10pt] {$2n$ half \NS } (10.2,-1);
	\end{tikzpicture}
	}
	\label{eq:branes_nM5_on_Dk_pulled_D8}
\end{align}
and the next step lies in suspending as many \Ds\ branes between \De\ 
branes as possible. Since half \NS\ branes cannot leave the \Os\ plane in the 
finite 
coupling regime, the strategy is as follows: firstly, transition the outermost 
half \NS\ brane through seven \De\ branes. The reasoning is as in Section 
\ref{sec:1M5_generic_k}, from the $k$ full \Ds\ branes that are suspended 
between one of the outer-most \NS\ and a \De\ brane, one can consider $(k-4)$ 
of them as going through the \NS\ and only $4$ of them as being frozen between 
the \NS\ and \De. Frozen branes do not contribute to the Higgs branch and can 
be eliminated by brane-annihilation \eqref{eq:brane_creation_all} when the \NS\ 
passes through half \De\ branes. 

Secondly, the remaining $(n-1)$ pairs of half \NS\ branes are considered as having 
$8$ half \Ds\ suspended between them, while the other $2(k-4)$ half \Ds\ branes 
are suspended between the pulled in \De\ branes. The brane configuration looks 
like
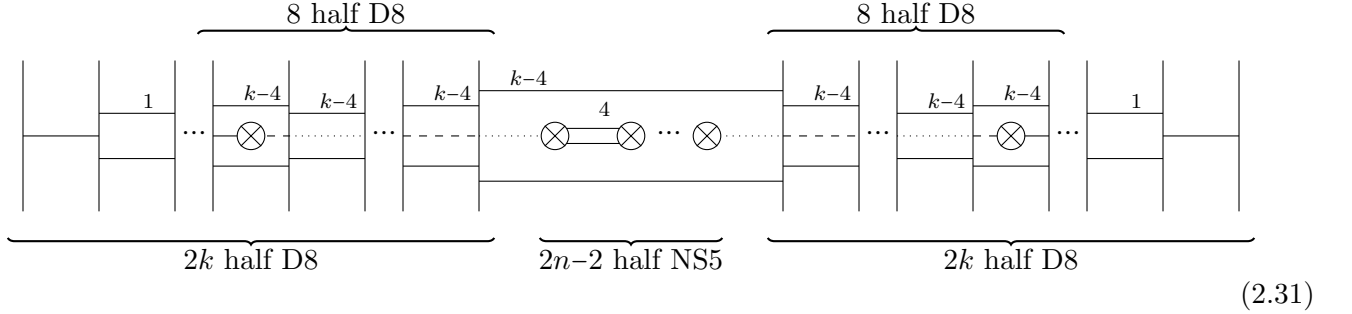
\begin{align}
 \raisebox{-.5\height}{
\begin{tikzpicture}
        \DeightMany{1}{-1}{1}
        \Dbrane{-1,0}{0,0}
		\DeightMany{1}{0}{1}
		\DsixOMinusFree{$1$}{0}{0}
		\DeightMany{1}{1}{1}
        \draw (1.25,0) node {$\cdots$};
        \DeightMany{1}{1.5}{1}
        \DsixOMinusFree{$k{-}4$}{1.5}{1}
        \Dbrane{1.5,0}{2,0}
        \OPlusTilde{2,0}{2.5,0}
		\DeightMany{1}{2.5}{1}
		\DsixOPlusFree{$k{-}4$}{2.5}{0}
		\DeightMany{1}{3.5}{1}
        \ns{2,0}
        \draw (3.75,0) node {$\cdots$};
        \DeightMany{1}{4}{1}
        \DeightMany{1}{5}{1}
        \DsixOPlusTildeFree{$k{-}4$}{4}{1}
        \DsixOPlusFree{$k{-}4$}{5}{3}
        \DsixOMinus{$4$}{6}
        \draw (7.5,0) node {$\cdots$};
     \draw (6,0.6)--(9,0.6);
     \draw (6,-0.6)--(9,-0.6);
     \OPlus{8,0}{9,0}
        \ns{6,0}
        \ns{7,0}
        \ns{8,0}
        \DeightMany{1}{9}{1}
        \DsixOPlusTildeFree{$k{-}4$}{9}{1}
        \DeightMany{1}{10}{1}
        \draw (10.25,0) node {$\cdots$};
        \DeightMany{1}{10.5}{1}
		\DsixOPlusFree{$k{-}4$}{10.5}{0}
		\DeightMany{1}{11.5}{1}
		\DsixOMinusFree{$k{-}4$}{11.5}{1}
		\OPlusTilde{11.5,0}{12,0}
        \Dbrane{12,0}{12.5,0}
        \DeightMany{1}{12.5}{1}
        \ns{12,0}
        \draw (12.75,0) node {$\cdots$};
		\DeightMany{1}{13}{1}
		\DsixOMinusFree{$1$}{13}{0}
		\DeightMany{1}{14}{1}
		\Dbrane{14,0}{15,0}
		\DeightMany{1}{15}{1}
\draw[decoration={brace,mirror,raise=10pt},decorate,thick]
  (-1.2,-1) -- node[below=10pt] {$2k$ half \De } (5.2,-1);
  \draw[decoration={brace,mirror,raise=10pt},decorate,thick]
  (5.2,1) -- node[above=10pt] {$8$ half \De } (1.5-0.2,1);
\draw[decoration={brace,mirror,raise=10pt},decorate,thick]
  (9-0.2,-1) -- node[below=10pt] {$2k$ half \De } (15.2,-1);
\draw[decoration={brace,mirror,raise=10pt},decorate,thick]
  (12.7,1) -- node[above=10pt] {$8$ half \De } (9-0.2,1);
\draw[decoration={brace,mirror,raise=10pt},decorate,thick]
  (6-0.2,-1) -- node[below=10pt] {$2n{-}2$ half \NS } (8.2,-1);
	\end{tikzpicture}
	}
	\label{eq:branes_nM5_on_Dk_finite_coupling}
\end{align}
In the centre, one observes $(n-1)$ pairs of half \NS\ branes with $8$ half 
\Ds\ 
branes suspended in between, and there is no way to suspend these \Ds\ between 
\De\ branes. 
Consequently, these \Ds\ do not contribute to the Higgs branch either and as 
such one considers them as contributing flavour nodes to the magnetic quiver. 
Employing the conversion to \emph{magnetic orientifolds} of Table 
\ref{tab:orientifold}, one reads off the magnetic quiver to be 
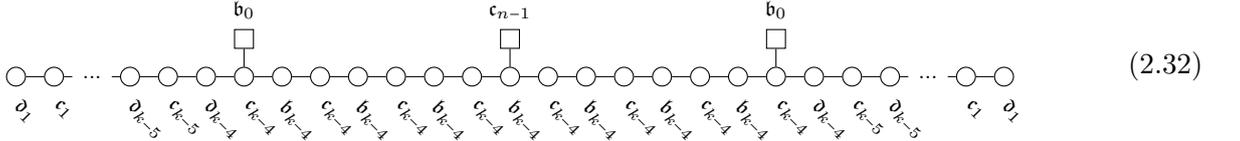
\begin{align}
 \raisebox{-.5\height}{
 	\begin{tikzpicture}
 	\tikzset{node distance = 0.5cm};
	\tikzstyle{gauge} = [circle, draw,inner sep=2.5pt];
	\tikzstyle{flavour} = [regular polygon,regular polygon sides=4,inner 
sep=2.5pt, draw];
	\node (g2) [gauge,label={[rotate=-45]below right:{\dd{1}}}] 
{};
	\node (g3) [gauge,right of=g2,label={[rotate=-45]below right:{\cc{1}}}] {};
	\node (g4) [right of=g3] {$\scriptstyle{\ldots}$};
	\node (g5) [gauge,right of=g4,label={[rotate=-45]below 
right:{\dd{k{-}5}}}] {};
	\node (g6) [gauge,right of=g5,label={[rotate=-45]below right:{\cc{k{-}5}}}] 
{};
    \node (g7) [gauge,right of=g6,label={[rotate=-45]below 
right:{\dd{k{-}4}}}] {};
	\node (g8) [gauge,right of=g7,label={[rotate=-45]below right:{\cc{k{-}4}}}] 
{};
    \node (g9) [gauge,right of=g8,label={[rotate=-45]below 
right:{\bb{k{-}4}}}] {};
	\node (g10) [gauge,right of=g9,label={[rotate=-45]below 
right:{\cc{k{-}4}}}] {};
    \node (g11) [gauge,right of=g10,label={[rotate=-45]below 
right:{\bb{k{-}4}}}] {};
	\node (g12) [gauge,right of=g11,label={[rotate=-45]below 
right:{\cc{k{-}4}}}] {};
    \node (g13) [gauge,right of=g12,label={[rotate=-45]below 
right:{\bb{k{-}4}}}] {};
	\node (g14) [gauge,right of=g13,label={[rotate=-45]below 
right:{\cc{k{-}4}}}] {};
    \node (g15) [gauge,right of=g14,label={[rotate=-45]below 
right:{\bb{k{-}4}}}] {};
	\node (g16) [gauge,right of=g15,label={[rotate=-45]below 
right:{\cc{k{-}4}}}] {};
    \node (g17) [gauge,right of=g16,label={[rotate=-45]below 
right:{\bb{k{-}4}}}] {};
	\node (g18) [gauge,right of=g17,label={[rotate=-45]below 
right:{\cc{k{-}4}}}] {};
    \node (g19) [gauge,right of=g18,label={[rotate=-45]below 
right:{\bb{k{-}4}}}] {};
	\node (g20) [gauge,right of=g19,label={[rotate=-45]below 
right:{\cc{k{-}4}}}] {};
    \node (g21) [gauge,right of=g20,label={[rotate=-45]below 
right:{\bb{k{-}4}}}] {};
	\node (g22) [gauge,right of=g21,label={[rotate=-45]below 
right:{\cc{k{-}4}}}] 
{};
    \node (g23) [gauge,right of=g22,label={[rotate=-45]below 
right:{\dd{k{-}4}}}] {};
	\node (g24) [gauge,right of=g23,label={[rotate=-45]below 
right:{\cc{k{-}5}}}] 
{};
	\node (g25) [gauge,right of=g24,label={[rotate=-45]below 
right:{\dd{k{-}5}}}] {};
	\node (g26) [right of=g25] {$\scriptstyle{\ldots}$};
	\node (g27) [gauge,right of=g26,label={[rotate=-45]below right:{\cc{1}}}] 
{};
	\node (g28) [gauge,right of=g27,label={[rotate=-45]below right:{\dd{1}}}] 
{};
	\node (f8) [flavour,above of=g8,label=above:{\bb{0}}] {};
	\node (f22) [flavour,above of=g22,label=above:{\bb{0}}] {};
	\node (b0) [flavour,above of=g15,label=above:{\cc{n-1}}] {};
	\draw (g2)--(g3) (g3)--(g4) (g4)--(g5) (g5)--(g6) (g6)--(g7) 
(g7)--(g8) (g8)--(g9) (g9)--(g10) (g10)--(g11) (g11)--(g12) (g12)--(g13)
(g13)--(g14) (g14)--(g15) (g15)--(g16) (g16)--(g17) (g17)--(g18) (g18)--(g19)
(g19)--(g20) (g20)--(g21) (g21)--(g22) (g22)--(g23) (g23)--(g24) (g24)--(g25) 
(g25)--(g26) (g26)--(g27) (g27)--(g28) 
(g15)--(b0) 
(g8)--(f8) (g22)--(f22);
	\end{tikzpicture}
	}
    \label{eq:k_nM5_magQuiver_finite}
\end{align}
The Coulomb branch dimension of \eqref{eq:k_nM5_magQuiver_finite} is readily 
computed
\begin{equation}
\begin{aligned}
 \dim_\HH \ \Coulomb^{3d} \left( \magQuivText \eqref{eq:k_nM5_magQuiver_finite} 
\right)
&=2\cdot \sum_{i=1}^{k-4} \left( \dim\ \calg_i  + \dim\ \dalg_i \right) 
  + 7\cdot \dim\ \balg_{k-4}   + 6\cdot \dim\ \calg_{k-4} \\
&= \dim\ \sorm(2k) - \dim\ \sorm(8) \,.
\end{aligned} 
\label{eq:dim_Coulomb_finite}
\end{equation}
To compute the Higgs branch dimension of \eqref{eq:electric_quiver_n_M5_Dk}, 
one needs to recall that there is no complete Higgsing of the $\sorm(2k)$ gauge 
nodes; instead, there is partial Higgsing $\sorm(2k) \to \sorm(8)$ such that 
one computes 
\begin{align}
 \dim_\HH \ \Higgs^{6d}_{\mathrm{finite}} 
\left(\elecTheory \eqref{eq:electric_quiver_n_M5_Dk} \right)
 &= n_h -n_v 
 = \dim \ \sorm(2k)  - \dim\ \sorm(8)
 \label{eq:dim_Higgs_finite} \\
 n_h &=  \frac{1}{2} \cdot 2k \cdot (2k-8) \cdot 2n \notag \\
n_v&= n \cdot \dim (\sprm(k-4)) 
+ (n-1) \cdot\left( \dim (\sorm(2k))  - \dim (\sorm(8)) \right)
\notag  \,,
\end{align}
which is independent of $n$, and confirms that the Higgs branch is 
trivial for $k=4$. One observes that both dimensions 
\eqref{eq:dim_Coulomb_finite} and 
\eqref{eq:dim_Higgs_finite} agree. Moreover, a computation of the topological 
symmetry of \eqref{eq:k_nM5_magQuiver_finite} reveals
\begin{align}
 G_J = \sorm(2k) \times \sorm(2k)
\end{align}
because the central $\balg_{k-4}$ node is never balanced for $k>1$, but always 
a 
\emph{good} in the sense of Appendix \ref{app:global_sym}. The Coulomb branch 
symmetry agrees with that of the Higgs branch of 
\eqref{eq:1M5_electric_quiver}. Therefore, the significance of this derivations 
is that 
\begin{align}
 \Coulomb^{3\diff} \left( \magQuivText \eqref{eq:k_nM5_magQuiver_finite} \right)
 = 
 \Higgs_\fin^{6\diff} \left( \elecTheory \eqref{eq:electric_quiver_n_M5_Dk} 
\right) \,.
\end{align}
The challenge in computing the Higgs branch of 
\eqref{eq:electric_quiver_n_M5_Dk} lies in non-complete Higgsing, and current 
techniques are not suitable or able to overcome the difficulties. Therefore, 
the magnetic quiver \eqref{eq:k_nM5_magQuiver_finite} provides a
prediction for the Higgs branch description.

\paragraph{One infinite gauge coupling.}
Next, one can proceed to one of the infinite gauge coupling phases. 
As indicated above, there are $(2n-1)$ tensor multiplets, i.e.\ $(2n-1)$ 
different order parameters that can be tuned. 
Moreover, recall that tuning a gauge 
coupling to infinity means that the associated pair of half \NS\ has to become 
coincident along $x^6$. By charge conservation, the numbers of \Ds\ branes on 
the left and right of a pair of half \NS\ branes are 
identical, as long as no \De\ branes are involved. Thus, the pair of half \NS\ 
branes can leave the orientifold in 
transverse $x^{7,8,9}$ direction and the \Ds\ branes from the left and right of 
the pair reconnect. 
As apparent from 
\eqref{eq:branes_nM5_on_Dk_finite_coupling}, 
there are two 
different types of pairs that can become coincident: 
\begin{compactenum}[(i)]
\item either a pair of half \NS\ branes with $4$ full \Ds\ branes in between,
\item or a pair of half \NS\ branes with no \Ds\ branes in between.
\end{compactenum}
To begin with, consider the left 
pair of half \NS\ branes with $8$ half \Ds\ branes in between as in brane 
configuration \eqref{eq:branes_nM5_on_Dk_finite_coupling}, then this pair can 
leave the \Os\ as explained above and one arrives at   
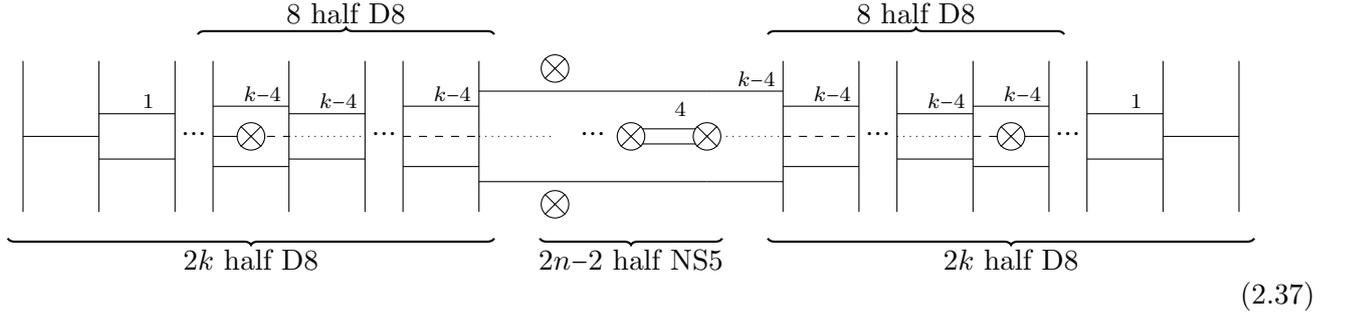
\begin{align}
 \raisebox{-.5\height}{
\begin{tikzpicture}
        \DeightMany{1}{-1}{1}
        \Dbrane{-1,0}{0,0}
		\DeightMany{1}{0}{1}
		\DsixOMinusFree{$1$}{0}{0}
		\DeightMany{1}{1}{1}
        \draw (1.25,0) node {$\cdots$};
        \DeightMany{1}{1.5}{1}
        \DsixOMinusFree{$k{-}4$}{1.5}{1}
        \Dbrane{1.5,0}{2,0}
        \OPlusTilde{2,0}{2.5,0}
		\DeightMany{1}{2.5}{1}
		\DsixOPlusFree{$k{-}4$}{2.5}{0}
		\DeightMany{1}{3.5}{1}
        \ns{2,0}
        \draw (3.75,0) node {$\cdots$};
        \DeightMany{1}{4}{1}
        \DeightMany{1}{5}{1}
        \DsixOPlusTildeFree{$k{-}4$}{4}{1}
        \DsixOPlusFree{$k{-}4$}{8}{3}
        \DsixOMinus{$4$}{7}
        \draw (6.5,0) node {$\cdots$};
     \draw (5,0.6)--(8,0.6);
     \draw (5,-0.6)--(8,-0.6);
     \OPlus{5,0}{6,0}
        \ns{6,0.9}
        \ns{6,-0.9}
        \ns{7,0}
        \ns{8,0}
        \DeightMany{1}{9}{1}
        \DsixOPlusTildeFree{$k{-}4$}{9}{1}
        \DeightMany{1}{10}{1}
        \draw (10.25,0) node {$\cdots$};
        \DeightMany{1}{10.5}{1}
		\DsixOPlusFree{$k{-}4$}{10.5}{0}
		\DeightMany{1}{11.5}{1}
		\DsixOMinusFree{$k{-}4$}{11.5}{1}
		\OPlusTilde{11.5,0}{12,0}
        \Dbrane{12,0}{12.5,0}
        \DeightMany{1}{12.5}{1}
        \ns{12,0}
        \draw (12.75,0) node {$\cdots$};
		\DeightMany{1}{13}{1}
		\DsixOMinusFree{$1$}{13}{0}
		\DeightMany{1}{14}{1}
		\Dbrane{14,0}{15,0}
		\DeightMany{1}{15}{1}
\draw[decoration={brace,mirror,raise=10pt},decorate,thick]
  (-1.2,-1) -- node[below=10pt] {$2k$ half \De } (5.2,-1);
  \draw[decoration={brace,mirror,raise=10pt},decorate,thick]
  (5.2,1) -- node[above=10pt] {$8$ half \De } (1.5-0.2,1);
\draw[decoration={brace,mirror,raise=10pt},decorate,thick]
  (9-0.2,-1) -- node[below=10pt] {$2k$ half \De } (15.2,-1);
\draw[decoration={brace,mirror,raise=10pt},decorate,thick]
  (12.7,1) -- node[above=10pt] {$8$ half \De } (9-0.2,1);
\draw[decoration={brace,mirror,raise=10pt},decorate,thick]
  (6-0.2,-1) -- node[below=10pt] {$2n{-}2$ half \NS } (8.2,-1);
	\end{tikzpicture}
	}
	\label{eq:branes_k_nM5_dim1_transition}
\end{align}
The difference to \eqref{eq:branes_nM5_on_Dk_finite_coupling} is that the left 
pair of half \NS\ now contributes gauge degrees of freedom to the magnetic 
quiver, because the \NS\ branes are free to move along $x^{7,8,9}$. 
Put differently, one can non-trivially suspend virtual \Df\ branes between the 
half \NS\ and its mirror image. The action of the orientifold leads to a 
magnetic vector multiplet of a symplectic gauge group. 
Consequently, the magnetic quiver is read off as 
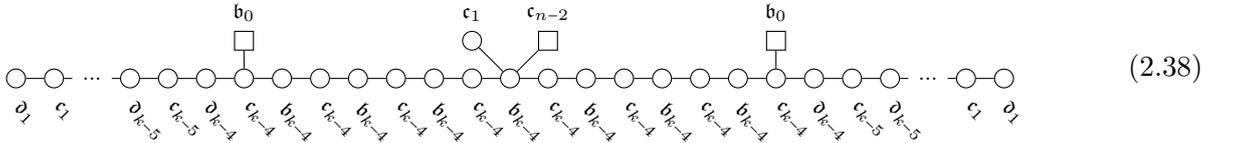
\begin{align}
 \raisebox{-.5\height}{
 	\begin{tikzpicture}
 	\tikzset{node distance = 0.5cm};
	\tikzstyle{gauge} = [circle, draw,inner sep=2.5pt];
	\tikzstyle{flavour} = [regular polygon,regular polygon sides=4,inner 
sep=2.5pt, draw];
	\node (g2) [gauge,label={[rotate=-45]below right:{\dd{1}}}] 
{};
	\node (g3) [gauge,right of=g2,label={[rotate=-45]below right:{\cc{1}}}] {};
	\node (g4) [right of=g3] {$\scriptstyle{\ldots}$};
	\node (g5) [gauge,right of=g4,label={[rotate=-45]below 
right:{\dd{k{-}5}}}] {};
	\node (g6) [gauge,right of=g5,label={[rotate=-45]below right:{\cc{k{-}5}}}] 
{};
    \node (g7) [gauge,right of=g6,label={[rotate=-45]below 
right:{\dd{k{-}4}}}] {};
	\node (g8) [gauge,right of=g7,label={[rotate=-45]below right:{\cc{k{-}4}}}] 
{};
    \node (g9) [gauge,right of=g8,label={[rotate=-45]below 
right:{\bb{k{-}4}}}] {};
	\node (g10) [gauge,right of=g9,label={[rotate=-45]below 
right:{\cc{k{-}4}}}] {};
    \node (g11) [gauge,right of=g10,label={[rotate=-45]below 
right:{\bb{k{-}4}}}] {};
	\node (g12) [gauge,right of=g11,label={[rotate=-45]below 
right:{\cc{k{-}4}}}] {};
    \node (g13) [gauge,right of=g12,label={[rotate=-45]below 
right:{\bb{k{-}4}}}] {};
	\node (g14) [gauge,right of=g13,label={[rotate=-45]below 
right:{\cc{k{-}4}}}] {};
    \node (g15) [gauge,right of=g14,label={[rotate=-45]below 
right:{\bb{k{-}4}}}] {};
	\node (g16) [gauge,right of=g15,label={[rotate=-45]below 
right:{\cc{k{-}4}}}] {};
    \node (g17) [gauge,right of=g16,label={[rotate=-45]below 
right:{\bb{k{-}4}}}] {};
	\node (g18) [gauge,right of=g17,label={[rotate=-45]below 
right:{\cc{k{-}4}}}] {};
    \node (g19) [gauge,right of=g18,label={[rotate=-45]below 
right:{\bb{k{-}4}}}] {};
	\node (g20) [gauge,right of=g19,label={[rotate=-45]below 
right:{\cc{k{-}4}}}] {};
    \node (g21) [gauge,right of=g20,label={[rotate=-45]below 
right:{\bb{k{-}4}}}] {};
	\node (g22) [gauge,right of=g21,label={[rotate=-45]below 
right:{\cc{k{-}4}}}] {};
    \node (g23) [gauge,right of=g22,label={[rotate=-45]below 
right:{\dd{k{-}4}}}] {};
	\node (g24) [gauge,right of=g23,label={[rotate=-45]below 
right:{\cc{k{-}5}}}] {};
	\node (g25) [gauge,right of=g24,label={[rotate=-45]below 
right:{\dd{k{-}5}}}] {};
	\node (g26) [right of=g25] {$\scriptstyle{\ldots}$};
	\node (g27) [gauge,right of=g26,label={[rotate=-45]below right:{\cc{1}}}] 
{};
	\node (g28) [gauge,right of=g27,label={[rotate=-45]below right:{\dd{1}}}] 
{};
	\node (f8) [flavour,above of=g8,label=above:{\bb{0}}] {};
	\node (f22) [flavour,above of=g22,label=above:{\bb{0}}] {};
	\node (b1) [gauge,above of=g14,label=above:{\cc{1}}] {};
    \node (b2) [flavour,above of=g16,label=above:{\cc{n{-}2}}] {};
	\draw (g2)--(g3) (g3)--(g4) (g4)--(g5) (g5)--(g6) (g6)--(g7) 
(g7)--(g8) (g8)--(g9) (g9)--(g10) (g10)--(g11) (g11)--(g12) (g12)--(g13)
(g13)--(g14) (g14)--(g15) (g15)--(g16) (g16)--(g17) (g17)--(g18) (g18)--(g19)
(g19)--(g20) (g20)--(g21) (g21)--(g22) (g22)--(g23) (g23)--(g24) (g24)--(g25) 
(g25)--(g26) (g26)--(g27) (g27)--(g28) 
(g15)--(b1) (g15)--(b2)
(g8)--(f8) (g22)--(f22);
	\end{tikzpicture}
	}
    \label{eq:k_nM5_magQuiver_infinite_+1}
\end{align}
In the electric theory \eqref{eq:electric_quiver_n_M5_Dk}, the changes 
are easily kept track off. Before the transition, there are $(2n-1)$ tensor 
multiplets and $(n-1)$ $\sorm(8)$ vector multiplets, while after the 
transition both numbers are reduced by one. In order to satisfy the anomaly 
cancellation 
condition \eqref{eq:anomaly_cancellation}, the number of hypermultiplets has 
to change as follows
\begin{align}
 n'_h -n_h = 29 \cdot (n_t -n'_t) - (n_v - n'_v) = 1 \,.
\end{align}
In other words, the simultaneous loss of one tensor multiplet and one 
$\sorm(8)$ vector multiplet has to be compensated by one new hypermultiplet.

The second option, for tuning one gauge coupling to infinity, is to choose a 
pair of half \NS\ branes with no \Ds\ branes in between, see configuration 
\eqref{eq:branes_nM5_on_Dk_finite_coupling}. By the same arguments as above, 
the pair becomes coincident along $x^6$, the \Ds\ branes on the left and right 
of the pair reconnect, and the pair of half \NS\ branes can leave the 
orientifold in transverse $x^{7,8,9}$ direction. Hence, the brane configuration 
becomes
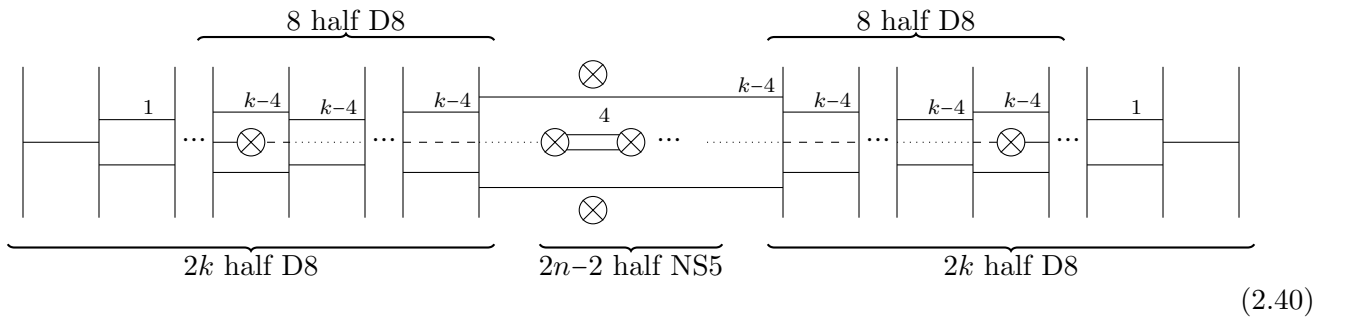
\begin{align}
  \raisebox{-.5\height}{
\begin{tikzpicture}
        \DeightMany{1}{-1}{1}
        \Dbrane{-1,0}{0,0}
		\DeightMany{1}{0}{1}
		\DsixOMinusFree{$1$}{0}{0}
		\DeightMany{1}{1}{1}
        \draw (1.25,0) node {$\cdots$};
        \DeightMany{1}{1.5}{1}
        \DsixOMinusFree{$k{-}4$}{1.5}{1}
        \Dbrane{1.5,0}{2,0}
        \OPlusTilde{2,0}{2.5,0}
		\DeightMany{1}{2.5}{1}
		\DsixOPlusFree{$k{-}4$}{2.5}{0}
		\DeightMany{1}{3.5}{1}
        \ns{2,0}
        \draw (3.75,0) node {$\cdots$};
        \DeightMany{1}{4}{1}
        \DeightMany{1}{5}{1}
        \DsixOPlusTildeFree{$k{-}4$}{4}{1}
        \DsixOPlusFree{$k{-}4$}{8}{3}
        \DsixOMinus{$4$}{6}
        \draw (7.5,0) node {$\cdots$};
     \draw (5,0.6)--(8,0.6);
     \draw (5,-0.6)--(8,-0.6);
     \OPlus{5,0}{6,0}
        \ns{6.5,0.9}
        \ns{6.5,-0.9}
        \ns{6,0}
        \ns{7,0}
        \DeightMany{1}{9}{1}
        \DsixOPlusTildeFree{$k{-}4$}{9}{1}
        \DeightMany{1}{10}{1}
        \draw (10.25,0) node {$\cdots$};
        \DeightMany{1}{10.5}{1}
		\DsixOPlusFree{$k{-}4$}{10.5}{0}
		\DeightMany{1}{11.5}{1}
		\DsixOMinusFree{$k{-}4$}{11.5}{1}
		\OPlusTilde{11.5,0}{12,0}
        \Dbrane{12,0}{12.5,0}
        \DeightMany{1}{12.5}{1}
        \ns{12,0}
        \draw (12.75,0) node {$\cdots$};
		\DeightMany{1}{13}{1}
		\DsixOMinusFree{$1$}{13}{0}
		\DeightMany{1}{14}{1}
		\Dbrane{14,0}{15,0}
		\DeightMany{1}{15}{1}
\draw[decoration={brace,mirror,raise=10pt},decorate,thick]
  (-1.2,-1) -- node[below=10pt] {$2k$ half \De } (5.2,-1);
  \draw[decoration={brace,mirror,raise=10pt},decorate,thick]
  (5.2,1) -- node[above=10pt] {$8$ half \De } (1.5-0.2,1);
\draw[decoration={brace,mirror,raise=10pt},decorate,thick]
  (9-0.2,-1) -- node[below=10pt] {$2k$ half \De } (15.2,-1);
\draw[decoration={brace,mirror,raise=10pt},decorate,thick]
  (12.7,1) -- node[above=10pt] {$8$ half \De } (9-0.2,1);
\draw[decoration={brace,mirror,raise=10pt},decorate,thick]
  (6-0.2,-1) -- node[below=10pt] {$2n{-}2$ half \NS } (8.2,-1);
	\end{tikzpicture}
	}
	\label{eq:branes_k_nM5_dim1_transition_var2}
\end{align}
The associated magnetic quiver is read off by the same logic as before. The 
\Ds\ 
branes suspended in intervals between half \De\ branes contribute magnetic 
gauge nodes 
according to the orientifold, see Table \ref{tab:orientifold}. The pairs of 
half \NS\ branes on the orientifold with $4$ full \Ds\ branes  suspended do 
contribute as flavours. In contrast, the pair of half \NS\ that left the 
orientifold contributes as magnetic vector multiplet. To see how, one suspends 
virtual \Df\ branes between the \NS s, and observes that the magnetic 
orientifold of an \Osm\ plan is again an \Osm , resulting in a symplectic gauge 
node.
In addition, virtual \Df\ branes can be suspended between the half 
\NS s that left the orientifold and the \Ds s in between the \NS s on the 
orientifold. Since these \Ds\ branes are not Higgs branch moduli, the \Df\ 
branes lead to a flavour $\sorm(8)$ node attached to the symplectic magnetic 
gauge node. Thus, the magnetic quiver associated to 
\eqref{eq:branes_k_nM5_dim1_transition_var2} becomes 
\begin{align}
 \raisebox{-.5\height}{
 	\begin{tikzpicture}
 	\tikzset{node distance = 0.5cm};
	\tikzstyle{gauge} = [circle, draw,inner sep=2.5pt];
	\tikzstyle{flavour} = [regular polygon,regular polygon sides=4,inner 
sep=2.5pt, draw];
	\node (g2) [gauge,label={[rotate=-45]below right:{\dd{1}}}] 
{};
	\node (g3) [gauge,right of=g2,label={[rotate=-45]below right:{\cc{1}}}] {};
	\node (g4) [right of=g3] {$\scriptstyle{\ldots}$};
	\node (g5) [gauge,right of=g4,label={[rotate=-45]below 
right:{\dd{k{-}5}}}] {};
	\node (g6) [gauge,right of=g5,label={[rotate=-45]below right:{\cc{k{-}5}}}] 
{};
    \node (g7) [gauge,right of=g6,label={[rotate=-45]below 
right:{\dd{k{-}4}}}] {};
	\node (g8) [gauge,right of=g7,label={[rotate=-45]below right:{\cc{k{-}4}}}] 
{};
    \node (g9) [gauge,right of=g8,label={[rotate=-45]below 
right:{\bb{k{-}4}}}] {};
	\node (g10) [gauge,right of=g9,label={[rotate=-45]below 
right:{\cc{k{-}4}}}] {};
    \node (g11) [gauge,right of=g10,label={[rotate=-45]below 
right:{\bb{k{-}4}}}] {};
	\node (g12) [gauge,right of=g11,label={[rotate=-45]below 
right:{\cc{k{-}4}}}] {};
    \node (g13) [gauge,right of=g12,label={[rotate=-45]below 
right:{\bb{k{-}4}}}] {};
	\node (g14) [gauge,right of=g13,label={[rotate=-45]below 
right:{\cc{k{-}4}}}] {};
    \node (g15) [gauge,right of=g14,label={[rotate=-45]below 
right:{\bb{k{-}4}}}] {};
	\node (g16) [gauge,right of=g15,label={[rotate=-45]below 
right:{\cc{k{-}4}}}] {};
    \node (g17) [gauge,right of=g16,label={[rotate=-45]below 
right:{\bb{k{-}4}}}] {};
	\node (g18) [gauge,right of=g17,label={[rotate=-45]below 
right:{\cc{k{-}4}}}] {};
    \node (g19) [gauge,right of=g18,label={[rotate=-45]below 
right:{\bb{k{-}4}}}] {};
	\node (g20) [gauge,right of=g19,label={[rotate=-45]below 
right:{\cc{k{-}4}}}] {};
    \node (g21) [gauge,right of=g20,label={[rotate=-45]below 
right:{\bb{k{-}4}}}] {};
	\node (g22) [gauge,right of=g21,label={[rotate=-45]below 
right:{\cc{k{-}4}}}] {};
    \node (g23) [gauge,right of=g22,label={[rotate=-45]below 
right:{\dd{k{-}4}}}] {};
	\node (g24) [gauge,right of=g23,label={[rotate=-45]below 
right:{\cc{k{-}5}}}] {};
	\node (g25) [gauge,right of=g24,label={[rotate=-45]below 
right:{\dd{k{-}5}}}] {};
	\node (g26) [right of=g25] {$\scriptstyle{\ldots}$};
	\node (g27) [gauge,right of=g26,label={[rotate=-45]below right:{\cc{1}}}] 
{};
	\node (g28) [gauge,right of=g27,label={[rotate=-45]below right:{\dd{1}}}] 
{};
	\node (f8) [flavour,above of=g8,label=above:{\bb{0}}] {};
	\node (f22) [flavour,above of=g22,label=above:{\bb{0}}] {};
	\node (b1) [gauge,above of=g14,label=left:{\cc{1}}] {};
    \node (fb1) [flavour,above of=b1,label=above:{\dd{4}}] {};
    \node (b2) [flavour,above of=g16,label=above:{\cc{n{-}2}}] {};
	\draw (g2)--(g3) (g3)--(g4) (g4)--(g5) (g5)--(g6) (g6)--(g7) 
(g7)--(g8) (g8)--(g9) (g9)--(g10) (g10)--(g11) (g11)--(g12) (g12)--(g13)
(g13)--(g14) (g14)--(g15) (g15)--(g16) (g16)--(g17) (g17)--(g18) (g18)--(g19)
(g19)--(g20) (g20)--(g21) (g21)--(g22) (g22)--(g23) (g23)--(g24) (g24)--(g25) 
(g25)--(g26) (g26)--(g27) (g27)--(g28) 
(g15)--(b1) (g15)--(b2)
(g8)--(f8) (g22)--(f22)
(b1)--(fb1);
	\end{tikzpicture}
	}
    \label{eq:k_nM5_magQuiver_infinite_+1_new}
\end{align}
which compared to \eqref{eq:k_nM5_magQuiver_infinite_+1} has the same moduli 
space dimension.

In fact, the physical transition from the $6$d perspective appears to be
identical to the first case. During the transition, the number of tensor 
multiplets and the 
number of $\sorm(8)$ vector multiplets are simultaneously reduced by one such 
that the 
anomaly cancellation condition \eqref{eq:anomaly_cancellation} enforces the 
appearance of one additional hypermultiplet. As such, this confirms 
the observation that both types of transitions 
\eqref{eq:branes_k_nM5_dim1_transition},  
\eqref{eq:branes_k_nM5_dim1_transition_var2} are one-dimensional.
As a consequence, one may also consider a transition from brane 
configuration \eqref{eq:branes_k_nM5_dim1_transition} to 
\eqref{eq:branes_k_nM5_dim1_transition_var2}. In other words, moving a pair of 
half \NS\ branes that left the orientifold along the $x^6$ direction across at 
least one half \NS\ brane. Following the brane configuration, as well as the 
associated magnetic quiver, leads to the \emph{prediction} that there is a 
discrete change of the Higgs branch whenever a pair of \NS\ branes outside the 
orientifold crosses a half \NS\ brane on the orientfold.

Moreover, one can go back to configuration 
\eqref{eq:branes_nM5_on_Dk_finite_coupling} and consider any pair of 
neighbouring half \NS\ branes. According to the above arguments, for any pair, 
the infinite gauge coupling transition for this pair is of the form
\begin{align}
 1\text{ tensor } + 1 \; \sorm(8) \text{ vector } \to 
 1 \text{ hyper } \,,
 \label{eq:1d_transition}
\end{align}
but the resulting magnetic quiver is either 
\eqref{eq:k_nM5_magQuiver_infinite_+1} or 
\eqref{eq:k_nM5_magQuiver_infinite_+1_new}, depending on which pair is chosen. 
In total, there are exactly 
$(2n-1) $ of these one-dimensional transitions.
Although different gauge couplings of \eqref{eq:electric_quiver_n_M5_Dk} are 
taken to infinity, the resulting moduli spaces fall into two classes, given by 
\eqref{eq:k_nM5_magQuiver_infinite_+1} or 
\eqref{eq:k_nM5_magQuiver_infinite_+1_new}. In addition, the transition between 
both is physically described by a change of $x^6$ position of a pair of \NS\ 
branes.
\paragraph{More infinite gauge couplings.}
Form the $(2n-2)$ half \NS\ branes in the centre of the brane configuration 
\eqref{eq:branes_nM5_on_Dk_pulled_D8}, one can form at most $(n-1)$ pairs that 
can under-go transition 
\eqref{eq:1d_transition}. An arbitrary intermediate stage is given by $l$ pairs 
of half \NS\ branes undergoing the transition 
\eqref{eq:1d_transition}, with $l_0$ pairs of the from 
\eqref{eq:branes_k_nM5_dim1_transition} and $l_1$ pairs of the form 
\eqref{eq:branes_k_nM5_dim1_transition_var2} such 
that $l=l_0+l_1$, and 
remaining separated along $x^6$. For $0\leq l \leq (n-1)$, the resulting 
magnetic quiver becomes
\begin{align}
 \raisebox{-.5\height}{
 	\begin{tikzpicture}
 	\tikzset{node distance = 0.5cm};
	\tikzstyle{gauge} = [circle, draw,inner sep=2.5pt];
	\tikzstyle{flavour} = [regular polygon,regular polygon sides=4,inner 
sep=2.5pt, draw];
	\node (g2) [gauge,label={[rotate=-45]below right:{\dd{1}}}] 
{};
	\node (g3) [gauge,right of=g2,label={[rotate=-45]below right:{\cc{1}}}] {};
	\node (g4) [right of=g3] {$\scriptstyle{\ldots}$};
	\node (g5) [gauge,right of=g4,label={[rotate=-45]below 
right:{\dd{k{-}5}}}] {};
	\node (g6) [gauge,right of=g5,label={[rotate=-45]below right:{\cc{k{-}5}}}] 
{};
    \node (g7) [gauge,right of=g6,label={[rotate=-45]below 
right:{\dd{k{-}4}}}] {};
	\node (g8) [gauge,right of=g7,label={[rotate=-45]below right:{\cc{k{-}4}}}] 
{};
    \node (g9) [gauge,right of=g8,label={[rotate=-45]below 
right:{\bb{k{-}4}}}] {};
	\node (g10) [gauge,right of=g9,label={[rotate=-45]below 
right:{\cc{k{-}4}}}] {};
    \node (g11) [gauge,right of=g10,label={[rotate=-45]below 
right:{\bb{k{-}4}}}] {};
	\node (g12) [gauge,right of=g11,label={[rotate=-45]below 
right:{\cc{k{-}4}}}] {};
    \node (g13) [gauge,right of=g12,label={[rotate=-45]below 
right:{\bb{k{-}4}}}] {};
	\node (g14) [gauge,right of=g13,label={[rotate=-45]below 
right:{\cc{k{-}4}}}] {};
    \node (g15) [gauge,right of=g14,label={[rotate=-45]below 
right:{\bb{k{-}4}}}] {};
	\node (g16) [gauge,right of=g15,label={[rotate=-45]below 
right:{\cc{k{-}4}}}] {};
    \node (g17) [gauge,right of=g16,label={[rotate=-45]below 
right:{\bb{k{-}4}}}] {};
	\node (g18) [gauge,right of=g17,label={[rotate=-45]below 
right:{\cc{k{-}4}}}] {};
    \node (g19) [gauge,right of=g18,label={[rotate=-45]below 
right:{\bb{k{-}4}}}] {};
	\node (g20) [gauge,right of=g19,label={[rotate=-45]below 
right:{\cc{k{-}4}}}] {};
    \node (g21) [gauge,right of=g20,label={[rotate=-45]below 
right:{\bb{k{-}4}}}] {};
	\node (g22) [gauge,right of=g21,label={[rotate=-45]below 
right:{\cc{k{-}4}}}] 
{};
    \node (g23) [gauge,right of=g22,label={[rotate=-45]below 
right:{\dd{k{-}4}}}] {};
	\node (g24) [gauge,right of=g23,label={[rotate=-45]below 
right:{\cc{k{-}5}}}] 
{};
	\node (g25) [gauge,right of=g24,label={[rotate=-45]below 
right:{\dd{k{-}5}}}] {};
	\node (g26) [right of=g25] {$\scriptstyle{\ldots}$};
	\node (g27) [gauge,right of=g26,label={[rotate=-45]below right:{\cc{1}}}] 
{};
	\node (g28) [gauge,right of=g27,label={[rotate=-45]below right:{\dd{1}}}] 
{};
	\node (f8) [flavour,above of=g8,label=above:{\bb{0}}] {};
	\node (f22) [flavour,above of=g22,label=above:{\bb{0}}] {};
	\node (b1) [gauge,above of=g12,label=above:{\cc{1}}] {};
    \node (b2) [gauge,above of=g14,label=above:{\cc{1}}] {};
    \node (b0) [above of=g13] {$\scriptstyle{\cdots}$};
    \node (ghost) [above of=g15] {};
    \node (b3) [flavour,above of=ghost,label=above:{\cc{n{-}l{-}1}}] {};
	\node (b4) [gauge,above of=g16,label=above:{\cc{1}}] {};
    \node (b5) [gauge,above of=g18,label=above :{\cc{1}}] {};
    \node (fb4) [flavour,above right of=b4,label=above:{\dd{4}}] {};
    \node (fb5) [flavour,above right of=b5,label=above:{\dd{4}}] {};
    \node (b6) [above of=g17] {$\scriptstyle{\cdots}$};
    \draw[decoration={brace,mirror,raise=30pt},decorate,thick]
  (g14) -- node[above=30pt] {\tiny{$l_0$}} (g12);
    \draw[decoration={brace,mirror,raise=15pt},decorate,thick]
  (fb5) -- node[above=15pt] {\tiny{$l_1$}} (fb4);
	\draw (g2)--(g3) (g3)--(g4) (g4)--(g5) (g5)--(g6) (g6)--(g7) 
(g7)--(g8) (g8)--(g9) (g9)--(g10) (g10)--(g11) (g11)--(g12) (g12)--(g13)
(g13)--(g14) (g14)--(g15) (g15)--(g16) (g16)--(g17) (g17)--(g18) (g18)--(g19)
(g19)--(g20) (g20)--(g21) (g21)--(g22) (g22)--(g23) (g23)--(g24) (g24)--(g25) 
(g25)--(g26) (g26)--(g27) (g27)--(g28) 
(g15)--(b1) (g15)--(b2) (g15)--(b3)
(g15)--(b4) (g15)--(b5)
(g8)--(f8) (g22)--(f22)
(b4)--(fb4) (b5)--(fb5);
	\end{tikzpicture}
	}
     \label{eq:k_nM5_magQuiver_infinite_+n-1}
\end{align}
and the Coulomb branch dimension has increased by $l$ quaternionic units in 
comparison to \eqref{eq:k_nM5_magQuiver_finite}. 
\paragraph{Discrete gauging.}
Consider the case in which all possible $(n-1)$ pairs of half \NS\ under-go the 
transition \eqref{eq:1d_transition}, then the brane configuration 
becomes
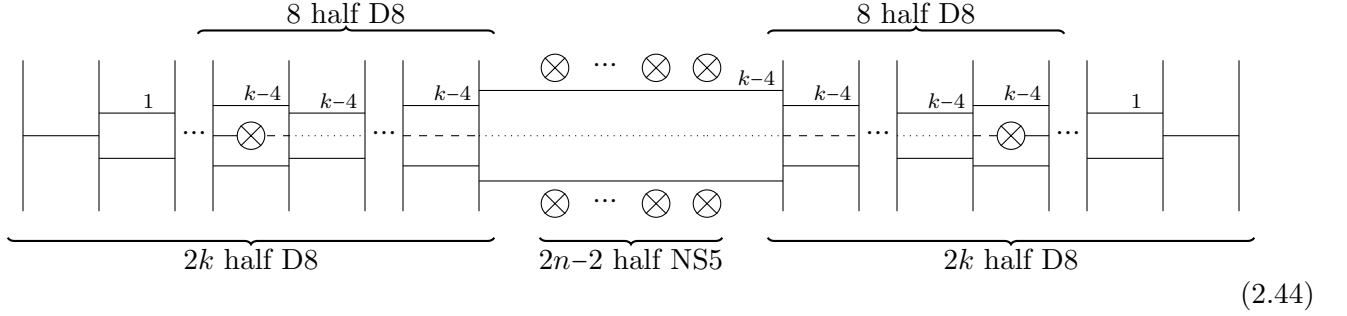
\begin{align}
  \raisebox{-.5\height}{
\begin{tikzpicture}
        \DeightMany{1}{-1}{1}
        \Dbrane{-1,0}{0,0}
		\DeightMany{1}{0}{1}
		\DsixOMinusFree{$1$}{0}{0}
		\DeightMany{1}{1}{1}
        \draw (1.25,0) node {$\cdots$};
        \DeightMany{1}{1.5}{1}
        \DsixOMinusFree{$k{-}4$}{1.5}{1}
        \Dbrane{1.5,0}{2,0}
        \OPlusTilde{2,0}{2.5,0}
		\DeightMany{1}{2.5}{1}
		\DsixOPlusFree{$k{-}4$}{2.5}{0}
		\DeightMany{1}{3.5}{1}
        \ns{2,0}
        \draw (3.75,0) node {$\cdots$};
        \DeightMany{1}{4}{1}
        \DeightMany{1}{5}{1}
        \DsixOPlusTildeFree{$k{-}4$}{4}{1}
        \DsixOPlusFree{$k{-}4$}{8}{3}
        \draw (6.66,0.9) node {$\cdots$};
        \draw (6.66,-0.9) node {$\cdots$};
     \draw (5,0.6)--(8,0.6);
     \draw (5,-0.6)--(8,-0.6);
     \OPlus{5,0}{8,0}
        \ns{6,0.9}
        \ns{6,-0.9}
        \ns{7.33,0.9}
        \ns{7.33,-0.9}
        \ns{8,0.9}
        \ns{8,-0.9}
        \DeightMany{1}{9}{1}
        \DsixOPlusTildeFree{$k{-}4$}{9}{1}
        \DeightMany{1}{10}{1}
        \draw (10.25,0) node {$\cdots$};
        \DeightMany{1}{10.5}{1}
		\DsixOPlusFree{$k{-}4$}{10.5}{0}
		\DeightMany{1}{11.5}{1}
		\DsixOMinusFree{$k{-}4$}{11.5}{1}
		\OPlusTilde{11.5,0}{12,0}
        \Dbrane{12,0}{12.5,0}
        \DeightMany{1}{12.5}{1}
        \ns{12,0}
        \draw (12.75,0) node {$\cdots$};
		\DeightMany{1}{13}{1}
		\DsixOMinusFree{$1$}{13}{0}
		\DeightMany{1}{14}{1}
		\Dbrane{14,0}{15,0}
		\DeightMany{1}{15}{1}
\draw[decoration={brace,mirror,raise=10pt},decorate,thick]
  (-1.2,-1) -- node[below=10pt] {$2k$ half \De } (5.2,-1);
  \draw[decoration={brace,mirror,raise=10pt},decorate,thick]
  (5.2,1) -- node[above=10pt] {$8$ half \De } (1.5-0.2,1);
\draw[decoration={brace,mirror,raise=10pt},decorate,thick]
  (9-0.2,-1) -- node[below=10pt] {$2k$ half \De } (15.2,-1);
\draw[decoration={brace,mirror,raise=10pt},decorate,thick]
  (12.7,1) -- node[above=10pt] {$8$ half \De } (9-0.2,1);
\draw[decoration={brace,mirror,raise=10pt},decorate,thick]
  (6-0.2,-1) -- node[below=10pt] {$2n{-}2$ half \NS } (8.2,-1);
	\end{tikzpicture}
	}
	\label{eq:branes_k_nM5_diml}
\end{align}
and the $x^6$ distance between the neighbouring pairs still corresponds to 
tensor multiplet, i.e.\ an inverse gauge coupling. 
By the rules establish so far, the magnetic quiver reads as follows:
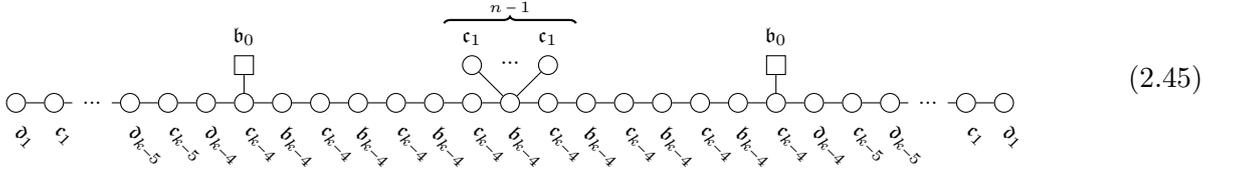
\begin{align}
 \raisebox{-.5\height}{
 	\begin{tikzpicture}
 	\tikzset{node distance = 0.5cm};
	\tikzstyle{gauge} = [circle, draw,inner sep=2.5pt];
	\tikzstyle{flavour} = [regular polygon,regular polygon sides=4,inner 
sep=2.5pt, draw];
	\node (g2) [gauge,label={[rotate=-45]below right:{\dd{1}}}] 
{};
	\node (g3) [gauge,right of=g2,label={[rotate=-45]below right:{\cc{1}}}] {};
	\node (g4) [right of=g3] {$\scriptstyle{\ldots}$};
	\node (g5) [gauge,right of=g4,label={[rotate=-45]below 
right:{\dd{k{-}5}}}] {};
	\node (g6) [gauge,right of=g5,label={[rotate=-45]below right:{\cc{k{-}5}}}] 
{};
    \node (g7) [gauge,right of=g6,label={[rotate=-45]below 
right:{\dd{k{-}4}}}] {};
	\node (g8) [gauge,right of=g7,label={[rotate=-45]below right:{\cc{k{-}4}}}] 
{};
    \node (g9) [gauge,right of=g8,label={[rotate=-45]below 
right:{\bb{k{-}4}}}] {};
	\node (g10) [gauge,right of=g9,label={[rotate=-45]below 
right:{\cc{k{-}4}}}] {};
    \node (g11) [gauge,right of=g10,label={[rotate=-45]below 
right:{\bb{k{-}4}}}] {};
	\node (g12) [gauge,right of=g11,label={[rotate=-45]below 
right:{\cc{k{-}4}}}] {};
    \node (g13) [gauge,right of=g12,label={[rotate=-45]below 
right:{\bb{k{-}4}}}] {};
	\node (g14) [gauge,right of=g13,label={[rotate=-45]below 
right:{\cc{k{-}4}}}] {};
    \node (g15) [gauge,right of=g14,label={[rotate=-45]below 
right:{\bb{k{-}4}}}] {};
	\node (g16) [gauge,right of=g15,label={[rotate=-45]below 
right:{\cc{k{-}4}}}] {};
    \node (g17) [gauge,right of=g16,label={[rotate=-45]below 
right:{\bb{k{-}4}}}] {};
	\node (g18) [gauge,right of=g17,label={[rotate=-45]below 
right:{\cc{k{-}4}}}] {};
    \node (g19) [gauge,right of=g18,label={[rotate=-45]below 
right:{\bb{k{-}4}}}] {};
	\node (g20) [gauge,right of=g19,label={[rotate=-45]below 
right:{\cc{k{-}4}}}] {};
    \node (g21) [gauge,right of=g20,label={[rotate=-45]below 
right:{\bb{k{-}4}}}] {};
	\node (g22) [gauge,right of=g21,label={[rotate=-45]below 
right:{\cc{k{-}4}}}] 
{};
    \node (g23) [gauge,right of=g22,label={[rotate=-45]below 
right:{\dd{k{-}4}}}] {};
	\node (g24) [gauge,right of=g23,label={[rotate=-45]below 
right:{\cc{k{-}5}}}] 
{};
	\node (g25) [gauge,right of=g24,label={[rotate=-45]below 
right:{\dd{k{-}5}}}] {};
	\node (g26) [right of=g25] {$\scriptstyle{\ldots}$};
	\node (g27) [gauge,right of=g26,label={[rotate=-45]below right:{\cc{1}}}] 
{};
	\node (g28) [gauge,right of=g27,label={[rotate=-45]below right:{\dd{1}}}] 
{};
	\node (f8) [flavour,above of=g8,label=above:{\bb{0}}] {};
	\node (f22) [flavour,above of=g22,label=above:{\bb{0}}] {};
	\node (b1) [gauge,above of=g14,label=above:{\cc{1}}] {};
    \node (b2) [gauge,above  of=g16,label=above:{\cc{1}}] {};
    \node (b0) [above of=g15] {$\scriptstyle{\cdots}$};
    \draw[decoration={brace,mirror,raise=30pt},decorate,thick]
  (g17) -- node[above=30pt] {\tiny{$n-1$}} (g13);
	\draw (g2)--(g3) (g3)--(g4) (g4)--(g5) (g5)--(g6) (g6)--(g7) 
(g7)--(g8) (g8)--(g9) (g9)--(g10) (g10)--(g11) (g11)--(g12) (g12)--(g13)
(g13)--(g14) (g14)--(g15) (g15)--(g16) (g16)--(g17) (g17)--(g18) (g18)--(g19)
(g19)--(g20) (g20)--(g21) (g21)--(g22) (g22)--(g23) (g23)--(g24) (g24)--(g25) 
(g25)--(g26) (g26)--(g27) (g27)--(g28) 
(g15)--(b1) (g15)--(b2)
(g8)--(f8) (g22)--(f22);
	\end{tikzpicture}
	}
     \label{eq:k_nM5_magQuiver_infinite_intermediate}
\end{align}
In particular, once all $(n-1)$ pairs of half \NS s in the centre of 
the brane configuration \eqref{eq:branes_nM5_on_Dk_finite_coupling} have left 
the orientifold, there is only one type of $\calg_{1}$ gauge node in the 
magnetic quiver.

Focusing on two neighbouring 
pairs, one could suspend half \Dtwo\ branes between the half \NS\ branes. 
Sending the $x^6$ distance to zero creates tensionless strings on the 
\Dtwo s. 
The analogous effect for \Mf\ branes on an A-type singularity has been 
considered in \cite{Hanany:2018vph} and argued to be a \emph{discrete gauging} 
of a permutation group acting on the (pairs of) \NS\ branes. Here, the argument 
applies to $n$ mirror pairs of \NS s in the presence an \Os\ plane.
The possibilities for the pairs to become coincident along $x^6$ are labeled by 
partitions $\{n_i\}_{i=1,\ldots,l}$ of $(n-1)$, meaning that $n_i$ of all pairs 
coincide in definite $x^6$ position and so on and so forth. Hence, one gauges a 
$\prod_{i=1}^l 
S_{n_i}$ discrete group and $(n-1-l)$ gauge couplings have been send to 
infinity. The brane configuration looks like
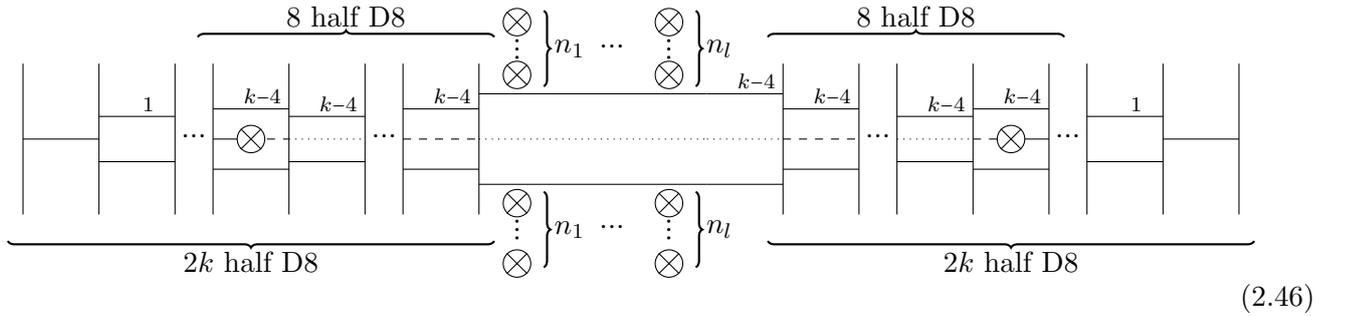
\begin{align}
  \raisebox{-.5\height}{
\begin{tikzpicture}
        \DeightMany{1}{-1}{1}
        \Dbrane{-1,0}{0,0}
		\DeightMany{1}{0}{1}
		\DsixOMinusFree{$1$}{0}{0}
		\DeightMany{1}{1}{1}
        \draw (1.25,0) node {$\cdots$};
        \DeightMany{1}{1.5}{1}
        \DsixOMinusFree{$k{-}4$}{1.5}{1}
        \Dbrane{1.5,0}{2,0}
        \OPlusTilde{2,0}{2.5,0}
		\DeightMany{1}{2.5}{1}
		\DsixOPlusFree{$k{-}4$}{2.5}{0}
		\DeightMany{1}{3.5}{1}
        \ns{2,0}
        \draw (3.75,0) node {$\cdots$};
        \DeightMany{1}{4}{1}
        \DeightMany{1}{5}{1}
        \DsixOPlusTildeFree{$k{-}4$}{4}{1}
        \DsixOPlusFree{$k{-}4$}{8}{3}
        \draw (6.75,1.2) node {$\cdots$};
        \draw (6.75,-1.2) node {$\cdots$};
     \draw (5,0.6)--(8,0.6);
     \draw (5,-0.6)--(8,-0.6);
     \OPlus{5,0}{8,0}
        \ns{5.5,0.85}
        \draw (5.5,1.2) node {$\vdots$};
        \ns{5.5,1.55}
        \draw[decoration={brace,mirror,raise=10pt},decorate,thick]
  (5.5,0.7) -- node[right=10pt] {$n_1$} (5.5,1.7);
        \ns{5.5,-0.85}
        \draw (5.5,-1.2) node {$\vdots$};
        \ns{5.5,-1.65}
        \draw[decoration={brace,mirror,raise=10pt},decorate,thick]
  (5.5,-1.7) -- node[right=10pt] {$n_1$} (5.5,-0.7);
        \ns{7.5,0.85}
        \draw (7.5,1.2) node {$\vdots$};
        \ns{7.5,1.55}
        \draw[decoration={brace,mirror,raise=10pt},decorate,thick]
  (7.5,0.7) -- node[right=10pt] {$n_l$} (7.5,1.7);
        \ns{7.5,-0.85}
        \draw (7.5,-1.2) node {$\vdots$};
        \ns{7.5,-1.65}
        \draw[decoration={brace,mirror,raise=10pt},decorate,thick]
  (7.5,-1.7) -- node[right=10pt] {$n_l$} (7.5,-0.7);
        \DeightMany{1}{9}{1}
        \DsixOPlusTildeFree{$k{-}4$}{9}{1}
        \DeightMany{1}{10}{1}
        \draw (10.25,0) node {$\cdots$};
        \DeightMany{1}{10.5}{1}
		\DsixOPlusFree{$k{-}4$}{10.5}{0}
		\DeightMany{1}{11.5}{1}
		\DsixOMinusFree{$k{-}4$}{11.5}{1}
		\OPlusTilde{11.5,0}{12,0}
        \Dbrane{12,0}{12.5,0}
        \DeightMany{1}{12.5}{1}
        \ns{12,0}
        \draw (12.75,0) node {$\cdots$};
		\DeightMany{1}{13}{1}
		\DsixOMinusFree{$1$}{13}{0}
		\DeightMany{1}{14}{1}
		\Dbrane{14,0}{15,0}
		\DeightMany{1}{15}{1}
\draw[decoration={brace,mirror,raise=10pt},decorate,thick]
  (-1.2,-1) -- node[below=10pt] {$2k$ half \De } (5.2,-1);
  \draw[decoration={brace,mirror,raise=10pt},decorate,thick]
  (5.2,1) -- node[above=10pt] {$8$ half \De } (1.5-0.2,1);
\draw[decoration={brace,mirror,raise=10pt},decorate,thick]
  (9-0.2,-1) -- node[below=10pt] {$2k$ half \De } (15.2,-1);
\draw[decoration={brace,mirror,raise=10pt},decorate,thick]
  (12.7,1) -- node[above=10pt] {$8$ half \De } (9-0.2,1);
	\end{tikzpicture}
	}
	\label{eq:branes_k_nM5_dim1_trans_generic}
\end{align}
Focusing on a stack of $n_i$ \NS\ branes in configuration 
\eqref{eq:branes_k_nM5_dim1_trans_generic}, which can be depicted as 
displaced 
along $x^{7,8,9}$, then \Df\ branes suspended between the \NS\ branes 
contribute to the massless degrees of freedom. Analogous to a stack of branes 
that is half BPS, 
the contribution lies in a gauge group and one additional hypermultiplet. Due 
to 
the presence of the  \Osp\ plane, which becomes a magnetic \Osmt\ plane, there 
is a non-trivial projection which reduces the gauge group to a symplectic group 
and the additional hypermultiplet transforms in the traceless second 
anti-symmetric representation $\Lambda^2$ of the symplectic gauge group. Since 
this vanished 
for $\calg_1$, it has not been detailed so far. Collecting all contributions 
for the brane configuration \eqref{eq:branes_k_nM5_dim1_trans_generic}, the 
resulting magnetic quiver reads as follows:
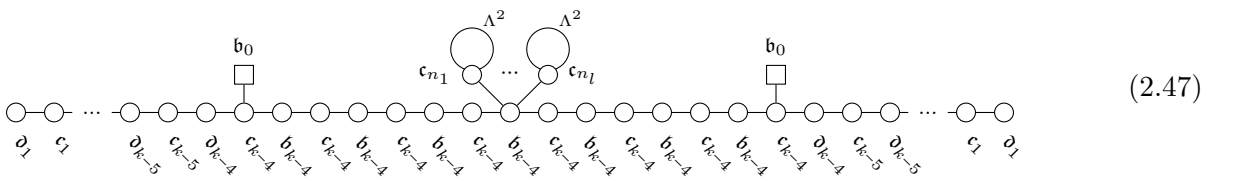
\begin{align}
 \raisebox{-.5\height}{
 	\begin{tikzpicture}
 	\tikzset{node distance = 0.5cm};
	\tikzstyle{gauge} = [circle, draw,inner sep=2.5pt];
	\tikzstyle{flavour} = [regular polygon,regular polygon sides=4,inner 
sep=2.5pt, draw];
	\node (g2) [gauge,label={[rotate=-45]below right:{\dd{1}}}] 
{};
	\node (g3) [gauge,right of=g2,label={[rotate=-45]below right:{\cc{1}}}] {};
	\node (g4) [right of=g3] {$\scriptstyle{\ldots}$};
	\node (g5) [gauge,right of=g4,label={[rotate=-45]below 
right:{\dd{k{-}5}}}] {};
	\node (g6) [gauge,right of=g5,label={[rotate=-45]below right:{\cc{k{-}5}}}] 
{};
    \node (g7) [gauge,right of=g6,label={[rotate=-45]below 
right:{\dd{k{-}4}}}] {};
	\node (g8) [gauge,right of=g7,label={[rotate=-45]below right:{\cc{k{-}4}}}] 
{};
    \node (g9) [gauge,right of=g8,label={[rotate=-45]below 
right:{\bb{k{-}4}}}] {};
	\node (g10) [gauge,right of=g9,label={[rotate=-45]below 
right:{\cc{k{-}4}}}] {};
    \node (g11) [gauge,right of=g10,label={[rotate=-45]below 
right:{\bb{k{-}4}}}] {};
	\node (g12) [gauge,right of=g11,label={[rotate=-45]below 
right:{\cc{k{-}4}}}] {};
    \node (g13) [gauge,right of=g12,label={[rotate=-45]below 
right:{\bb{k{-}4}}}] {};
	\node (g14) [gauge,right of=g13,label={[rotate=-45]below 
right:{\cc{k{-}4}}}] {};
    \node (g15) [gauge,right of=g14,label={[rotate=-45]below 
right:{\bb{k{-}4}}}] {};
	\node (g16) [gauge,right of=g15,label={[rotate=-45]below 
right:{\cc{k{-}4}}}] {};
    \node (g17) [gauge,right of=g16,label={[rotate=-45]below 
right:{\bb{k{-}4}}}] {};
	\node (g18) [gauge,right of=g17,label={[rotate=-45]below 
right:{\cc{k{-}4}}}] {};
    \node (g19) [gauge,right of=g18,label={[rotate=-45]below 
right:{\bb{k{-}4}}}] {};
	\node (g20) [gauge,right of=g19,label={[rotate=-45]below 
right:{\cc{k{-}4}}}] {};
    \node (g21) [gauge,right of=g20,label={[rotate=-45]below 
right:{\bb{k{-}4}}}] {};
	\node (g22) [gauge,right of=g21,label={[rotate=-45]below 
right:{\cc{k{-}4}}}] 
{};
    \node (g23) [gauge,right of=g22,label={[rotate=-45]below 
right:{\dd{k{-}4}}}] {};
	\node (g24) [gauge,right of=g23,label={[rotate=-45]below 
right:{\cc{k{-}5}}}] 
{};
	\node (g25) [gauge,right of=g24,label={[rotate=-45]below 
right:{\dd{k{-}5}}}] {};
	\node (g26) [right of=g25] {$\scriptstyle{\ldots}$};
	\node (g27) [gauge,right of=g26,label={[rotate=-45]below right:{\cc{1}}}] 
{};
	\node (g28) [gauge,right of=g27,label={[rotate=-45]below right:{\dd{1}}}] 
{};
	\node (f8) [flavour,above of=g8,label=above:{\bb{0}}] {};
	\node (f22) [flavour,above of=g22,label=above:{\bb{0}}] {};
	\node (b1) [gauge,above of=g14,label= left:{\cc{n_1}}] {};
    \node (b2) [gauge,above of=g16,label= right:{\cc{n_l}}] {};
    \node (b0) [above of=g15] {$\scriptstyle{\cdots}$};
    	\draw (6.3,1.25) node {\tiny{$\Lambda^2$}};
	\draw [-] (6+0.095,0.575) arc (-70:90:8pt);
	\draw [-] (6+-0.095,0.575) arc (250:90:8pt);
	\draw (7.3,1.25) node {\tiny{$\Lambda^2$}};
	\draw [-] (7+0.095,0.575) arc (-70:90:8pt);
	\draw [-] (7+-0.095,0.575) arc (250:90:8pt);
	\draw (g2)--(g3) (g3)--(g4) (g4)--(g5) (g5)--(g6) (g6)--(g7) 
(g7)--(g8) (g8)--(g9) (g9)--(g10) (g10)--(g11) (g11)--(g12) (g12)--(g13)
(g13)--(g14) (g14)--(g15) (g15)--(g16) (g16)--(g17) (g17)--(g18) (g18)--(g19)
(g19)--(g20) (g20)--(g21) (g21)--(g22) (g22)--(g23) (g23)--(g24) (g24)--(g25) 
(g25)--(g26) (g26)--(g27) (g27)--(g28) 
(g15)--(b1) (g15)--(b2)
(g8)--(f8) (g22)--(f22);
	\end{tikzpicture}
	}
     \label{eq:k_multiple_M5_magQuiver_dim=l_generic}
\end{align}
%
The question is now, whether there is a relation between the Coulomb branches 
of \eqref{eq:k_nM5_magQuiver_infinite_intermediate} and 
\eqref{eq:k_multiple_M5_magQuiver_dim=l_generic}. Physically, the Coulomb 
branch of \eqref{eq:branes_k_nM5_dim1_trans_generic} describes the 
Higgs branch of the phase where the maximal number of transitions of the type 
\eqref{eq:1d_transition} have occurred. Hence, $(n-1)$ gauge couplings are 
infinite. In contrast, \eqref{eq:k_multiple_M5_magQuiver_dim=l_generic} 
starts 
from the phase \eqref{eq:branes_k_nM5_diml} and tunes further $(n-1-l)$ 
gauge 
couplings to infinity. Again, the transition is due to a discrete gauging 
\cite{Hanany:2018vph} of the permutation subgroup $\prod_{i=1}^l S_{n_i}$ of 
the 
full permutation group $S_{n-1}$ acting on the $(n-1)$ pairs of half \NS\ 
branes 
in \eqref{eq:branes_k_nM5_diml}.
Gauging a discrete permutation group on the Higgs branch of the electric 
theory corresponds to a quotient of the permutation group on 
the Coulomb branch of the magnetic theory. As shown in \cite[Sec.\ 
2.2]{Hanany:2018cgo}, the discrete quotient on the Coulomb branch translates 
into an simple operation on the $\calg_1$ bouquet of 
\eqref{eq:k_nM5_magQuiver_infinite_intermediate} that results in 
\eqref{eq:k_multiple_M5_magQuiver_dim=l_generic}. Thus, the relation between 
the moduli spaces is
\begin{align}
 \Coulomb^{3d} \left( \magQuivText 
\eqref{eq:k_multiple_M5_magQuiver_dim=l_generic} \right)
= 
\Coulomb^{3d} \left( \magQuivText 
\eqref{eq:k_nM5_magQuiver_infinite_+n-1} \right) \slash 
\prod_{i=1}^{l} S_{n_i} \,.
\label{eq:discrete_gauging_k_intermediate}
\end{align}
As a remark, this \emph{discrete gauging} transition can occure in any of the 
intermediate phases described by 
\eqref{eq:k_nM5_magQuiver_infinite_intermediate}. 
There, one would label all possible 
cases by partitions of $l$ instead. Since the discussion is analogous to the 
one just presented, it is not further detailed.

Likewise, one may consider discrete gauging in the phase 
\eqref{eq:k_nM5_magQuiver_infinite_+n-1}. Without loss of generality, one can 
assume that all pairs of \NS\ branes that underwent transition 
\eqref{eq:branes_k_nM5_dim1_transition} (or 
\eqref{eq:branes_k_nM5_dim1_transition_var2}) are in the same $x^6$ interval 
defined by two half \NS\ on the orientifold. Then, one can consider either 
family of \NS\ pairs becoming coincident along $x^6$, i.e.\ discrete gauging. 
For the pure $\calg_1$ bouquet of size $l_0$, the resulting effects is the same 
as above due to \cite[Sec.\ 2.2]{Hanany:2018cgo}. For the ($\calg_1  \circ - 
\Box \dalg_4$) bouquet of size $l_1$, the discrete quotient effect on the 
magnetic quiver is a straightforward extension of \cite[Sec.\ 
2.2]{Hanany:2018cgo}, i.e.\ one obtains an $\calg_{l_1}$ gauge node with a 
$\dalg_4$ flavour node and an additional traceless 2nd rank anti-symmetric 
hypermultiplet.
\paragraph{Small instanton transition.}
Return to the brane configuration \eqref{eq:branes_k_nM5_diml}, and consider 
how to take the separation of the two half \NS s that remain on the 
orientifold to zero. This is the next logical question, because by the previous 
paragraphs 
one knows how to take all other gauge couplings to infinity. 
In order to take the last remaining gauge coupling to infinity, one has to 
reunite the 
remaining two \NS\ on the orientifold and then remove them from the \Os\ plane. 
By transitioning the two outermost half \NS\ branes through the half \De\ 
branes, one creates \Ds\ branes according to rules in 
\eqref{eq:brane_creation_all}. At the instance during which the \NS\ become 
coincident and leave the \Os\ plane, the \Ds\ brane reconnect such that the 
resulting brane configurations becomes
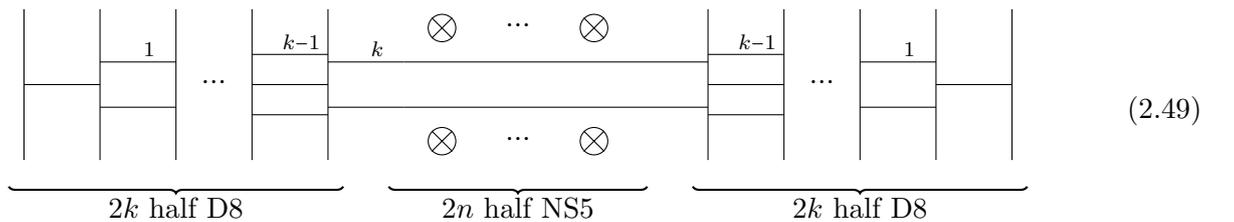
\begin{align}
 \raisebox{-.5\height}{
\begin{tikzpicture}
        \DeightMany{1}{0}{1}
        \Dbrane{0,0}{1,0}
		\DeightMany{1}{1}{1}
		\DsixOMinusFree{$1$}{1}{0}
		\DeightMany{1}{2}{1}
        \draw (2.5,0) node {$\cdots$};
        \DeightMany{1}{3}{1}
        \DsixOMinusTildeFree{$k{-}1$}{3}{1}
        \DeightMany{1}{4}{1}
        \DsixOMinusFree{$k$}{4}{0}
        \Dbrane{5,0.3}{9,0.3}
        \Dbrane{5,-0.3}{9,-0.3}
        \ns{5.5,0.75}
        \ns{5.5,-0.75}
        \draw (6.5,0.75) node {$\cdots$};
        \draw (6.5,-0.75) node {$\cdots$};
        \ns{7.5,0.75}
        \ns{7.5,-0.75}
        \DeightMany{1}{9}{1}
        \DsixOMinusTildeFree{$k{-}1$}{9}{1}
        \DeightMany{1}{10}{1}
        \draw (10.5,0) node {$\cdots$};
        \DeightMany{1}{11}{1}
		\DsixOMinusFree{$1$}{11}{0}
		\DeightMany{1}{12}{1}
        \Dbrane{12,0}{13,0}
        \DeightMany{1}{13}{1}
\draw[decoration={brace,mirror,raise=10pt},decorate,thick]
  (-0.2,-1) -- node[below=10pt] {$2k$ half \De } (4.2,-1);
\draw[decoration={brace,mirror,raise=10pt},decorate,thick]
  (9-0.2,-1) -- node[below=10pt] {$2k$ half \De } (13.2,-1);
\draw[decoration={brace,mirror,raise=10pt},decorate,thick]
  (5-0.2,-1) -- node[below=10pt] {$2n$ half \NS } (8.2,-1);
	\end{tikzpicture}
	}
	\label{eq:branes_nM5_on_Dk_bouquet}
\end{align}
and, here, all \NS\ pairs are separated along $x^6$. 
By the arguments presented above, the magnetic quiver for 
\eqref{eq:branes_nM5_on_Dk_bouquet} is readily read off to be
\begin{align}
 \raisebox{-.5\height}{
 	\begin{tikzpicture}
 	\tikzset{node distance = 0.5cm};
	\tikzstyle{gauge} = [circle, draw,inner sep=2.5pt];
	\tikzstyle{flavour} = [regular polygon,regular polygon sides=4,inner 
sep=2.5pt, draw];
	\node (g2) [gauge,label={[rotate=-45]below right:{\dd{1}}}] 
{};
	\node (g3) [gauge,right of=g2,label={[rotate=-45]below right:{\cc{1}}}] {};
\	\node (g4) [gauge,right of=g3,label={[rotate=-45]below 
right:{\dd{2}}}] {};
	\node (g5) [gauge,right of=g4,label={[rotate=-45]below right:{\cc{2}}}] 
{};
	\node (g6) [right of=g5] {$\scriptstyle{\ldots}$};
    \node (g7) [gauge,right of=g6,label={[rotate=-45]below 
right:{\dd{k{-}1}}}] {};
	\node (g8) [gauge,right of=g7,label={[rotate=-45]below right:{\cc{k{-}1}}}] 
{};
    \node (g9) [gauge,right of=g8,label={[rotate=-45]below 
right:{\dd{k}}}] {};
	\node (g10) [gauge,right of=g9,label={[rotate=-45]below 
right:{\cc{k{-}1}}}] 
{};
    \node (g11) [gauge,right of=g10,label={[rotate=-45]below 
right:{\dd{k{-}1}}}] {};
    \node (g12) [right of=g11] {$\scriptstyle{\ldots}$};
	\node (g13) [gauge,right of=g12,label={[rotate=-45]below right:{\cc{2}}}] 
{};
    \node (g14) [gauge,right of=g13,label={[rotate=-45]below right:{\dd{2}}}] 
{};
	\node (g15) [gauge,right of=g14,label={[rotate=-45]below right:{\cc{1}}}] 
{};
	\node (g16) [gauge,right of=g15,label={[rotate=-45]below right:{\dd{1}}}] 
{};
	\node (b1) [gauge,above of=g8,label=above:{\cc{1}}] {};
    \node (b2) [gauge,above of=g10,label=above:{\cc{1}}] {};
    \node (b0) [above of=g9] {$\scriptstyle{\cdots}$};
    \draw[decoration={brace,mirror,raise=30pt},decorate,thick]
  (g11) -- node[above=30pt] {\tiny{$n$}} (g7);
	\draw  (g2)--(g3) (g3)--(g4) (g4)--(g5) (g5)--(g6) (g6)--(g7) 
(g7)--(g8) (g8)--(g9) (g9)--(g10) (g10)--(g11) (g11)--(g12) (g12)--(g13)
(g13)--(g14) (g14)--(g15) (g15)--(g16)  (g9)--(b1) (g9)--(b2);
	\end{tikzpicture}
	} 
	\,,
		\label{eq:k_magQuiver_infinite_separated}
\end{align}
and its Coulomb branch describes a Higgs branch phase of 
\eqref{eq:electric_quiver_n_M5_Dk} with $n$ gauge couplings tuned to infinity.

The nature of this last transition can be deduced in multiple ways. On the one 
hand, the starting point \eqref{eq:branes_k_nM5_diml} describes one 
remaining \Mf\ that fractionated on the D-type singularity. Taking it off the 
singularity corresponds to the small $E_8$ instanton transition as discussed 
above. Put differently, before the transition there existed one extra tensor 
multiplet, which is lost afterwards. Since the number of 
vector 
multiplets 
has not changed, there need to be $29$ additional hypermultiplets to satisfy 
\eqref{eq:anomaly_cancellation}.
On the other hand, one can apply quiver subtraction to 
\eqref{eq:k_magQuiver_infinite_separated} and 
\eqref{eq:k_nM5_magQuiver_infinite_+n-1} and deduce that the 
difference 
quiver is precisely \eqref{eq:magQuiver_k=4_infinite}. 
As detailed in Section \ref{sec:Hasse}, the transverse slice of the Coulomb 
branch of 
\eqref{eq:k_nM5_magQuiver_infinite_+n-1} inside the Coulomb branch 
of \eqref{eq:k_magQuiver_infinite_separated}
is the closure of the minimal nilpotent orbit of $E_8$. 

As discussed above, the $x^6$ separation between the pairs of \NS\ branes in 
\eqref{eq:branes_nM5_on_Dk_bouquet} corresponds to tensor multiplets. The 
possibilities of taking different subsets of gauge couplings to infinity are, 
again, labeled by partitions $\{n_i\}_{i=1,\ldots,l}$ of $n$, meaning 
$n_i$ pairs of half \NS\ brane coincide 
along $x^6$, with $\sum_{i=1}^l n_i = n$.
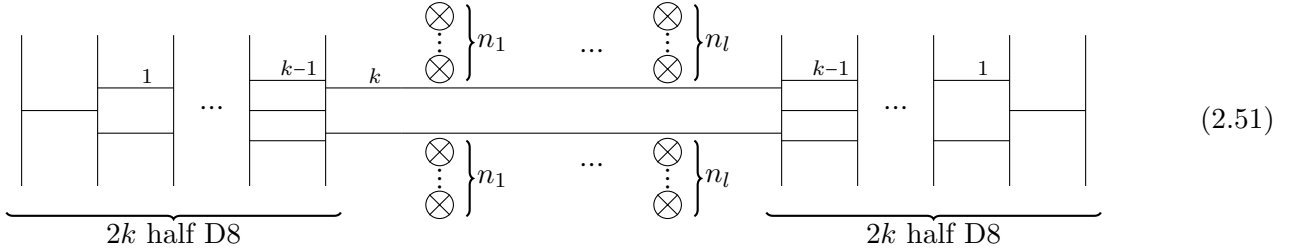
\begin{align}
 \raisebox{-.5\height}{
\begin{tikzpicture}
        \DeightMany{1}{0}{1}
        \Dbrane{0,0}{1,0}
		\DeightMany{1}{1}{1}
		\DsixOMinusFree{$1$}{1}{0}
		\DeightMany{1}{2}{1}
        \draw (2.5,0) node {$\cdots$};
        \DeightMany{1}{3}{1}
        \DsixOMinusTildeFree{$k{-}1$}{3}{1}
        \DeightMany{1}{4}{1}
        \DsixOMinusFree{$k$}{4}{0}
        \Dbrane{5,0.3}{10,0.3}
        \Dbrane{5,-0.3}{10,-0.3}
        \ns{5.5,0.55}
        \draw (5.5,0.9) node {$\vdots$};
        \ns{5.5,1.25}
        \draw[decoration={brace,mirror,raise=10pt},decorate,thick]
  (5.5,0.4) -- node[right=10pt] {$n_1$} (5.5,1.4);
        \ns{5.5,-0.55}
        \draw (5.5,-0.9) node {$\vdots$};
        \ns{5.5,-1.25}
        \draw[decoration={brace,mirror,raise=10pt},decorate,thick]
  (5.5,-1.4) -- node[right=10pt] {$n_1$} (5.5,-0.4);
        \draw (7.5,0.75) node {$\cdots$};
        \draw (7.5,-0.75) node {$\cdots$};
        \ns{8.5,0.55}
        \draw (8.5,0.9) node {$\vdots$};
        \ns{8.5,1.25}
        \draw[decoration={brace,mirror,raise=10pt},decorate,thick]
  (8.5,0.4) -- node[right=10pt] {$n_l$} (8.5,1.4);
        \ns{8.5,-0.55}
        \draw (8.5,-0.9) node {$\vdots$};
        \ns{8.5,-1.25}
        \draw[decoration={brace,mirror,raise=10pt},decorate,thick]
  (8.5,-1.4) -- node[right=10pt] {$n_l$} (8.5,-0.4);
        \DeightMany{1}{10}{1}
        \DsixOMinusTildeFree{$k{-}1$}{10}{1}
        \DeightMany{1}{11}{1}
        \draw (11.5,0) node {$\cdots$};
        \DeightMany{1}{12}{1}
		\DsixOMinusFree{$1$}{12}{1}
		\DeightMany{1}{13}{1}
        \Dbrane{13,0}{14,0}
        \DeightMany{1}{14}{1}
\draw[decoration={brace,mirror,raise=10pt},decorate,thick]
  (-0.2,-1) -- node[below=10pt] {$2k$ half \De } (4.2,-1);
\draw[decoration={brace,mirror,raise=10pt},decorate,thick]
  (10-0.2,-1) -- node[below=10pt] {$2k$ half \De } (14.2,-1);
	\end{tikzpicture}
	}
\end{align}
The logic is the same as in 
\eqref{eq:k_multiple_M5_magQuiver_dim=l_generic}. Therefore, the magnetic 
quiver becomes 
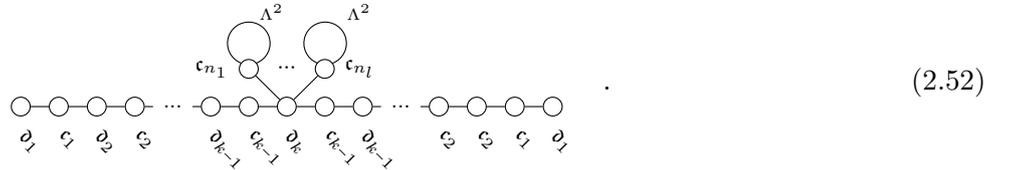
\begin{align}
 \raisebox{-.5\height}{
 	\begin{tikzpicture}
 	\tikzset{node distance = 0.5cm};
	\tikzstyle{gauge} = [circle, draw,inner sep=2.5pt];
	\tikzstyle{flavour} = [regular polygon,regular polygon sides=4,inner 
sep=2.5pt, draw];
	\node (g2) [gauge,label={[rotate=-45]below right:{\dd{1}}}] 
{};
	\node (g3) [gauge,right of=g2,label={[rotate=-45]below right:{\cc{1}}}] {};
	\node (g4) [gauge,right of=g3,label={[rotate=-45]below 
right:{\dd{2}}}] {};
	\node (g5) [gauge,right of=g4,label={[rotate=-45]below right:{\cc{2}}}] 
{};
	\node (g6) [right of=g5] {$\scriptstyle{\ldots}$};
    \node (g7) [gauge,right of=g6,label={[rotate=-45]below 
right:{\dd{k{-}1}}}] {};
	\node (g8) [gauge,right of=g7,label={[rotate=-45]below right:{\cc{k{-}1}}}] 
{};
    \node (g9) [gauge,right of=g8,label={[rotate=-45]below 
right:{\dd{k}}}] {};
	\node (g10) [gauge,right of=g9,label={[rotate=-45]below 
right:{\cc{k{-}1}}}] 
{};
    \node (g11) [gauge,right of=g10,label={[rotate=-45]below 
right:{\dd{k{-}1}}}] {};
    \node (g12) [right of=g11] {$\scriptstyle{\ldots}$};
	\node (g13) [gauge,right of=g12,label={[rotate=-45]below right:{\cc{2}}}] 
{};
    \node (g14) [gauge,right of=g13,label={[rotate=-45]below right:{\cc{2}}}] 
{};
	\node (g15) [gauge,right of=g14,label={[rotate=-45]below right:{\cc{1}}}] 
{};
	\node (g16) [gauge,right of=g15,label={[rotate=-45]below right:{\dd{1}}}] 
{};
	\node (b1) [gauge,above of=g8,label= left:{\cc{n_1}}] {};
    \node (b2) [gauge,above  of=g10,label= right:{\cc{n_l}}] {};
    \node (b0) [above of=g9] {$\scriptstyle{\cdots}$};
    \draw (3.3,1.25) node {\tiny{$\Lambda^2$}};
	\draw [-] (3+0.095,0.575) arc (-70:90:8pt);
	\draw [-] (3+-0.095,0.575) arc (250:90:8pt);
    \draw (4.45,1.25) node {\tiny{$\Lambda^2$}};
	\draw [-] (4.15-0.145+0.095,0.575) arc (-70:90:8pt);
	\draw [-] (4.15-0.145+-0.095,0.575) arc (250:90:8pt);
	\draw  (g2)--(g3) (g3)--(g4) (g4)--(g5) (g5)--(g6) (g6)--(g7) 
(g7)--(g8) (g8)--(g9) (g9)--(g10) (g10)--(g11) (g11)--(g12) (g12)--(g13)
(g13)--(g14) (g14)--(g15) (g15)--(g16)  (g9)--(b1) (g9)--(b2);
	\end{tikzpicture}
	} 
	\,.
		\label{eq:k_magQuiver_infinite_generic}
\end{align}
It is important to recall that the Coulomb branch of 
\eqref{eq:k_magQuiver_infinite_generic} describes a Higgs branch 
phase where $(2n-l)$ gauge couplings are tuned to infinity.
According to \cite[Sec.\ 2.2]{Hanany:2018cgo}, the moduli spaces are related via
\begin{align}
 \Coulomb^{3\diff} 
 \left( \magQuivText \eqref{eq:k_magQuiver_infinite_generic}  \right) 
 = 
 \Coulomb^{3\diff} 
 \left(\magQuivText \eqref{eq:k_magQuiver_infinite_separated} \right)
 \slash \prod_i S_{n_i}
 \,.
\label{eq:discrete_gauging_k}
\end{align}
Physically, there exists a discrete $S_n$ 
action, or of its subgroups, on the pairs of half \NS\ branes, which is gauged 
when all pairs become coincident. 

The Coulomb branch symmetry of 
\eqref{eq:k_magQuiver_infinite_generic} is
\begin{align}
 G_J = \sorm(2k) \times \sorm(2k)
\end{align}
because the central $\dalg_k$ nodes is always \emph{good}, but never balanced 
for $n>1$. This symmetry agrees with the Higgs branch symmetry of 
\eqref{eq:electric_quiver_n_M5_Dk} at the origin of the tensor branch.
In addition, there is discrete Coulomb branch symmetry factor which corresponds 
to the symmetry of the magnetic quiver.
Next, the Coulomb branch dimension of \eqref{eq:k_magQuiver_infinite_generic} 
is readily computed
\begin{equation}
\begin{aligned}
 \dim_\HH \ \Coulomb^{3\diff}\left( \magQuivText 
\eqref{eq:k_magQuiver_infinite_generic} \right)
&= 2\cdot \sum_{i=1}^{k-1}\left( \dim\ \calg_i + \dim\ \dalg_i \right) + 
\dim\ \dalg_k + \sum_{i=1}^{l} \dim\ \calg_{n_i}\\
&=n+\dim\ \sorm(2k) \,.
\end{aligned} 
\label{eq:dim_Coulomb_infinite}
\end{equation}
\paragraph{Origin of tensor branch.}
Lastly, the origin of the tensor branch is reached when all half \NS s have 
left the orientifold pairwise and all pairs are coincident; hence, partition 
$\{n\}$ and the brane configuration becomes
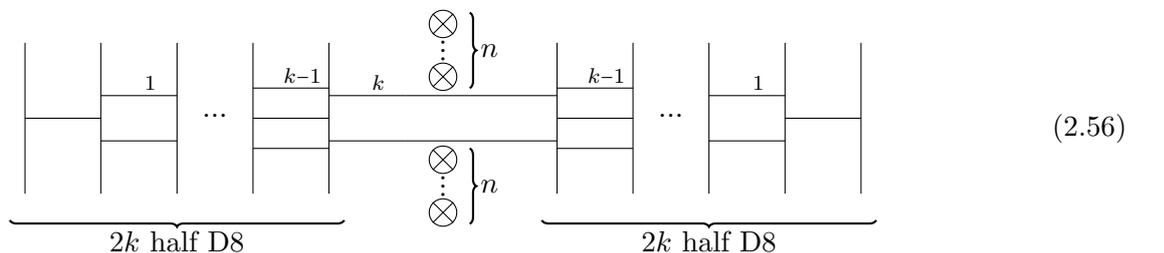
\begin{align}
 \raisebox{-.5\height}{
\begin{tikzpicture}
        \DeightMany{1}{0}{1}
        \Dbrane{0,0}{1,0}
		\DeightMany{1}{1}{1}
		\DsixOMinusFree{$1$}{1}{0}
		\DeightMany{1}{2}{1}
        \draw (2.5,0) node {$\cdots$};
        \DeightMany{1}{3}{1}
        \DsixOMinusTildeFree{$k{-}1$}{3}{1}
        \DeightMany{1}{4}{1}
        \DsixOMinusFree{$k$}{4}{0}
        \Dbrane{5,0.3}{7,0.3}
        \Dbrane{5,-0.3}{7,-0.3}
        \ns{5.5,0.55}
        \draw (5.5,0.9) node {$\vdots$};
        \ns{5.5,1.25}
        \draw[decoration={brace,mirror,raise=10pt},decorate,thick]
  (5.5,0.4) -- node[right=10pt] {$n$} (5.5,1.4);
        \ns{5.5,-0.55}
        \draw (5.5,-0.9) node {$\vdots$};
        \ns{5.5,-1.25}
        \draw[decoration={brace,mirror,raise=10pt},decorate,thick]
  (5.5,-1.4) -- node[right=10pt] {$n$} (5.5,-0.4);
        \DeightMany{1}{7}{1}
        \DsixOMinusTildeFree{$k{-}1$}{7}{1}
        \DeightMany{1}{8}{1}
        \draw (8.5,0) node {$\cdots$};
        \DeightMany{1}{9}{1}
		\DsixOMinusFree{$1$}{9}{0}
		\DeightMany{1}{10}{1}
        \Dbrane{10,0}{11,0}
        \DeightMany{1}{11}{1}
\draw[decoration={brace,mirror,raise=10pt},decorate,thick]
  (-0.2,-1) -- node[below=10pt] {$2k$ half \De } (4.2,-1);
\draw[decoration={brace,mirror,raise=10pt},decorate,thick]
  (7-0.2,-1) -- node[below=10pt] {$2k$ half \De } (11.2,-1);
	\end{tikzpicture}
	}
\end{align}
such that the corresponding magnetic quiver, using Table \ref{tab:orientifold}, 
is read off to be
\begin{align}
 \raisebox{-.5\height}{
 	\begin{tikzpicture}
 	\tikzset{node distance = 0.5cm};
	\tikzstyle{gauge} = [circle, draw,inner sep=2.5pt];
	\tikzstyle{flavour} = [regular polygon,regular polygon sides=4,inner 
sep=2.5pt, draw];
	\node (g2) [gauge,label={[rotate=-45]below right:{\dd{1}}}] {};
	\node (g3) [gauge,right of=g2,label={[rotate=-45]below right:{\cc{1}}}] {};
	\node (g4) [gauge,right of=g3,label={[rotate=-45]below right:{\dd{2}}}] {};
	\node (g5) [gauge,right of=g4,label={[rotate=-45]below right:{\cc{2}}}] 
{};
	\node (g6) [right of=g5] {$\scriptstyle{\ldots}$};
    \node (g7) [gauge,right of=g6,label={[rotate=-45]below 
right:{\dd{k{-}1}}}] 
{};
	\node (g8) [gauge,right of=g7,label={[rotate=-45]below right:{\cc{k{-}1}}}] 
{};
    \node (g9) [gauge,right of=g8,label={[rotate=-45]below right:{\dd{k}}}] {};
	\node (g10) [gauge,right of=g9,label={[rotate=-45]below 
right:{\cc{k{-}1}}}] {};
    \node (g11) [gauge,right of=g10,label={[rotate=-45]below 
right:{\dd{k{-}1}}}] {};
    \node (g12) [right of=g11] {$\scriptstyle{\ldots}$};
	\node (g13) [gauge,right of=g12,label={[rotate=-45]below right:{\cc{2}}}] 
{};
    \node (g14) [gauge,right of=g13,label={[rotate=-45]below right:{\dd{2}}}] 
{};
	\node (g15) [gauge,right of=g14,label={[rotate=-45]below right:{\cc{1}}}] 
{};
	\node (g16) [gauge,right of=g15,label={[rotate=-45]below right:{\dd{1}}}] 
{};
	\node (g0) [gauge,above of=g9,label=right:{\cc{n}}] {};
	\draw (3.9,1.15) node {\tiny{$\Lambda^2$}};
	\draw [-] (3.5+0.095,0.58) arc (-70:90:8pt);
	\draw [-] (3.5+-0.095,0.58) arc (250:90:8pt);
	\draw  (g2)--(g3) (g3)--(g4) (g4)--(g5) (g5)--(g6) (g6)--(g7) 
(g7)--(g8) (g8)--(g9) (g9)--(g10) (g10)--(g11) (g11)--(g12) (g12)--(g13)
(g13)--(g14) (g14)--(g15) (g15)--(g16)  (g9)--(g0);
	\end{tikzpicture}
	} 
	\,,
		\label{eq:k_magQuiver_infinite_coincident}
\end{align}
which had been \emph{conjectured} in \cite{Hanany:2018uhm} as a description for 
the Higgs branch at infinite coupling. Here, the magnetic quiver has been 
\emph{derived} from a brane system. The Coulomb branch dimension 
\eqref{eq:dim_Coulomb_infinite} and symmetry are the same as above.

The Higgs branch dimension at infinite coupling has been 
computed in \cite{Mekareeya:2017sqh} to be
\begin{align}
 \dim_{\HH}\  \Higgs^{6\diff}_{\infty} 
 \left(\elecTheory \eqref{eq:electric_quiver_n_M5_Dk} \right)
 &= 29\cdot n_t +n_h -n_v = n + 
\dim\ \sorm(2k) \,,\\
n_t&= n  \,, \notag \\
n_h&= \frac{1}{2} \cdot 2k \cdot (2k-8) \cdot 2n \,, \notag \\
n_v&= n \cdot \dim (\sprm(k-4)) + (n-1) \cdot \dim (\sorm(2k)) \,. \notag 
\label{eq:dim_Higgs_infinite}
\end{align}
Using the formalism of \emph{magnetic quivers}, one is now able to explain the 
jump in moduli space dimension
\begin{align}
 \dim_{\HH}  \Higgs^{6\diff}_{\infty} - \dim_\HH \ 
\Higgs^{6\diff}_{\fin} 
 = n+ 28 = (n-1) +29
\end{align}
in more detail. As the theory has $(2n-1)$ tensor multiplets, there 
are $(2n-1)$ order parameters that can be tuned and, as such, one expects 
$(2n-1)$ distinct 
phase transitions. The above analysis demonstrates the following:
\begin{compactenum}[(i)]
    \item There are $(n-1)$ transitions of the form
    \begin{align}
 1 \text{ tensor }+ 1\; \sorm(8) \text{ vector } \rightarrow 1 \text{ hyper }
    \end{align}
such that the moduli space jumps by one quaternionic unit. This will be called 
a \emph{$D_4$ transition}.
    \item There is precisely one small $E_8$ instanton transition 
    \begin{align}
 1 \text{ tensor } \rightarrow 29 \text{ hypers }
    \end{align}
    and the dimension has to jump by $29$.
    \item There are $(n-1)$ discrete gauging transitions in which the Higgs 
branch does not jump in dimension.
\end{compactenum}

Geometrically, the magnetic quivers also allow to study the transverse slices. 
As in \eqref{eq:quiver_subtraction_1M5}, one can take the difference between 
the 
magnetic quivers \eqref{eq:k_nM5_magQuiver_infinite_+n-1} and 
\eqref{eq:k_magQuiver_infinite_separated}, describing the phase before and 
after the final transition. By the results of \cite{Cabrera:2018ann}, the 
Coulomb branch of this \emph{difference quiver} describes the transverse slice. 
Inspecting the relevant theories reveals
\begin{align}
 \magQuivText \eqref{eq:k_magQuiver_infinite_separated}
 -
 \magQuivText \eqref{eq:k_nM5_magQuiver_infinite_+n-1}
 =
 \magQuivText \eqref{eq:k>4_magQuiver_infinite}
\end{align}
such that the transverse slice is again the closure of the minimal nilpotent 
orbit of $E_8$, as suspected for an $E_8$ transition.
%
%
%
%
\subsection{Derivation rules}
\label{sec:rules}
Having discussed the various transitions between the different Higgs branch 
phases and how to derive their associated magnetic quivers, one can summarise 
and formalise the rules as follows:
\begin{myConj}[Magnetic quiver]
For a \Ds-\De-\NS\ brane system in the presence of \Os\ orientifold planes, 
cf.\ Table \ref{tab:directions}, in which all 
\Ds\ branes are suspended between \De\ branes, the massless BPS states, deduced 
from stretching virtual \Df\ branes, arise from the following configurations:
\begin{compactenum}[(i)]
 \item Stack of $m$ full \Ds\ branes on top of a \Os\ plane suspended 
between two \De s in a finite $x^6$ 
interval:  the vertical motion along the $x^7$, $x^8$, $x^9$ directions gives 
rise to a magnetic vector multiplet due to \Df s stretched between them. 
Depending on the type of magnetic \Os\ plane, the magnetic gauge group is
\begin{subequations}
\begin{align}
&\raisebox{-.5\height}{
	\begin{tikzpicture}
	\tikzstyle{gauge} = [circle, draw,inner sep=3pt];
        \DeightMany{1}{0}{1}
        \DeightMany{1}{1}{1}
		\draw[decoration={brace,raise=3pt},decorate,thick](0,-0.5) -- 
node[left=6pt] { $m$ \Ds\ $\&$ \Osm } (0,0.5);
        \DsixOMinus{m}{0}
		\ArrowMagnetic{3}{0}
		\node (g) at (6,0) [gauge,label=below:{$\dalg_m$}] {};
  		\draw (0,-1.3) node {\De };
	\end{tikzpicture}
	}
	\\
	&\raisebox{-.5\height}{
	\begin{tikzpicture}
	\tikzstyle{gauge} = [circle, draw,inner sep=3pt];
        \DeightMany{1}{0}{1}
        \DeightMany{1}{1}{1}
		\draw[decoration={brace,raise=3pt},decorate,thick](0,-0.5) -- 
node[left=6pt] { $m$ \Ds\ $\&$ \Osmt } (0,0.5);
        \DsixOMinusTilde{m}{0}
		\ArrowMagnetic{3}{0}
		\node (g) at (6,0) [gauge,label=below:{$\calg_m$}] {};
  		\draw (0,-1.3) node {\De };
	\end{tikzpicture}
	}
	\\
&\raisebox{-.5\height}{
	\begin{tikzpicture}
	\tikzstyle{gauge} = [circle, draw,inner sep=3pt];
        \DeightMany{1}{0}{1}
        \DeightMany{1}{1}{1}
		\draw[decoration={brace,raise=3pt},decorate,thick](0,-0.5) -- 
node[left=6pt] { $m$ \Ds\ $\&$ \Osp } (0,0.5);
        \DsixOPlus{m}{0}
		\ArrowMagnetic{3}{0}
		\node (g) at (6,0) [gauge,label=below:{$\balg_{m}$}] {};
  		\draw (0,-1.3) node {\De };
	\end{tikzpicture}
	}
	\\
	&\raisebox{-.5\height}{
	\begin{tikzpicture}
	\tikzstyle{gauge} = [circle, draw,inner sep=3pt];
        \DeightMany{1}{0}{1}
        \DeightMany{1}{1}{1}
		\draw[decoration={brace,raise=3pt},decorate,thick](0,-0.5) -- 
node[left=6pt] { $m$ \Ds\ $\&$ \Ospt } (0,0.5);
        \DsixOPlusTilde{m}{0}
		\ArrowMagnetic{3}{0}
		\node (g) at (6,0) [gauge,label=below:{$\calg_m$}] {};
  		\draw (0,-1.3) node {\De };
	\end{tikzpicture}
	}
	\label{eq:rule_magQuiv_1}
\end{align}
\end{subequations}
see also Table \ref{tab:orientifold}.
\item Stacks of $m$ full \Ds\ on some \Os\ plane and $l$ full \Ds\ 
branes (on some other \Os\ plane) in adjacent \De\ intervals along the 
$x^6$ direction: the \Df\ branes suspended between \Ds s of different intervals 
induce a 
magnetic half hypermultiplet transforming as bifundamentals in the 
corresponding magnetic gauge groups.
\item Stack of $m$ full \NS\ branes above an \Osm\ or \Osp\ orientifold 
at coincident $x^6$ position: the vertical 
motion along the $x^7$, $x^8$, $x^9$ 
directions gives rise to a 
$\calg_m$ magnetic vector multiplet due to \Df s stretched between.
Since the \NS s are free to move along the $x^6$ direction, there is an 
additional hypermultiplet transforming in the traceless second anti-symmetric 
representation $\Lambda^2$ of $\calg_m$. Put differently, virtual \Df\ 
branes suspended between the half \NS s and their mirrors furnish the 
anti-symmetric representation due to the orientifold action.
\begin{subequations}
\begin{align}
\raisebox{-.5\height}{
	\begin{tikzpicture}
	\tikzstyle{gauge} = [circle, draw,inner sep=3pt];
		\DeightMany{1}{0}{1}
		\DeightMany{1}{2}{1}
        \draw[decoration={brace,raise=3pt},decorate,thick](0,-0.5) -- 
node[left=6pt] { \Osm } (0,0.5);
        \ns{1,0.35}
        \draw (1,0.7) node {$\vdots$};
        \ns{1,1.05}
        \draw[decoration={brace,mirror,raise=10pt},decorate,thick]
  (1,0.2) -- node[right=10pt] {$m$} (1,1.2);
        \ns{1,-0.35}
        \draw (1,-0.7) node {$\vdots$};
        \ns{1,-1.05}
        \draw[decoration={brace,mirror,raise=10pt},decorate,thick]
  (1,-1.2) -- node[right=10pt] {$m$} (1,-0.2);
		\ArrowMagnetic{4}{0}
		\node (g) at (7,0) [gauge,label=below:{$\calg_m$}] {};
	\node (l1) [above right of=g]{$\Lambda^2$};
	\draw [-] (7+0.11,0.105) arc (-70:90:10pt);
	\draw [-] (7+-0.11,0.105) arc (250:90:10pt);
  		\draw (0,-1.3) node {\De };
	\end{tikzpicture}
	}
\\
\raisebox{-.5\height}{
	\begin{tikzpicture}
	\tikzstyle{gauge} = [circle, draw,inner sep=3pt];
		\DeightMany{1}{0}{1}
		\DeightMany{1}{2}{1}
        \OPlus{0,0}{2,0}
        \draw[decoration={brace,raise=3pt},decorate,thick](0,-0.5) -- 
node[left=6pt] { \Osp } (0,0.5);
        \ns{1,0.35}
        \draw (1,0.7) node {$\vdots$};
        \ns{1,1.05}
        \draw[decoration={brace,mirror,raise=10pt},decorate,thick]
  (1,0.2) -- node[right=10pt] {$m$} (1,1.2);
        \ns{1,-0.35}
        \draw (1,-0.7) node {$\vdots$};
        \ns{1,-1.05}
        \draw[decoration={brace,mirror,raise=10pt},decorate,thick]
  (1,-1.2) -- node[right=10pt] {$m$} (1,-0.2);
		\ArrowMagnetic{4}{0}
		\node (g) at (7,0) [gauge,label=below:{$\calg_m$}] {};
	\node (l1) [above right of=g]{$\Lambda^2$};
	\draw [-] (7+0.11,0.105) arc (-70:90:10pt);
	\draw [-] (7+-0.11,0.105) arc (250:90:10pt);
  		\draw (0,-1.3) node {\De };
	\end{tikzpicture}
	}
	\label{eq:rule_magQuiv_3}
\end{align}
\end{subequations}
\item Stack of $m$ full \NS\ branes above an \Osm\  
orientifold at coincident $x^6$ positions, in between two half \NS\ branes that 
are stuck on the orientifold and have $4$ \Ds\ branes suspended in-between. In 
addition to the symplectic magnetic vector multiplet and the additional 
magnetic anti-symmetric hypermultiplet, there is a $\dalg_4$ magnetic flavour 
node due to virtual \Df\ branes that can be stretched between the \NS s and the 
\Ds s. 
\begin{align}
\raisebox{-.5\height}{
	\begin{tikzpicture}
	\tikzstyle{gauge} = [circle, draw,inner sep=3pt];
	\tikzstyle{flavour} = [regular polygon,regular polygon sides=4,inner 
sep=3pt, draw];
\DsixOMinus{}{0}
\DsixOMinus{$4$}{1}
\ns{0,0}
\ns{2,0}
        \draw[decoration={brace,raise=3pt},decorate,thick](-0.1,-0.5) -- 
node[left=6pt] { \Osm } (-0.1,0.5);
        \ns{1,0.45}
        \draw (1,0.8) node {$\vdots$};
        \ns{1,1.15}
        \draw[decoration={brace,mirror,raise=10pt},decorate,thick]
  (1,0.3) -- node[right=10pt] {$m$} (1,1.3);
        \ns{1,-0.45}
        \draw (1,-0.8) node {$\vdots$};
        \ns{1,-1.15}
        \draw[decoration={brace,mirror,raise=10pt},decorate,thick]
  (1,-1.3) -- node[right=10pt] {$m$} (1,-0.3);
		\ArrowMagnetic{4}{0}
		\node (g) at (7,0) [gauge,label=below:{$\calg_m$}] {};
		\node (f) at (8,0) [flavour,label=below:{$\dalg_4$}] {};
		\draw (g)--(f);
	\node (l1) [above right of=g]{$\Lambda^2$};
	\draw [-] (7+0.11,0.105) arc (-70:90:10pt);
	\draw [-] (7+-0.11,0.105) arc (250:90:10pt);
	\end{tikzpicture}
	}
	\label{eq:rule_magQuiv_3prim}
\end{align}
\item Stacks of $l$ full \Ds\ and $m$ full \NS\ branes between two 
\De\ in a finite $x^6$ interval: the vertical distance in the $x^7$, $x^8$, 
$x^9$ directions leads to a magnetic half hypermultiplet transforming as 
bifundamentals in the corresponding magnetic gauge groups. 
    \item Suppose a single half \NS\ is stuck on the \Os\ plane in an \De\ 
interval with $k$ full \Ds\ branes suspended between the \De\ branes. 
Since the \NS\ is not free to move, it does not contribute a 
magnetic degree of freedom. Put differently, since the \NS\ is on the 
orientifold and has no mirror image, there are no \Df\ branes that induce a
magnetic vector multiplet.
Nevertheless, the stuck half \NS\ brane contributes an $\balg_0$ flavour 
to the 
magnetic gauge multiplet in that finite \De\ segment. The magnetic 
bifundamentals are associated to virtual \Df\ branes stretched between the 
stuck half \NS\ and the \Ds\ branes.
\begin{align}
\raisebox{-.5\height}{
	\begin{tikzpicture}
	\tikzstyle{gauge} = [circle, draw,inner sep=3pt];
	 \tikzstyle{flavour} = [regular polygon,regular polygon sides=4,inner 
sep=3pt, draw];
		\DeightMany{1}{0}{1}
		\DeightMany{1}{2}{1}
        \DsixOMinusTildeFree{$k$}{0}{1}
        \OPlusTilde{1,0}{2,0}
        \Dbrane{1,0.4}{2,0.4}
        \Dbrane{1,-0.4}{2,-0.4}
        \ns{1,0}
		\ArrowMagnetic{4}{0}
		\node (g) at (7,-0.5) [gauge,label=right:{$\calg_k$}] {};
		\node (f) at (7,0.5) [flavour,label=right:{$\balg_0$}] {};
		\draw (g)--(f);
  		\draw (0,-1.3) node {\De };
	\end{tikzpicture}
	}
\end{align}

    \item Suppose a pair of half \NS\ branes is stuck on the orientifold 
between two \De\ branes. Again, as they have no freedom to move, each pair of 
half \NS\ contributes as $\calg_1$ flavour node, due to virtual \Df\ branes 
ending on them.
The $4$ full \Ds\ branes are not Higgs branch degrees of freedom, simply 
because the \Ds\ cannot be suspended between \De\ branes. 
\begin{align}
\raisebox{-.5\height}{
	\begin{tikzpicture}
	\tikzstyle{gauge} = [circle, draw,inner sep=3pt];
    \tikzstyle{flavour} = [regular polygon,regular polygon sides=4,inner 
sep=3pt, draw];
	\DeightMany{1}{0}{1}
	\DsixOPlusFree{$k$}{0}{3}
	\DsixOMinus{$4$}{1}
    \OPlus{2,0}{3,0}
    \Dbrane{1,0.6}{3,0.6}
    \Dbrane{1,-0.6}{3,-0.6}
	\DeightMany{1}{3}{1}
		\ns{1,0}
		\ns{2,0}
		\ArrowMagnetic{4}{0}
		\node (g1) at (7,-0.5) [gauge,label=right:{$\balg_k$}] {};
		\node (g2) at (7,0.5) [flavour,label=right:{$\calg_1$}] {};
		\draw (g1)--(g2);
  		\draw (0,-1.3) node {\De };
	\end{tikzpicture}
		}
	\label{eq:rule_magQuiv_4}
\end{align}
\end{compactenum}
The massless degrees of freedom can be encoded in a quiver diagram in the 
familiar way.
\label{conj:magnetic_quiver}
\end{myConj}
\paragraph{Remark.}
In view of the other types of bouquets discussed in \cite{Hanany:2018cgo}, one 
may wonder if these can appear in this set-up.
Due to the dual Type IIA description of an $D_k$ singularity in M-theory, there 
is always an even number of half \Ds\ branes. Therefore, one has to pull in an 
even number of half \De\ branes from $x^6 = \pm \infty$. It follows that the 
central orientifold is 
either \Osm\ or \Osp\ (before the $E_8$ transition); hence, the 
magnetic orientifold is \Osm\ or \Osmt , respectively. Consequently, the 
pairs of half \NS\ branes lifted from the 
orientifold will always lead to $\calg_1$-type bouquets.
%
%
\subsection{Phase diagram}
In the above sections, many different transitions have been discussed by using 
the magnetic quiver. In Table \ref{tab:phases} the entire phase structure is 
presented, graded according to quaternionic dimension of the moduli space and 
the number of infinite gauge couplings. 
For simplicity and readability, all $D_4$ transitions are assumed to be of 
the form \eqref{eq:branes_k_nM5_dim1_transition}, because any transition 
resulting from \eqref{eq:branes_k_nM5_dim1_transition_var2} can be converted 
into this form by moving a pair of half \NS\ branes along $x^6$.
Summarising the above, 
the three 
transitions have the following impact:
\begin{compactitem}
 \item The $D_4$ transitions increase the quaternionic dimension as well 
as the number of infinite couplings by one. Hence, Higgs branch phases along 
the diagonal in 
Table \ref{tab:phases} can be related by $D_4$ transitions. For instance,
\begin{equation}
\Higgs_{\scriptscriptstyle{p{\times}D_4}}^{\scriptscriptstyle{\{n_i\}}} 
 \xrightarrow[\;\text{transition}\;]{D_4}  
\Higgs_{\scriptscriptstyle{(p{+}1){\times}D_4}}^{ \scriptscriptstyle{ 
\{n'_j\} }}  
\end{equation}
where $\{n_i\}_{i=1,\ldots,l}$ is a partition of $p$, i.e.\ $\sum_{i=1}^l n_i 
=p$, and $\{n'_j\}_{j=1,\ldots,l+1}$ is a partition of $(p{+}1)$ that is 
obtained by appending a $1$ to partition $\{n_i\}$, i.e.\ $\{n'_j\} \equiv 
\{n_1,\ldots,n_l,1\} $ such that $\sum_{i=1}^{l+1} n'_i = p+1$.
 \item The discrete gauging transitions do not increase the quaternionic 
dimension, but the number of infinite couplings increases depending on the 
length of partition. Thus, Higgs branch phases along the vertical direction are 
related by discrete gauging. In detail, for a partition $\{n_i\}$ of $p$
\begin{equation}
 \Higgs_{\scriptscriptstyle{p{\times}D_4}}^{\scriptscriptstyle{\{1^p\}}} 
 \xrightarrow[\;\text{gauging}\;]{\prod_i S_{n_i}}  
\Higgs_{\scriptscriptstyle{p{\times}D_4}}^{ \scriptscriptstyle{ 
\{n_i\} }}  
\end{equation}
where the discrete gauging of $\prod_i S_{n_i}$ increases the number of 
infinite couplings by $\sum_{i=1}^l (n_i-1)= n-l$, with $l=$ length of the 
partition. The identical statement holds for $\Higgs_{\substack{ 
\scriptscriptstyle{ (n{-}1){\times} D_4} \\ 
\scriptscriptstyle{ 1 {\times} E_8} }}^{\scriptscriptstyle{\{n_i\}}}$ with 
$\{n_i\}$ being a partition of $n$.
 \item The small $E_8$ instanton transition increases the quaternionic 
dimension by $29$; however, the number of infinite couplings increases only by 
one. This transition relates Higgs branch phases in the last two columns of 
Table \ref{tab:phases}, which means
\begin{align}
 \Higgs_{\scriptscriptstyle{(n{-}1){\times}D_4}}^{ \scriptscriptstyle{ 
\{n_i\} }} 
\xrightarrow[\;\text{transition}\;]{E_8 \text{ instanton}} 
\Higgs_{\substack{ 
\scriptscriptstyle{ (n{-}1){\times} D_4} \\ 
\scriptscriptstyle{ 1 {\times} E_8} }}^{\scriptscriptstyle{\{n'_j\}}}
\end{align}
where $\{n_i\}_{i=1,\ldots,l}$ is a partition of $(n-1)$ and $\{n'_j\}$ is a 
partition of $n$ obtained via appending a single $1$ to $\{n_i\}$, i.e.\ 
$\{n'_j\}=\{n_1,\ldots,n_l,1\}$.
\end{compactitem}

\begin{table}[t]
\centering
\begin{tabular}{c|*{9}{c}}
\toprule
\multirow{2}{*}{$\substack{\text{\# infinite} \\ \text{gauge } \\ 
\text{couplings}} $} & \multicolumn{9}{c}{$\HH$-dim of moduli space} \\
  & $d$ & $d{+}1$ & $d{+}2$ & $d{+}3$ & $d{+}4$ & $\ldots$ & $d{+}(n{-}1)$ &  & 
$d{+}(n{+}28)$ \\ \midrule
$0 $ & $\Higgs_\fin$ \\
$1$ &  & $\Higgs_{\scriptscriptstyle{1{\times} 
D_4}}^{\scriptscriptstyle{\{1\}}}$ \\
$2$ &  &  & $\Higgs_{\scriptscriptstyle{2{\times} 
D_4}}^{\scriptscriptstyle{\{1^2\}}}$ &  \\
$3$ &  & & $\Higgs_{\scriptscriptstyle{2{\times} 
D_4}}^{\scriptscriptstyle{\{2\}}}$  & 
$\Higgs_{\scriptscriptstyle{3{\times} D_4}}^{\scriptscriptstyle{\{1^3\}}}$
\\
$4$ &  &  & &$\Higgs_{\scriptscriptstyle{3{\times} 
D_4}}^{\scriptscriptstyle{\{2,1\}}}$ & 
$\Higgs_{\scriptscriptstyle{4{\times} D_4}}^{\scriptscriptstyle{\{1^4\}}}$ 
 
\\
$5$ &  &  & &$\Higgs_{\scriptscriptstyle{3{\times} 
D_4}}^{\scriptscriptstyle{\{3\}}}$ & 
$\Higgs_{\scriptscriptstyle{4{\times} 
D_4}}^{\scriptscriptstyle{\{2,1^2\}}}$   \\
$6$ &  &  & & & $\vdots$  \\
$7$ &  &  & & & $\Higgs_{\scriptscriptstyle{ 4{\times} 
D_4}}^{\scriptscriptstyle{\{4\}}}$  \\
$\vdots$ &  &  & & & & $\ddots$ \\
$n{-}1 $  &  &  & & & & & $\Higgs_{\scriptscriptstyle{(n{-}1)\times 
D_4}}^{\scriptscriptstyle{\{1^{n-1}\}}}$  \\
$n$ &  &  & & & & & $\Higgs_{\scriptscriptstyle{(n{-}1){\times} 
D_4}}^{\scriptscriptstyle{\{2,1^{n-2}\}}}$ & & 
$\Higgs_{\substack{ \scriptscriptstyle{ (n{-}1){\times} D_4} \\ 
\scriptscriptstyle{ 1 {\times} E_8} }}^{\scriptscriptstyle{\{1^n\}}}$  \\
\multirow{2}{*}{$\vdots$} &  &  & & & & & $\vdots$ & & 
$\Higgs_{\substack{ \scriptscriptstyle{ (n{-}1){\times} D_4} \\
\scriptscriptstyle{ 1 {\times} E_8} }}^{\scriptscriptstyle{\{2,1^{n-1}\}}}$ \\
 & \\
$2n{-}2 $ &  &  & & & & & $\Higgs_{\scriptscriptstyle{ (n{-}1)\times 
D_4} }^{\scriptscriptstyle{\{n-1\}}}$ & & 
$\vdots$ \\
$2n{-}1$ &  &  & & & & &  & & 
$\Higgs_{\substack{\scriptscriptstyle{ (n{-}1) {\times} D_4} \\
\scriptscriptstyle{1 {\times} E_8} }}^{\scriptscriptstyle{\{n\}}}\equiv 
\Higgs_\infty$ \\
\bottomrule
\end{tabular}
\caption{The multitude of Higgs branch phases for $n$ \Mf s on a $\C^2 \slash 
\D_{k-2}$ singularity. The subscript $p\times D_4$ indicates $p$ 
$D_4$ transitions, while $1\times E_8$ indicate the single small instanton 
transitions. The superscript $\{n_i\}$ denotes a partition of $p$ 
indicating the discrete gauging of a permutation (sub-)groups 
$\prod_i S_{n_i}$. At finite coupling, the Higgs branch dimension is $d=\dim\ 
\sorm(2k)- \dim\ \sorm(8)$. }
\label{tab:phases}
\end{table}

  \section{Hasse diagram}
\label{sec:Hasse}
In Section \ref{sec:magnetic_quiver} the different Higgs branches of theories 
corresponding to $n$ \Mf s on a D-type singularity have been described via 
magnetic quivers. Besides providing the Higgs branch description, one can 
moreover attempt to analyse the Higgs branch geometries understood as 
symplectic singularities \cite{beauville2000symplectic}. As put forward in 
\cite{Bourget:2019aer}, the singularity structure can be encoded in a Hasse 
diagram. In many cases, the Hasse diagram can be derived either from the brane 
configuration using Kraft-Procesi transitions 
\cite{Cabrera:2016vvv,Cabrera:2017njm} or from the magnetic quiver description 
via quiver subtraction \cite{Cabrera:2018ann,Hanany:2018uhm}.

For Lagrangian theories, the Hasse diagram is intimately related to 
the Higgs mechanism. In more detail, consider an electric theory 
with 
gauge group $G$ and matter fields transforming in some (finite dimensional) 
representation 
$\mathcal{R}$, which renders the theory anomaly-free. Suppose there exists a 
subgroup $H\subset G$ such that the matter representation $\mathcal{R}$ and the 
adjoint representation $\mathrm{Adj}^{G}$ decompose into irreducible $H$ 
representations $\mathbf{r}_i$ as follows:
\begin{align}
 \mathcal{R}\big|_H &= \bigoplus_{i} a_i \mathbf{r}_i 
 \;, \; a_i \in \NN \cup \{0\}
 \quad 
 \mathrm{and}
 \qquad
 \mathrm{Adj}^{G}\big|_H = \mathrm{Adj}^H \oplus \bigoplus_{i} b_i 
\mathbf{r}_i
\;, \; b_i \in \NN \cup \{0\}
\, ,
\end{align}
where the infinite summation is taken over all irreducible 
representations $\{\mathbf{r}_i\}_i$ of $H$. However, only a finite number of 
the multiplicities $a_i$, $b_i$ is non-trivial, because $\mathcal{R}$ is 
finite dimensional.  
An assignment of vacuum 
expectation values breaks $G\to H$ consistently only if the 
multiplicities $a_i$, $b_i$ satisfy finitely many constraints: 
\begin{align}
 a_i\geq b_i \; , \; \forall i \;.
 \label{eq:Higgs_constraint}
\end{align}
The resulting $H$ gauge theory has matter content transforming as 
$\mathcal{R}'=\oplus_i (a_i-b_i)\mathbf{r}_i$, which is assumed to be 
anomaly-free. 
More specifically, $\mathcal{R}'$ may contain the trivial representation 
$i=\mathrm{triv}$, such that the $H$ gauge theory has non-trivially 
charged matter 
$\mathcal{R}''=\oplus_{i\neq \mathrm{triv}} (a_i-b_i)\mathbf{r}_i$ 
alongside with
$(a_{\mathrm{triv}} - b_{\mathrm{triv}})\geq0$ massless 
gauge singlets.

Returning to the $G$ gauge theory, its Higgs branch $\Higgs_G$ admits a 
foliation $\{\mathcal{L}_\kappa\}$ in which a leaf $\mathcal{L}_\kappa$ 
corresponds to the set of vacuum expectation values that break $G\to H_\kappa$. 
The closure of a leaf $\mathcal{L}_\kappa$ is a symplectic singularity 
parameterised by the $(a_{\mathrm{triv}} - b_{\mathrm{triv}})$ massless 
states 
that appear as singlets in the Higgsing process. The leaves themselves admit a 
partial order via inclusion: $\mathcal{L}_\kappa < \mathcal{L}_\lambda$ if and 
only if $\mathcal{L}_\kappa \subset \overline{\mathcal{L}}_\lambda$. As argued 
in \cite{Bourget:2019aer}, the partial order of leaves is in one-to-one 
correspondence with the partial order among the set of subgroups 
$\{H_\kappa\}$, such that the $G$ gauge theory can be Higgsed to the 
corresponding $H_\kappa$ gauge theory, satisfying \eqref{eq:Higgs_constraint}.
In other words, $\mathcal{L}_\kappa < \mathcal{L}_\lambda$ if and only if  
$H_\kappa > H_\lambda$, i.e. $H_\kappa \supset H_{\lambda}$. 
To any ordered pair of leaves $(\mathcal{L}_\kappa,\mathcal{L}_\lambda)$, with 
$\mathcal{L}_\kappa < \mathcal{L}_\lambda $, there exists an associated 
transverse slice $\mathcal{S}_{\kappa,\lambda}$, meaning that the 
space transverse to a point in $\mathcal{L}_\kappa$ inside the closure 
$\overline{\mathcal{L}}_\lambda$ equals $\mathcal{S}_{\kappa,\lambda}$. As 
example, the transverse slice to the pair $(\{0\}\equiv 
\mathcal{L}_{\mathrm{triv}},\Higgs_G)$ is just $\Higgs_G$ itself. Likewise, the 
pair $( \mathcal{L}_{\kappa},\Higgs_G)$ has a transverse slice given by the 
Higgs branch of the $H_\kappa$ gauge theory with matter content 
$\mathcal{R}''$. This is physically intuitive, as the unbroken gauge theory at 
any point of $\mathcal{L}_\kappa$ has  gauge group $H_\kappa$ with 
corresponding matter fields.
Moreover, the commutant $C_\kappa$ of $H_\kappa$ inside $G$ is a group of 
dimension $b_{\mathrm{triv}}$. There are $a_{\mathrm{triv}}$ many 
hypermulitplets transforming under $C_\kappa$ as $\mathcal{F}$ which, 
in general, is a sum of irreducible $C_\kappa$ representations. Consequently, 
the closure $\overline{\mathcal{L}}_\kappa$ is described by the Higgs branch of 
the $C_\kappa$ gauge theory with matter content $\mathcal{F}$.

As summary, the Hasse diagram encodes the decomposition of the Higgs 
branch into symplectic leaves. The closures of the leaves correspond to 
massless states appearing as gauge singlets, if a Higgs mechanism description 
is available. More generally, the leaf closures are described by magnetic 
quivers, as exemplified below. Moreover, the transverse slices correspond to 
Higgs branches of gauge theories, accessible via partial Higgsing (if 
applicable). For a Higgs branch which does not originate from a Lagrangian 
theory, the decomposition into symplectic leaves still exists and can be 
summarised in a Hasse diagram, but there is no description via the Higgs 
mechanism.

Considering the simplest theories relevant for this paper --- $6$d $\Ncal=(1,0)$ 
$\sprm(k-4)$ gauge theory with $\sorm(4k)$ flavour --- the Hasse 
diagram of the Higgs branch of \eqref{eq:1M5_electric_quiver} at finite and 
infinite gauge coupling is detailed in 
\cite[Tab.\ 8]{Bourget:2019aer}, based on the magnetic quiver realisation with 
unitary gauge groups \cite{Cabrera:2019izd}. Here, a complementary derivation 
is pursued 
from (i) the brane configuration with \Os\ orientifolds and  (ii) magnetic 
quivers with orthosymplectic gauge groups.
\subsection{From brane configuration}
\label{sec:Hasse_from_brane}
Recall that the brane configuration \eqref{eq:branes_1M5_on_Dk_infinite} 
describes the Higgs 
branch at infinite coupling of a single $\sprm(k-4)$ gauge group with 
$\sorm(4k)$ flavour node. To trace out the structure of the Higgs branch as a 
symplectic singularity, Kraft-Procesi transitions need to be performed.
Hence, one needs to find out which minimal transition is possible.
An important realisation is that a minimal transition is accomplished by 
moving a minimal set of 
\Ds\ suspended between \De\ branes to being suspended between \NS\ branes, see 
for instance \cite{Cabrera:2016vvv,Cabrera:2017njm} and also 
\cite[Sec.\ 2]{Bourget:2019aer}.

\paragraph{$e_8$ transition.}
In view of brane configuration \eqref{eq:branes_1M5_on_Dk_infinite}, the only 
way to achieve any such transition is to confine the \NS\ branes to the 
orientifold. 
Once the pair of half \NS s is on the \Osm\, the resulting full \NS\ brane 
cannot fractionate, because there are no \Ds\ branes attached from the left or 
right. 
Hence, to achieve a splitting of the full \NS , one has to move some \Ds\ 
branes onto the \NS\ brane, and split each \Ds\ brane to end on a half \De\ 
brane on one side and the \NS\ on the other. Respecting the S-rule and 
remembering that one needs $4$ full \Ds\ branes on the left and right of the 
full \NS\ for it to fractionate, the brane configuration becomes:
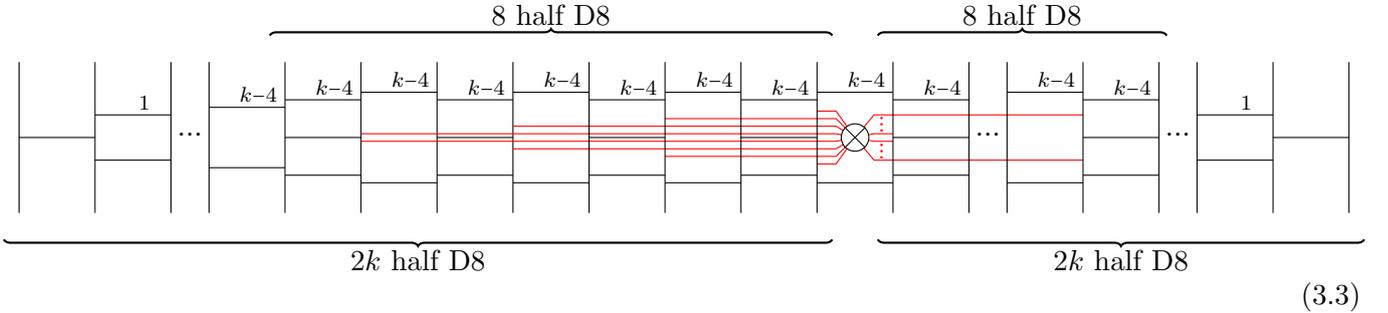
\begin{align}
 \raisebox{-.5\height}{
\begin{tikzpicture}
        \DeightMany{1}{-1.5}{1}
        \Dbrane{-1.5,0}{-0.5,0}
		\DeightMany{1}{-0.5}{1}
		\DsixOMinusFree{$1$}{-0.5}{0}
		\DeightMany{1}{0.5}{1}
        \draw (0.75,0) node {$\cdots$};
        \DeightMany{1}{1}{1}
        \DsixOMinusFree{$k{-}4$}{1}{1}
        \DeightMany{1}{2}{1}
        \DsixOMinusTildeFree{$k{-}4$}{2}{2}
		\DeightMany{1}{3}{1}
		\DsixOMinusFree{$k{-}4$}{3}{3}
		\DeightMany{1}{4}{1}
		\DsixOMinusTildeFree{$k{-}4$}{4}{2}
		\DeightMany{1}{5}{1}
%
\draw[red] (9.5,0)--(9.25,0.05)--(3,0.05);
\draw[red] (9.5,-0)--(9.25,-0.05)--(3,-0.05);
\draw[red] (9.5,0)--(9.25,0.15)--(5,0.15);
\draw[red] (9.5,-0)--(9.25,-0.15)--(5,-0.15);
\draw[red] (9.5,0)--(9.25,0.25)--(7,0.25);
\draw[red] (9.5,0)--(9.25,-0.25)--(7,-0.25);
\draw[red] (9.5,0)--(9.25,0.35)--(9,0.35);
\draw[red] (9.5,0)--(9.25,-0.35)--(9,-0.35);
%
\draw[red] (9.5,0)--(9.75,0.05)--(10,0.05);
\draw[red] (9.5,-0)--(9.75,-0.05)--(10,-0.05);
\draw[red] (9.5,0)--(9.75,0.3)--(12.5,0.3);
\draw[red] (9.5,0)--(9.75,-0.3)--(12.5,-0.3);
        \draw[red] (9.85,0.175) node {\scriptsize{$\vdots$}};
        \draw[red] (9.85,-0.175) node {\scriptsize{$\vdots$}};
        \ns{9.5,0}
        \DsixOMinusFree{$k{-}4$}{5}{3}
        \DeightMany{1}{6}{1}
        \DsixOMinusTildeFree{$k{-}4$}{6}{2}
        \DeightMany{1}{7}{1}
        \DsixOMinusFree{$k{-}4$}{7}{3}
        \DeightMany{1}{8}{1}
        \DsixOMinusTildeFree{$k{-}4$}{8}{2}
        \DeightMany{1}{9}{1}
        \DsixOMinusFree{$k{-}4$}{9}{3}
%
%
        \DeightMany{1}{10}{1}
        \DsixOMinusTildeFree{$k{-}4$}{10}{2}
        \DeightMany{1}{11}{1}
        \draw (11.25,0) node {$\cdots$};
        \DeightMany{1}{11.5}{1}
		\DsixOMinusFree{$k{-}4$}{11.5}{3}
		\DeightMany{1}{12.5}{1}
		\DsixOMinusTildeFree{$k{-}4$}{12.5}{2}
        \DeightMany{1}{13.5}{1}
        \draw (13.75,0) node {$\cdots$};
		\DeightMany{1}{14}{1}
		\DsixOMinusFree{$1$}{14}{0}
		\DeightMany{1}{15}{1}
		\Dbrane{15,0}{16,0}
		\DeightMany{1}{16}{1}
\draw[decoration={brace,mirror,raise=10pt},decorate,thick]
  (-1.7,-1) -- node[below=10pt] {$2k$ half \De } (9.2,-1);
  \draw[decoration={brace,mirror,raise=10pt},decorate,thick]
  (9.2,1) -- node[above=10pt] {$8$ half \De } (2-0.2,1);
\draw[decoration={brace,mirror,raise=10pt},decorate,thick]
  (10-0.2,-1) -- node[below=10pt] {$2k$ half \De } (16.2,-1);
\draw[decoration={brace,mirror,raise=10pt},decorate,thick]
  (13.6,1) -- node[above=10pt] {$8$ half \De } (10-0.2,1);
	\end{tikzpicture}
	}
\end{align}
Here, the \Ds\ branes that have been aligned to respect to S-rule are displayed 
in red; the number of freely moving \Ds\ branes has been adjusted accordingly.
Now, the full \NS\ brane can fractionate into two half \NS s, which are 
confined to the \Os\ plane. To reach an easier to read configuration, one can 
eliminate the frozen \Ds\ branes between the \NS\ and 
\De\ branes via a brane transition of the half \NS\ branes through 
enough half \De\ branes. Note that this is 
analogous to the discussion in Section \ref{sec:1M5_generic_k}. 
Taking care of brane annihilation \eqref{eq:brane_creation_all}, the brane 
configuration becomes
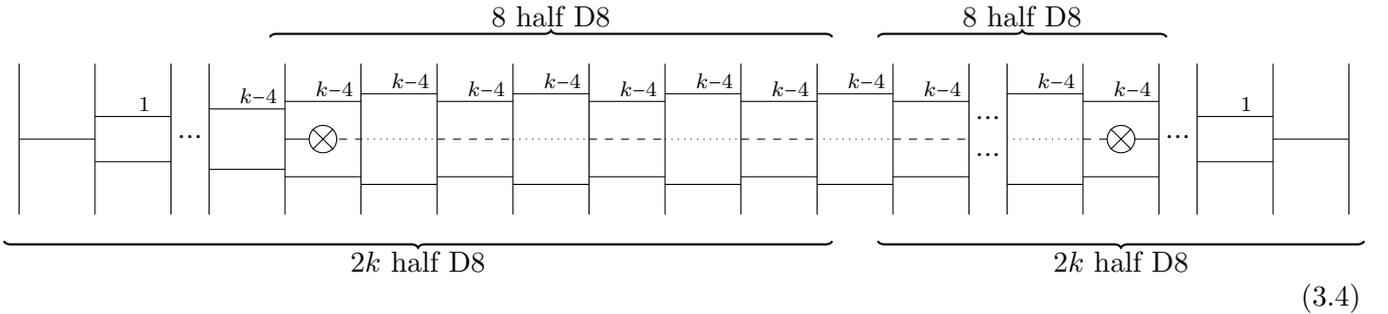
\begin{align}
 \raisebox{-.5\height}{
\begin{tikzpicture}
        \DeightMany{1}{-1.5}{1}
        \Dbrane{-1.5,0}{-0.5,0}
		\DeightMany{1}{-0.5}{1}
		\DsixOMinusFree{$1$}{-0.5}{0}
		\DeightMany{1}{0.5}{1}
        \draw (0.75,0) node {$\cdots$};
        \DeightMany{1}{1}{1}
        \DsixOMinusFree{$k{-}4$}{1}{1}
        \DeightMany{1}{2}{1}
        \DsixOMinusFree{$k{-}4$}{2}{2}
        \Dbrane{2,0}{2.5,0}
        \OPlusTilde{2.5,0}{3,0}
		\DeightMany{1}{3}{1}
		\DsixOPlusFree{$k{-}4$}{3}{3}
		\DeightMany{1}{4}{1}
		\DsixOPlusTildeFree{$k{-}4$}{4}{2}
		\DeightMany{1}{5}{1}
        \ns{2.5,0}
        \DsixOPlusFree{$k{-}4$}{5}{3}
        \DeightMany{1}{6}{1}
        \DsixOPlusTildeFree{$k{-}4$}{6}{2}
        \DeightMany{1}{7}{1}
        \DsixOPlusFree{$k{-}4$}{7}{3}
        \DeightMany{1}{8}{1}
        \DsixOPlusTildeFree{$k{-}4$}{8}{2}
        \DeightMany{1}{9}{1}
        \DsixOPlusFree{$k{-}4$}{9}{3}
%
%
        \DeightMany{1}{10}{1}
        \DsixOPlusTildeFree{$k{-}4$}{10}{2}
        \DeightMany{1}{11}{1}
        \draw (11.25,0.25) node {$\cdots$};
        \draw (11.25,-0.25) node {$\cdots$};
        \DeightMany{1}{11.5}{1}
		\DsixOPlusFree{$k{-}4$}{11.5}{3}
		\DeightMany{1}{12.5}{1}
		\DsixOMinusFree{$k{-}4$}{12.5}{2}
		\OPlusTilde{12.5,0}{13,0}
        \Dbrane{13,0}{13.5,0}
        \DeightMany{1}{13.5}{1}
        \ns{13,0}
        \draw (13.75,0) node {$\cdots$};
		\DeightMany{1}{14}{1}
		\DsixOMinusFree{$1$}{14}{0}
		\DeightMany{1}{15}{1}
		\Dbrane{15,0}{16,0}
		\DeightMany{1}{16}{1}
\draw[decoration={brace,mirror,raise=10pt},decorate,thick]
  (-1.7,-1) -- node[below=10pt] {$2k$ half \De } (9.2,-1);
  \draw[decoration={brace,mirror,raise=10pt},decorate,thick]
  (9.2,1) -- node[above=10pt] {$8$ half \De } (2-0.2,1);
\draw[decoration={brace,mirror,raise=10pt},decorate,thick]
  (10-0.2,-1) -- node[below=10pt] {$2k$ half \De } (16.2,-1);
\draw[decoration={brace,mirror,raise=10pt},decorate,thick]
  (13.6,1) -- node[above=10pt] {$8$ half \De } (10-0.2,1);
	\end{tikzpicture}
	}
	\label{eq:Hasse_after_e8}
\end{align}
and one recognises that the brane configuration of the remaining freely moving 
\Ds\ branes yields the finite coupling case of \eqref{eq:moving_NS}.

As a consistency check, one counts the loss in magnetic degrees of 
freedom: there are $28$ freely moving (full) \Ds\ segments lost during the 
transition and the half \NS\ branes are confined to the orientifold plane, 
marking another lost degree of freedom. Therefore, one recovers a loss of $29$ 
quaternionic dimension during the small $E_8$ instanton transition.

Moreover, one can read off the electric and magnetic theory of this 
configuration. Unsurprisingly, the magnetic theory is just the one derived in 
\eqref{eq:k>4_magQuiver_finite}. The electric theory is seen to be trivial, as 
there are no \Ds\ branes suspended between the half \NS\ branes. In terms of 
the electric theory \eqref{eq:1M5_electric_quiver}, the triviality of the 
electric theory in the phase \eqref{eq:Hasse_after_e8} is due the locus of the 
Higgs branch where the $\sprm(k-4)$ gauge group is completely broken to the 
trivial gauge group. 
\paragraph{$d_{10}$ transition.}
Moving on to the finite coupling Higgs branch, one needs to find all possible 
Kraft-Procesi transitions. Inspecting brane configuration 
\eqref{eq:Hasse_after_e8}, it is straightforward to see the next transition: 
moving the half \NS\ branes outwards through two half \De\ branes each and 
accounting for brane creation. In detail,
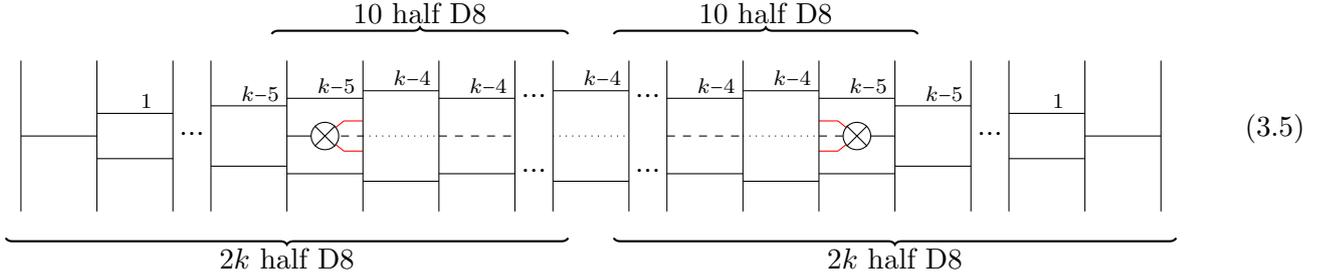
\begin{align}
 \raisebox{-.5\height}{
\begin{tikzpicture}
        \DeightMany{1}{-1.5}{1}
        \Dbrane{-1.5,0}{-0.5,0}
		\DeightMany{1}{-0.5}{1}
		\DsixOMinusFree{$1$}{-0.5}{0}
		\DeightMany{1}{0.5}{1}
        \draw (0.75,0) node {$\cdots$};
        \DeightMany{1}{1}{1}
        \DsixOMinusFree{$k{-}5$}{1}{1}
        \DeightMany{1}{2}{1}
        \DsixOMinusFree{$k{-}5$}{2}{2}
        \Dbrane{2,0}{2.5,0}
        \OPlusTilde{2.5,0}{3,0}
		\DeightMany{1}{3}{1}
		\DsixOPlusFree{$k{-}4$}{3}{3}
		\DeightMany{1}{4}{1}
		\DsixOPlusTildeFree{$k{-}4$}{4}{2}
		\DeightMany{1}{5}{1}
%
\draw[red] (2.5,0)--(2.75,0.2)--(3,0.2);
\draw[red] (2.5,0)--(2.75,-0.2)--(3,-0.2);
\draw[red] (9,0.2)--(9.25,0.2)--(9.5,0);
\draw[red] (9,-0.2)--(9.25,-0.2)--(9.5,0);
        \ns{2.5,0}
        \draw (5.25,0.5) node {$\cdots$};
        \draw (5.25,-0.5) node {$\cdots$};
        \DeightMany{1}{5.5}{1}
        \DsixOPlusFree{$k{-}4$}{5.5}{3}
        \DeightMany{1}{6.5}{1}
        \draw (6.75,0.5) node {$\cdots$};
        \draw (6.75,-0.5) node {$\cdots$};
        \DsixOPlusTildeFree{$k{-}4$}{7}{2}
        \DeightMany{1}{7}{1}
        \DsixOPlusFree{$k{-}4$}{8}{3}
        \DeightMany{1}{8}{1}
        \DeightMany{1}{9}{1}
		\DsixOMinusFree{$k{-}5$}{9}{2}
		\OPlusTilde{9,0}{9.5,0}
        \Dbrane{9.5,0}{10,0}
        \DeightMany{1}{10}{1}
        \DsixOMinusFree{$k{-}5$}{10}{1}
        \DeightMany{1}{11}{1}
        \ns{9.5,0}
        \draw (11.25,0) node {$\cdots$};
		\DeightMany{1}{11.5}{1}
		\DsixOMinusFree{$1$}{11.5}{0}
		\DeightMany{1}{12.5}{1}
		\Dbrane{12.5,0}{13.5,0}
		\DeightMany{1}{13.5}{1}
\draw[decoration={brace,mirror,raise=10pt},decorate,thick]
  (-1.7,-1) -- node[below=10pt] {$2k$ half \De } (5.7,-1);
  \draw[decoration={brace,mirror,raise=10pt},decorate,thick]
  (5.7,1) -- node[above=10pt] {$10$ half \De } (2-0.2,1);
\draw[decoration={brace,mirror,raise=10pt},decorate,thick]
  (6.3,-1) -- node[below=10pt] {$2k$ half \De } (13.7,-1);
\draw[decoration={brace,mirror,raise=10pt},decorate,thick]
  (10.3,1) -- node[above=10pt] {$10$ half \De } (6.3,1);
	\end{tikzpicture}
	}
\end{align}
and the created \Ds\ branes (displayed in red) indicate that the next 
KP-transition proceeds by aligning sufficiently many freely moving \Ds\ such 
that there is one full \Ds\ brane suspended between the two half \NS s. In the 
brane configuration, this becomes
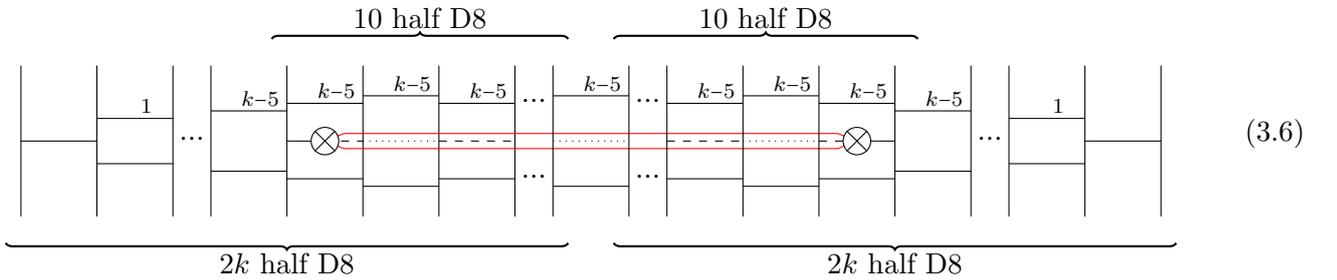
\begin{align}
 \raisebox{-.5\height}{
\begin{tikzpicture}
        \DeightMany{1}{-1.5}{1}
        \Dbrane{-1.5,0}{-0.5,0}
		\DeightMany{1}{-0.5}{1}
		\DsixOMinusFree{$1$}{-0.5}{0}
		\DeightMany{1}{0.5}{1}
        \draw (0.75,0) node {$\cdots$};
        \DeightMany{1}{1}{1}
        \DsixOMinusFree{$k{-}5$}{1}{1}
        \DeightMany{1}{2}{1}
        \DsixOMinusFree{$k{-}5$}{2}{2}
        \Dbrane{2,0}{2.5,0}
        \OPlusTilde{2.5,0}{3,0}
		\DeightMany{1}{3}{1}
		\DsixOPlusFree{$k{-}5$}{3}{3}
		\DeightMany{1}{4}{1}
		\DsixOPlusTildeFree{$k{-}5$}{4}{2}
		\DeightMany{1}{5}{1}
\draw[red] (2.5,0)--(2.75,0.1)--(9.25,0.1)--(9.5,0);
\draw[red] (2.5,0)--(2.75,-0.1)--(9.25,-0.1)--(9.5,0);
        \ns{2.5,0}
        \draw (5.25,0.5) node {$\cdots$};
        \draw (5.25,-0.5) node {$\cdots$};
        \DeightMany{1}{5.5}{1}
        \DsixOPlusFree{$k{-}5$}{5.5}{3}
        \DeightMany{1}{6.5}{1}
        \draw (6.75,0.5) node {$\cdots$};
        \draw (6.75,-0.5) node {$\cdots$};
        \DsixOPlusTildeFree{$k{-}5$}{7}{2}
        \DeightMany{1}{7}{1}
        \DsixOPlusFree{$k{-}5$}{8}{3}
        \DeightMany{1}{8}{1}
        \DeightMany{1}{9}{1}
		\DsixOMinusFree{$k{-}5$}{9}{2}
		\OPlusTilde{9,0}{9.5,0}
        \Dbrane{9.5,0}{10,0}
        \DeightMany{1}{10}{1}
        \DsixOMinusFree{$k{-}5$}{10}{1}
        \DeightMany{1}{11}{1}
        \ns{9.5,0}
        \draw (11.25,0) node {$\cdots$};
		\DeightMany{1}{11.5}{1}
		\DsixOMinusFree{$1$}{11.5}{0}
		\DeightMany{1}{12.5}{1}
		\Dbrane{12.5,0}{13.5,0}
		\DeightMany{1}{13.5}{1}
\draw[decoration={brace,mirror,raise=10pt},decorate,thick]
  (-1.7,-1) -- node[below=10pt] {$2k$ half \De } (5.7,-1);
  \draw[decoration={brace,mirror,raise=10pt},decorate,thick]
  (5.7,1) -- node[above=10pt] {$10$ half \De } (2-0.2,1);
\draw[decoration={brace,mirror,raise=10pt},decorate,thick]
  (6.3,-1) -- node[below=10pt] {$2k$ half \De } (13.7,-1);
\draw[decoration={brace,mirror,raise=10pt},decorate,thick]
  (10.3,1) -- node[above=10pt] {$10$ half \De } (6.3,1);
	\end{tikzpicture}
	}
\label{eq:Hasse_after_d10}
\end{align}
and the numbers of freely moving \Ds\ branes has been adjusted. One computes 
that the number of lost freely moving \Ds\ branes is $17$, and the next step is 
to figure out the nature of the transition.

Then the remaining magnetic theory is deduced from the freely moving \Ds\ 
branes, as before. The \Ds\ branes suspended between the half \NS\ branes do 
not contribute, while the \NS\ branes still induce flavour nodes. Therefore, 
the magnetic quiver becomes
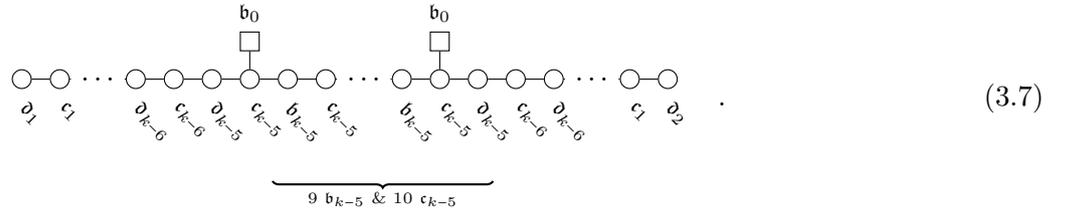
\begin{align}
 \raisebox{-.5\height}{
 	\begin{tikzpicture}
 	\tikzset{node distance = 0.5cm};
	\tikzstyle{gauge} = [circle, draw,inner sep=2.5pt];
	\tikzstyle{flavour} = [regular polygon,regular polygon sides=4,inner 
sep=2.5pt, draw];
\node (g2) [gauge,label={[rotate=-45]below right:{\dd{1}}}] {};
\node (g3) [gauge,right of=g2,label={[rotate=-45]below right:{\cc{1}}}] {};
\node (g4) [right of=g3] {$\ldots$};
\node (g5) [gauge,right of=g4,label={[rotate=-45]below right:{\dd{k{-}6}}}]{};
\node (g6) [gauge,right of=g5,label={[rotate=-45]below right:{\cc{k{-}6}}}]{};
\node (g7) [gauge,right of=g6,label={[rotate=-45]below right:{\dd{k{-}5}}}] 
{};
	\node (g8) [gauge,right of=g7,label={[rotate=-45]below right:{\cc{k{-}5}}}] 
{};
    \node (g9) [gauge,right of=g8,label={[rotate=-45]below 
right:{\bb{k{-}5}}}] {};
	\node (g10) [gauge,right of=g9,label={[rotate=-45]below 
right:{\cc{k{-}5}}}] 
{};
    \node (g11) [right of=g10] {$\ldots$};
    \node (g12) [gauge,right of=g11,label={[rotate=-45]below 
right:{\bb{k{-}5}}}] {};
	\node (g13) [gauge,right of=g12,label={[rotate=-45]below 
right:{\cc{k{-}5}}}] 
{};
    \node (g14) [gauge,right of=g13,label={[rotate=-45]below 
right:{\dd{k{-}5}}}] {};
	\node (g15) [gauge,right of=g14,label={[rotate=-45]below 
right:{\cc{k{-}6}}}] 
{};
	\node (g16) [gauge,right of=g15,label={[rotate=-45]below 
right:{\dd{k{-}6}}}] {};
	\node (g17) [right of=g16] {$\ldots$};
	\node (g18) [gauge,right of=g17,label={[rotate=-45]below right:{\cc{1}}}] 
{};
	\node (g19) [gauge,right of=g18,label={[rotate=-45]below right:{\dd{2}}}] 
{};
	\node (f8) [flavour,above of=g8,label=above:{\bb{0}}] {};
	\node (f13) [flavour,above of=g13,label=above:{\bb{0}}] {};
	\draw (g2)--(g3) (g3)--(g4) (g4)--(g5) (g5)--(g6) (g6)--(g7) 
(g7)--(g8) (g8)--(g9) (g9)--(g10) (g10)--(g11) (g11)--(g12) (g12)--(g13)
(g13)--(g14) (g14)--(g15) (g15)--(g16) (g16)--(g17) (g17)--(g18) (g18)--(g19) 
(g8)--(f8) (g13)--(f13);
\draw[decoration={brace,mirror,raise=10pt},decorate,thick]
  (3.5-0.2,-1) -- node[below=10pt] {\tiny{$9$} $\balg_{k{-5}}$ \tiny{\& $10$} 
$\calg_{k{-}5}$ } (6+0.2,-1);
	\end{tikzpicture}
	} 
	\,.
\end{align}
and the Coulomb branch dimension is reduced by $17$ in comparison to 
\eqref{eq:k>4_magQuiver_finite}.

Next, one reads off the electric theory in this configuration from the red 
brane subconfiguration, and finds an $\sprm(1)$ gauge theory with $\sorm(20)$ 
flavour. The Higgs branch thereof is the closure of the minimal nilpotent orbit 
of $\sorm(20)$, which has quaternionic dimension $17$. A transition of this 
type is called a $d_{10}$ transition.  
\paragraph{More $d_{2l}$ transitions.}
Lastly, the transition that led to \eqref{eq:Hasse_after_d10} can be 
iterated until all \Ds\ branes are suspended between the 
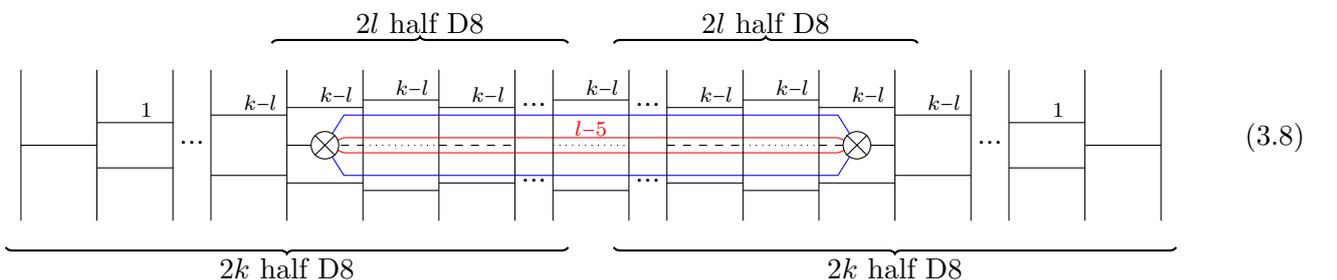
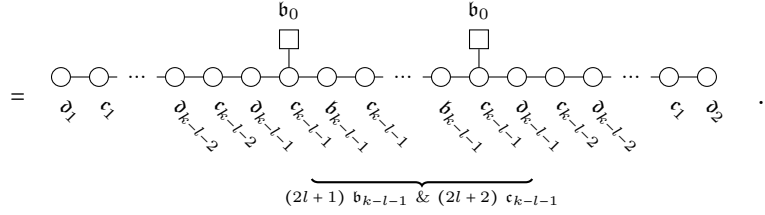
\NS\ branes. 

\begin{align}
 \raisebox{-.5\height}{
\begin{tikzpicture}
        \DeightMany{1}{-1.5}{1}
        \Dbrane{-1.5,0}{-0.5,0}
		\DeightMany{1}{-0.5}{1}
		\DsixOMinusFree{$1$}{-0.5}{0}
		\DeightMany{1}{0.5}{1}
        \draw (0.75,0) node {$\cdots$};
        \DeightMany{1}{1}{1}
        \DsixOMinusFree{$k{-}l$}{1}{1}
        \DeightMany{1}{2}{1}
        \DsixOMinusFree{$k{-}l$}{2}{2}
        \Dbrane{2,0}{2.5,0}
        \OPlusTilde{2.5,0}{3,0}
		\DeightMany{1}{3}{1}
		\DsixOPlusFree{$k{-}l$}{3}{3}
		\DeightMany{1}{4}{1}
		\DsixOPlusTildeFree{$k{-}l$}{4}{2}
		\DeightMany{1}{5}{1}
%
\draw[blue] (2.5,0)--(2.75,0.4)--(9.25,0.4)--(9.5,0);
\draw[blue] (2.5,-0)--(2.75,-0.4)--(9.25,-0.4)--(9.5,0);
\node[red,label={[red]above:{$\scriptstyle{l{-}5}$}}] at (6,-0.15) {};
\draw[red] (2.5,0)--(2.75,0.1)--(9.25,0.1)--(9.5,0);
\draw[red] (2.5,0)--(2.75,-0.1)--(9.25,-0.1)--(9.5,0);
        \ns{2.5,0}
        \draw (5.25,0.5) node {$\cdots$};
        \draw (5.25,-0.5) node {$\cdots$};
        \DeightMany{1}{5.5}{1}
        \DsixOPlusFree{$k{-}l$}{5.5}{3}
        \DeightMany{1}{6.5}{1}
        \draw (6.75,0.5) node {$\cdots$};
        \draw (6.75,-0.5) node {$\cdots$};
        \DsixOPlusTildeFree{$k{-}l$}{7}{2}
        \DeightMany{1}{7}{1}
        \DsixOPlusFree{$k{-}l$}{8}{3}
        \DeightMany{1}{8}{1}
        \DeightMany{1}{9}{1}
		\DsixOMinusFree{$k{-}l$}{9}{2}
		\OPlusTilde{9,0}{9.5,0}
        \Dbrane{9.5,0}{10,0}
        \DeightMany{1}{10}{1}
        \DsixOMinusFree{$k{-}l$}{10}{1}
        \DeightMany{1}{11}{1}
        \ns{9.5,0}
        \draw (11.25,0) node {$\cdots$};
		\DeightMany{1}{11.5}{1}
		\DsixOMinusFree{$1$}{11.5}{0}
		\DeightMany{1}{12.5}{1}
		\Dbrane{12.5,0}{13.5,0}
		\DeightMany{1}{13.5}{1}
\draw[decoration={brace,mirror,raise=10pt},decorate,thick]
  (-1.7,-1) -- node[below=10pt] {$2k$ half \De } (5.7,-1);
  \draw[decoration={brace,mirror,raise=10pt},decorate,thick]
  (5.7,1) -- node[above=10pt] {$2l$ half \De } (2-0.2,1);
\draw[decoration={brace,mirror,raise=10pt},decorate,thick]
  (6.3,-1) -- node[below=10pt] {$2k$ half \De } (13.7,-1);
\draw[decoration={brace,mirror,raise=10pt},decorate,thick]
  (10.3,1) -- node[above=10pt] {$2l$ half \De } (6.3,1);
	\end{tikzpicture}
	}
\label{eq:Hasse_after_some_d-transitions}
\end{align}
The magnetic theory is determined as before
\begin{align}
 \raisebox{-.5\height}{
 	\begin{tikzpicture}
 	\tikzset{node distance = 0.5cm};
	\tikzstyle{gauge} = [circle, draw,inner sep=2.5pt];
	\tikzstyle{flavour} = [regular polygon,regular polygon sides=4,inner 
sep=2.5pt, draw];
\node (g2) [gauge,label={[rotate=-45]below right:{\dd{1}}}] {};
\node (g3) [gauge,right of=g2,label={[rotate=-45]below right:{\cc{1}}}] {};
\node (g4) [right of=g3] {$\scriptstyle{\ldots}$};
\node (g5) [gauge,right of=g4,label={[rotate=-45]below 
right:{\dd{k{-}l{-}1}}}]{};
\node (g6) [gauge,right of=g5,label={[rotate=-45]below 
right:{\cc{k{-}l{-}1}}}]{};
\node (g7) [gauge,right of=g6,label={[rotate=-45]below right:{\dd{k{-}l}}}] 
{};
	\node (g8) [gauge,right of=g7,label={[rotate=-45]below right:{\cc{k{-}l}}}] 
{};
    \node (g9) [gauge,right of=g8,label={[rotate=-45]below 
right:{\bb{k{-}l}}}] {};
	\node (g10) [gauge,right of=g9,label={[rotate=-45]below 
right:{\cc{k{-}l}}}] 
{};
    \node (g11) [right of=g10] {$\scriptstyle{\ldots}$};
    \node (g12) [gauge,right of=g11,label={[rotate=-45]below 
right:{\bb{k{-}l}}}] {};
	\node (g13) [gauge,right of=g12,label={[rotate=-45]below 
right:{\cc{k{-}l}}}] 
{};
    \node (g14) [gauge,right of=g13,label={[rotate=-45]below 
right:{\dd{k{-}l}}}] {};
	\node (g15) [gauge,right of=g14,label={[rotate=-45]below 
right:{\cc{k{-}l{-}1}}}] 
{};
	\node (g16) [gauge,right of=g15,label={[rotate=-45]below 
right:{\dd{k{-}l{-}1}}}] {};
	\node (g17) [right of=g16] {$\scriptstyle{\ldots}$};
	\node (g18) [gauge,right of=g17,label={[rotate=-45]below right:{\cc{1}}}] 
{};
	\node (g19) [gauge,right of=g18,label={[rotate=-45]below right:{\dd{2}}}] 
{};
	\node (f8) [flavour,above of=g8,label=above:{\bb{0}}] {};
	\node (f13) [flavour,above of=g13,label=above:{\bb{0}}] {};
	\draw (g2)--(g3) (g3)--(g4) (g4)--(g5) (g5)--(g6) (g6)--(g7) 
(g7)--(g8) (g8)--(g9) (g9)--(g10) (g10)--(g11) (g11)--(g12) (g12)--(g13)
(g13)--(g14) (g14)--(g15) (g15)--(g16) (g16)--(g17) (g17)--(g18) (g18)--(g19) 
(g8)--(f8) (g13)--(f13);
\draw[decoration={brace,mirror,raise=10pt},decorate,thick]
  (3.5-0.2,-1) -- node[below=10pt] {\tiny{$(2l-1)$} $\balg_{k{-l}}$ \tiny{\& 
$2l$} 
$\calg_{k{-}l}$ } (6+0.2,-1);
	\end{tikzpicture}
	} 
	\,.
	\label{eq:magQuiver_after_d_2l}
\end{align}
The full electric theory, determined by all red and blue \Ds\ branes in this 
phase \eqref{eq:Hasse_after_some_d-transitions}, is an $\sprm(l{-}4)$ gauge 
theory with $\sorm(2l)$ flavour group. However, the electric theory 
corresponding to the KP-transition is giving by the blue brane subconfiguration 
and describes an $\sprm(1)$ gauge theory with $\sorm(2l)$ flavour. The Higgs 
branch of the latter is the closure of the minimal nilpotent orbit of 
$\sorm(2l)$. Therefore, the transition is of type $d_l$.  

The brane configuration for the last step is then straightforwardly deduced by 
setting $l=k$ in \eqref{eq:Hasse_after_some_d-transitions}.
For completeness, one verifies the electric and magnetic theory for this 
configuration. The magnetic theory is empty, as there are no magnetic degrees 
of freedom left. The electric theory, following from all \Ds\ branes 
suspended between \NS\ branes, is a $\sprm(k-4)$ gauge theory with $\sorm(4k)$ 
flavour. The transition is described by an $\sprm(1)$ gauge group with 
$\sorm(4k)$ flavour such that the Higgs branch thereof is the closures of the 
minimal nilpotent orbit of $\sorm(4k)$. Hence, one recovers a $d_{2k}$ 
transition.
\begin{figure}[t]
\centering
 	\begin{tikzpicture}
	\tikzstyle{Hasse} = [circle, draw,inner sep=1.5pt,fill=black];
	\node (g0) [Hasse, label=below right:{$e_8$}] {};
	\node (g1) [Hasse, below of =g0, label=below right:{$d_{10}$}] {};
	\node (g2) [Hasse, below of =g1, label=below right:{$d_{12}$}] {};
	\node (g3) [Hasse, below of =g2] {};
	\node (g4) [below of =g3] {$\vdots$};
	\node (g5) [Hasse, below of =g4, label=below right:{$d_{2l}$}] {};
	\node (g6) [Hasse, below of =g5] {};
	\node (g7) [below of =g6] {$\vdots$};
    \node (g8) [Hasse, below of =g7, label=below right:{$d_{2k}$}] {};
	\node (g9) [Hasse, below of =g8] {};
	\draw (g0)--(g1) (g1)--(g2) (g2)--(g3) (g3)--(g4) (g4)--(g5) (g5)--(g6) 
(g6)--(g7) (g7)--(g8) (g8)--(g9);
	\draw[decoration={brace,mirror,raise=10pt},decorate,thick]
  (10,-9.1) -- node[right=10pt] {$\scriptstyle{\Higgs_\infty} 
  \big($
  \raisebox{-.25\height}{ $
\overset{\overset{\Box}|}{ \circ} 
\overset{\overset{\scriptscriptstyle{\sorm(4k)}}{\phantom{|} } 
}{\scriptscriptstyle{ \sprm(k { - } 4)} } $ }
$ \big)$ } (10,0.1);
\draw[decoration={brace,mirror,raise=10pt},decorate,thick]
  (7,-9.1) -- node[right=10pt] {$\scriptstyle{\Higgs_\fin} \big( $
  \raisebox{-.25\height}{ $
\overset{\overset{\Box}|}{ \circ} 
\overset{\overset{\scriptscriptstyle{\sorm(4k)}}{\phantom{|} } 
}{\scriptscriptstyle{ \sprm(k { - } 4)} }  $ }
$\big)$ } (7,-1.0+0.1);
  	\draw[decoration={brace,mirror,raise=10pt},decorate,thick]
  (1,-6.1) -- node[right=10pt] {$\scriptstyle{\Higgs_\fin} \big( $
  \raisebox{-.25\height}{ $
\overset{\overset{\Box}|}{ \circ} 
\overset{\overset{\scriptscriptstyle{\sorm(4l)}}{\phantom{|} } 
}{\scriptscriptstyle{ \sprm(l { - } 4)} }  $ }
$\big)$ } (1,-1.0+0.1);
  	\draw[decoration={brace,mirror,raise=10pt},decorate,thick]
  (4,-6.1) -- node[right=10pt] {$\scriptstyle{\Higgs_\infty} \big( $
  \raisebox{-.25\height}{ $
\overset{\overset{\Box}|}{ \circ} 
\overset{\overset{\scriptscriptstyle{\sorm(4l)}}{\phantom{|} } 
}{\scriptscriptstyle{ \sprm(l { - } 4)} }  $ }
$\big)$ } (4,+0.1);
  	\draw[decoration={brace,mirror,raise=10pt},decorate,thick]
  (2,-9.1) -- node[right=10pt] {$\scriptstyle{\Coulomb} \big( 
\substack{\text{magnetic}\\ \text{quiver} } \; \eqref{eq:magQuiver_after_d_2l} 
\big)$ } (2,-6.1);
  	\draw[decoration={brace,mirror,raise=10pt},decorate,thick]
  (2,-1.1) -- node[right=10pt] {$\scriptstyle{\clorbit{\min}^{E_8}} $ } (2,0.1);
	\end{tikzpicture}
\caption{The Hasse diagram for the Higgs branch of 
\eqref{eq:1M5_electric_quiver} at infinite gauge coupling. There are two types 
of minimal transitions: firstly, the $e_8$ transition, i.e. the transverse 
slice is the closure of the minimal nilpotent orbit of $E_8$. Secondly, 
various $d_{2l}$ transitions, i.e.\ the transverse slice is the closure of the 
minimal nilpotent orbit of $\sorm(4l)$.}
\label{fig:Hasse}
\end{figure}

Summarising the findings, the Hasse diagram is displayed in Figure 
\ref{fig:Hasse}. From there, one deduces various geometric relationships such 
as: For a fixed $\sprm(k-4)$ gauge theory with $\sorm(4k)$ flavour, the 
transverse slice of the Higgs branch at finite gauge coupling inside the Higgs 
branch at infinite coupling is the minimal nilpotent orbit closure of $E_8$. In 
addition for $4\leq l<k$, the transverse slice of the Higgs branch of an 
$\sprm(l-4)$ theory at finite (or infinite) coupling  inside the Higgs branch of 
an $\sprm(k-4)$ theory at finite (or infinite) coupling is the Coulomb branch 
of the magnetic quiver \eqref{eq:magQuiver_after_d_2l}.
\subsection{From quiver subtraction}
The analysis can be repeated by means of quiver subtraction 
\cite{Cabrera:2018ann} that translates Kraft-Procesi transitions 
\cite{Cabrera:2016vvv,Cabrera:2017njm,Hanany:2018uhm} of the brane 
configurations into an operation on the magnetic quivers. 
Contrary to \cite{Bourget:2019aer}, the realisation of the KP transitions here 
requires orthosymplectic quivers.
As shown in Section \ref{sec:Hasse_from_brane}, the simplest case 
\eqref{eq:1M5_electric_quiver} only requires an orthosymplectic quiver for the 
$d_{2l}$ transitions \cite[Tab.\ 7]{Cabrera:2017njm} and for the $e_8$ 
transition 
\cite[Eq.\ (2.43)]{Hanany:2018uhm}. The rules for quiver subtraction of minimal 
transitions in orthosymplectic quivers can be summarised as follows: 
The two to-be-subtracted quivers are aligned along the common subquiver. One 
only subtracts gauge nodes of the same algebra type and the arithmetic works 
like:
\begin{align}
\balg_{n}-\balg_l = \balg_{n-l} \, , \qquad
\calg_n-\calg_l = \calg_{n-l} \, , \qquad
\dalg_n-\dalg_l = \balg_{n-l} \, , \quad \mathrm{for }\; n\geq l\,.
\end{align}
The resulting quiver needs to be rebalanced, analogously to 
\cite{Bourget:2019aer}.
\paragraph{$e_8$ transition.}
The small $E_8$ instanton transition has been discussed in Section 
\ref{sec:single_M5} in detail. Inspecting the magnetic quiver 
\eqref{eq:k>4_magQuiver_infinite} and knowing the orthosymplectic quiver 
realisation for the $e_8$ transition \eqref{eq:magQuiver_k=4_infinite}, one 
recognises the possibility of subtracting the $e_8$ quiver, because the quiver 
\eqref{eq:magQuiver_k=4_infinite} is a subquiver of 
\eqref{eq:k>4_magQuiver_infinite}. In detail, quiver subtraction yields
\begin{equation}
\begin{aligned}
 &\quad \quad \; \;
 \raisebox{-.5\height}{
 	\begin{tikzpicture}
 	\tikzset{node distance = 0.5cm};
	\tikzstyle{gauge} = [circle, draw,inner sep=2.5pt];
	\tikzstyle{flavour} = [regular polygon,regular polygon sides=4,inner 
sep=2.5pt, draw];
	\node (g2) [gauge,label={[rotate=-45]below right:{\dd{1}}}] 
{};
	\node (g3) [gauge,right of=g2,label={[rotate=-45]below right:{\cc{1}}}] {};
	\node (g4) [gauge,right of=g3,label={[rotate=-45]below 
right:{\dd{2}}}] {};
	\node (g5) [gauge,right of=g4,label={[rotate=-45]below right:{\cc{2}}}] 
{};
	\node (g6) [right of=g5] {$\scriptstyle{\ldots}$};
    \node (g7) [gauge,right of=g6,label={[rotate=-45]below 
right:{\dd{k{-}1}}}] {};
	\node (g8) [gauge,right of=g7,label={[rotate=-45]below right:{\cc{k{-}1}}}] 
{};
    \node (g9) [gauge,right of=g8,label={[rotate=-45]below 
right:{\dd{k}}}] {};
	\node (g10) [gauge,right of=g9,label={[rotate=-45]below 
right:{\cc{k{-}1}}}] 
{};
    \node (g11) [gauge,right of=g10,label={[rotate=-45]below 
right:{\dd{k{-}1}}}] {};
    \node (g12) [right of=g11] {$\scriptstyle{\ldots}$};
	\node (g13) [gauge,right of=g12,label={[rotate=-45]below right:{\cc{2}}}] 
{};
    \node (g14) [gauge,right of=g13,label={[rotate=-45]below right:{\dd{2}}}] 
{};
	\node (g15) [gauge,right of=g14,label={[rotate=-45]below right:{\cc{1}}}] 
{};
	\node (g16) [gauge,right of=g15,label={[rotate=-45]below right:{\dd{1}}}] 
{};
	\node (g0) [gauge,above of=g9,label=above:{\cc{1}}] {};
	\draw  (g2)--(g3) (g3)--(g4) (g4)--(g5) (g5)--(g6) (g6)--(g7) 
(g7)--(g8) (g8)--(g9) (g9)--(g10) (g10)--(g11) (g11)--(g12) (g12)--(g13)
(g13)--(g14) (g14)--(g15) (g15)--(g16)  (g9)--(g0);
	\end{tikzpicture}
	} 
	\\
&-   \qquad \quad
 \raisebox{-.5\height}{
 	\begin{tikzpicture}
 	\tikzset{node distance = 0.5cm};
	\tikzstyle{gauge} = [circle, draw,inner sep=2.5pt];
	\tikzstyle{flavour} = [regular polygon,regular polygon sides=4,inner 
sep=2.5pt, draw];
	\node (g2) [gauge,label={[rotate=-45]below right:{\dd{1}}}] 
{};
	\node (g3) [gauge,right of=g2,label={[rotate=-45]below right:{\cc{1}}}] {};
	\node (g4) [gauge,right of=g3,label={[rotate=-45]below 
right:{\dd{2}}}] {};
	\node (g5) [gauge,right of=g4,label={[rotate=-45]below right:{\cc{2}}}] 
{};
    \node (g6) [gauge,right of=g5,label={[rotate=-45]below 
right:{\dd{3}}}] {};
	\node (g7) [gauge,right of=g6,label={[rotate=-45]below right:{\cc{3}}}] 
{};
    \node (g8) [gauge,right of=g7,label={[rotate=-45]below 
right:{\dd{4}}}] {};
	\node (g9) [gauge,right of=g8,label={[rotate=-45]below 
right:{\cc{3}}}] 
{};
    \node (g10) [gauge,right of=g9,label={[rotate=-45]below 
right:{\dd{3}}}] {};
	\node (g11) [gauge,right of=g10,label={[rotate=-45]below right:{\cc{2}}}] 
{};
    \node (g12) [gauge,right of=g11,label={[rotate=-45]below right:{\dd{2}}}] 
{};
	\node (g13) [gauge,right of=g12,label={[rotate=-45]below right:{\cc{1}}}] 
{};
	\node (g14) [gauge,right of=g13,label={[rotate=-45]below right:{\dd{1}}}] 
{};
	\node (g0) [gauge,above of=g8,label=above:{\cc{1}}] {};
	\draw  (g2)--(g3) (g3)--(g4) (g4)--(g5) (g5)--(g6) (g6)--(g7) 
(g7)--(g8) (g8)--(g9) (g9)--(g10) (g10)--(g11) (g11)--(g12) (g12)--(g13)
(g13)--(g14)  (g8)--(g0) ;
	\end{tikzpicture}
	}
	\\
&=
\raisebox{-.5\height}{
 	\begin{tikzpicture}
 	\tikzset{node distance = 0.5cm};
	\tikzstyle{gauge} = [circle, draw,inner sep=2.5pt];
	\tikzstyle{flavour} = [regular polygon,regular polygon sides=4,inner 
sep=2.5pt, draw];
	\node (g2) [gauge,label={[rotate=-45]below right:{\dd{1}}}] 
{};
	\node (g3) [gauge,right of=g2,label={[rotate=-45]below right:{\cc{1}}}] {};
	\node (g4) [right of=g3] {$\scriptstyle{\ldots}$};
	\node (g5) [gauge,right of=g4,label={[rotate=-45]below 
right:{\dd{k{-}5}}}] {};
	\node (g6) [gauge,right of=g5,label={[rotate=-45]below right:{\cc{k{-}5}}}] 
{};
    \node (g7) [gauge,right of=g6,label={[rotate=-45]below 
right:{\dd{k{-}4}}}] {};
	\node (g8) [gauge,right of=g7,label={[rotate=-45]below right:{\cc{k{-}4}}}] 
{};
    \node (g9) [gauge,right of=g8,label={[rotate=-45]below 
right:{\bb{k{-}4}}}] {};
	\node (g10) [gauge,right of=g9,label={[rotate=-45]below 
right:{\cc{k{-}4}}}] 
{};
    \node (g11) [right of=g10] {$\scriptstyle{\ldots}$};
    \node (g12) [gauge,right of=g11,label={[rotate=-45]below 
right:{\bb{k{-}4}}}] {};
	\node (g13) [gauge,right of=g12,label={[rotate=-45]below 
right:{\cc{k{-}4}}}] 
{};
    \node (g14) [gauge,right of=g13,label={[rotate=-45]below 
right:{\dd{k{-}4}}}] {};
	\node (g15) [gauge,right of=g14,label={[rotate=-45]below 
right:{\cc{k{-}5}}}] 
{};
	\node (g16) [gauge,right of=g15,label={[rotate=-45]below 
right:{\dd{k{-}5}}}] {};
	\node (g17) [right of=g16] {$\scriptstyle{\ldots}$};
	\node (g18) [gauge,right of=g17,label={[rotate=-45]below right:{\cc{1}}}] 
{};
	\node (g19) [gauge,right of=g18,label={[rotate=-45]below right:{\dd{1}}}] 
{};
	\node (f8) [flavour,above of=g8,label=above:{\bb{0}}] {};
	\node (f13) [flavour,above of=g13,label=above:{\bb{0}}] {};
	\draw (g2)--(g3) (g3)--(g4) (g4)--(g5) (g5)--(g6) (g6)--(g7) 
(g7)--(g8) (g8)--(g9) (g9)--(g10) (g10)--(g11) (g11)--(g12) (g12)--(g13)
(g13)--(g14) (g14)--(g15) (g15)--(g16) (g16)--(g17) (g17)--(g18) (g18)--(g19) 
(g8)--(f8) (g13)--(f13);
\draw[decoration={brace,mirror,raise=10pt},decorate,thick]
  (3.5-0.2,-1) -- node[below=10pt] {\tiny{$7$} \bb{k{-4}} \tiny{\& $8$} 
\cc{k{-}4} } (6+0.2,-1);
	\end{tikzpicture}
	}
\end{aligned}
\label{eq:quiver_subtraction_1M5}
\end{equation}
such that the magnetic quiver for the finite coupling Higgs branch is obtained.
\paragraph{$d_{2l}$ transition.}
Given any of the  magnetic quivers \eqref{eq:magQuiver_after_d_2l}, the 
strategy is to identify subgraphs that 
correspond to KP transitions. Again, one needs to find 
possible subgraphs such that they can accommodate either a closure of a minimal 
nilpotent orbit or a Kleinian singularity. Inspecting the 
general case \eqref{eq:magQuiver_after_d_2l} 
and comparing to the known KP transitions of \cite[Tab.\ 6 \& 
7]{Cabrera:2017njm} one 
recognises that the $d_{2l}$ transitions is the only possibility. The 
subtraction becomes 
\begin{align}
&\; \raisebox{-.5\height}{
 	\begin{tikzpicture}
 	\tikzset{node distance = 0.5cm};
	\tikzstyle{gauge} = [circle, draw,inner sep=2.5pt];
	\tikzstyle{flavour} = [regular polygon,regular polygon sides=4,inner 
sep=2.5pt, draw];
\node (g2) [gauge,label={[rotate=-45]below right:{\dd{1}}}] {};
\node (g3) [gauge,right of=g2,label={[rotate=-45]below right:{\cc{1}}}] {};
\node (g4) [right of=g3] {$\scriptstyle{\ldots}$};
\node (g5) [gauge,right of=g4,label={[rotate=-45]below 
right:{\dd{k{-}l{-}1}}}]{};
\node (g6) [gauge,right of=g5,label={[rotate=-45]below 
right:{\cc{k{-}l{-}1}}}]{};
\node (g7) [gauge,right of=g6,label={[rotate=-45]below right:{\dd{k{-}l}}}] 
{};
	\node (g8) [gauge,right of=g7,label={[rotate=-45]below right:{\cc{k{-}l}}}] 
{};
    \node (g9) [gauge,right of=g8,label={[rotate=-45]below 
right:{\bb{k{-}l}}}] {};
	\node (g10) [gauge,right of=g9,label={[rotate=-45]below 
right:{\cc{k{-}l}}}] 
{};
    \node (g11) [right of=g10] {$\scriptstyle{\ldots}$};
    \node (g12) [gauge,right of=g11,label={[rotate=-45]below 
right:{\bb{k{-}l}}}] {};
	\node (g13) [gauge,right of=g12,label={[rotate=-45]below 
right:{\cc{k{-}l}}}] 
{};
    \node (g14) [gauge,right of=g13,label={[rotate=-45]below 
right:{\dd{k{-}l}}}] {};
	\node (g15) [gauge,right of=g14,label={[rotate=-45]below 
right:{\cc{k{-}l{-}1}}}] 
{};
	\node (g16) [gauge,right of=g15,label={[rotate=-45]below 
right:{\dd{k{-}l{-}1}}}] {};
	\node (g17) [right of=g16] {$\scriptstyle{\ldots}$};
	\node (g18) [gauge,right of=g17,label={[rotate=-45]below right:{\cc{1}}}] 
{};
	\node (g19) [gauge,right of=g18,label={[rotate=-45]below right:{\dd{2}}}] 
{};
	\node (f8) [flavour,above of=g8,label=above:{\bb{0}}] {};
	\node (f13) [flavour,above of=g13,label=above:{\bb{0}}] {};
	\draw (g2)--(g3) (g3)--(g4) (g4)--(g5) (g5)--(g6) (g6)--(g7) 
(g7)--(g8) (g8)--(g9) (g9)--(g10) (g10)--(g11) (g11)--(g12) (g12)--(g13)
(g13)--(g14) (g14)--(g15) (g15)--(g16) (g16)--(g17) (g17)--(g18) (g18)--(g19) 
(g8)--(f8) (g13)--(f13);
\draw[decoration={brace,mirror,raise=10pt},decorate,thick]
  (3.5-0.2,-1) -- node[below=10pt] {\tiny{$(2l-1)$} $\balg_{k{-l}}$ \tiny{\& 
$2l$} 
$\calg_{k{-}l}$ } (6+0.2,-1);
	\end{tikzpicture}
	} 
	\notag\\
	&-\qquad \qquad \qquad
	\raisebox{-.5\height}{
 	\begin{tikzpicture}
 	\tikzset{node distance = 0.5cm};
	\tikzstyle{gauge} = [circle, draw,inner sep=2.5pt];
	\tikzstyle{flavour} = [regular polygon,regular polygon sides=4,inner 
sep=2.5pt, draw];
    \node (g7) [gauge,label={[rotate=-45]below 
right:{\dd{1}}}] {};
	\node (g8) [gauge,right of=g7,label={[rotate=-45]below right:{\cc{1}}}] 
{};
    \node (g9) [gauge,right of=g8,label={[rotate=-45]below 
right:{\bb{1}}}] {};
	\node (g10) [gauge,right of=g9,label={[rotate=-45]below 
right:{\cc{1}}}] 
{};
    \node (g11) [right of=g10] {$\scriptstyle{\ldots}$};
    \node (g12) [gauge,right of=g11,label={[rotate=-45]below 
right:{\bb{1}}}] {};
	\node (g13) [gauge,right of=g12,label={[rotate=-45]below 
right:{\cc{1}}}] 
{};
    \node (g14) [gauge,right of=g13,label={[rotate=-45]below 
right:{\dd{1}}}] {};
	\node (f8) [flavour,above of=g8,label=above:{\bb{0}}] {};
	\node (f13) [flavour,above of=g13,label=above:{\bb{0}}] {};
	\draw (g7)--(g8) (g8)--(g9) (g9)--(g10) (g10)--(g11) (g11)--(g12) 
(g12)--(g13) (g13)--(g14)  (g8)--(f8) (g13)--(f13);
\draw[decoration={brace,mirror,raise=10pt},decorate,thick]
  (1-0.6,-0.4) -- node[below=10pt] {\tiny{$(2l{-}1)$} $\balg_{1}$ \tiny{\& 
$2l$} 
$\calg_{1}$ } (3.5-0.2,-0.4);
	\end{tikzpicture}
	} 
	\label{eq:quiver_subtraction_d_2l}
	\\
&= \raisebox{-.5\height}{
 	\begin{tikzpicture}
 	\tikzset{node distance = 0.5cm};
	\tikzstyle{gauge} = [circle, draw,inner sep=2.5pt];
	\tikzstyle{flavour} = [regular polygon,regular polygon sides=4,inner 
sep=2.5pt, draw];
\node (g2) [gauge,label={[rotate=-45]below right:{\dd{1}}}] {};
\node (g3) [gauge,right of=g2,label={[rotate=-45]below right:{\cc{1}}}] {};
\node (g4) [right of=g3] {$\scriptstyle{\ldots}$};
\node (g5) [gauge,right of=g4,label={[rotate=-45]below 
right:{\dd{k{-}l{-}2}}}]{};
\node (g6) [gauge,right of=g5,label={[rotate=-45]below 
right:{\cc{k{-}l{-}2}}}]{};
\node (g7) [gauge,right of=g6,label={[rotate=-45]below right:{\dd{k{-}l{-}1}}}] 
{};
	\node (g8) [gauge,right of=g7,label={[rotate=-45]below 
right:{\cc{k{-}l{-}1}}}] 
{};
    \node (g9) [gauge,right of=g8,label={[rotate=-45]below 
right:{\bb{k{-}l{-}1}}}] {};
	\node (g10) [gauge,right of=g9,label={[rotate=-45]below 
right:{\cc{k{-}l{-}1}}}] 
{};
    \node (g11) [right of=g10] {$\scriptstyle{\ldots}$};
    \node (g12) [gauge,right of=g11,label={[rotate=-45]below 
right:{\bb{k{-}l{-}1}}}] {};
	\node (g13) [gauge,right of=g12,label={[rotate=-45]below 
right:{\cc{k{-}l{-}1}}}] 
{};
    \node (g14) [gauge,right of=g13,label={[rotate=-45]below 
right:{\dd{k{-}l{-}1}}}] {};
	\node (g15) [gauge,right of=g14,label={[rotate=-45]below 
right:{\cc{k{-}l{-}2}}}] 
{};
	\node (g16) [gauge,right of=g15,label={[rotate=-45]below 
right:{\dd{k{-}l{-}2}}}] {};
	\node (g17) [right of=g16] {$\scriptstyle{\ldots}$};
	\node (g18) [gauge,right of=g17,label={[rotate=-45]below right:{\cc{1}}}] 
{};
	\node (g19) [gauge,right of=g18,label={[rotate=-45]below right:{\dd{2}}}] 
{};
	\node (f8) [flavour,above of=g8,label=above:{\bb{0}}] {};
	\node (f13) [flavour,above of=g13,label=above:{\bb{0}}] {};
	\draw (g2)--(g3) (g3)--(g4) (g4)--(g5) (g5)--(g6) (g6)--(g7) 
(g7)--(g8) (g8)--(g9) (g9)--(g10) (g10)--(g11) (g11)--(g12) (g12)--(g13)
(g13)--(g14) (g14)--(g15) (g15)--(g16) (g16)--(g17) (g17)--(g18) (g18)--(g19) 
(g8)--(f8) (g13)--(f13);
\draw[decoration={brace,mirror,raise=10pt},decorate,thick]
  (3.5-0.2,-1) -- node[below=10pt] {\tiny{$(2l+1)$} $\balg_{k{-l-1}}$ \tiny{\& 
$(2l+2)$} 
$\calg_{k{-}l{-}1}$ } (6+0.2,-1);
	\end{tikzpicture}
	}
	\notag \,.
\end{align}
In fact, the relevant $d_{2l}$ magnetic quiver can also be seen as a 
consequence of the brane considerations in Section \ref{sec:Hasse_from_brane}. 
There, one observes that the $d_{2l}$ transition is due to the electric theory 
$\sprm(1)$ with $\sorm(2l)$ flavour, and its magnetic quiver (or even its 
$3$d $\Ncal=4$ mirror) is exactly the quiver for the $d_{2l}$ quiver 
subtraction.

  \section{Conclusions and Outlook}
\label{sec:conclusion}
In this paper, the formalism of \emph{magnetic quivers} for $6$d $\Ncal=(1,0)$  
Higgs branches has been extended to orthogonal and symplectic gauge nodes. Most 
notably, the entire derivation is based on Type IIA brane configurations and 
can be summarised as in Conjecture \ref{conj:magnetic_quiver}. The main 
conceptual point lies in the generalisation of the S-duality rules of \Ot\ 
planes to the proposed \emph{magnetic orientifolds}, see Table 
\ref{tab:orientifold}. In contrast to the physical nature of S-duality for \Ot\ 
planes, the magnetic orientifolds are purely of conceptual nature. In other 
words, they are considered as tool that allows to derive the magnetic quivers 
for \Ds-\De-\NS\ brane configurations in the presences of \Os\ planes.

In this paper, \emph{all} Higgs branches of the $6$d 
$\Ncal=(1,0)$ theories coming from a single \Mf\ or multiple \Mf s on 
$\R\times\C^2\slash\D_{k-2}$ have been described with magnetic quivers. The 
concept of a \emph{magnetic quiver} has been reviewed in Section 
\ref{sec:el-mag_quiver}.

In case of a single \Mf, the magnetic quivers for the Higgs branch at finite 
\cite{Feng:2000eq} and infinite coupling \cite{Hanany:2018uhm} have been known 
before. The novel point discussed in Section 
\ref{sec:single_M5} is that these magnetic quiver can be derived from a brane 
configuration and, moreover, this brane construction correctly shows that the 
Higgs branch phase transition is a small $E_8$ instanton transition.  

In the case of $n$ \Mf\ branes, the magnetic quiver for the Higgs branch 
at the origin of the tensor branch had only been conjectured in 
\cite{Hanany:2018uhm}. As discussed in Section \ref{sec:multiple_M5}, the 
formalism allows to derive the magnetic quivers for the Higgs branches over 
every point in the tensor branch. In particular, the nature of the transitions 
to different singular loci of the tensor branch has been revealed. Generically, 
there are three type of transitions in order to reach the infinite coupling 
phase. (i) There are $(n-1)$ one-dimensional $D_4$ transitions in which one 
simultaneously trades one tensor multiplet and one $\sorm(8)$ vector multiplet 
for a single hypermultiplet. (ii) There is exactly one small $E_8$ instanton 
transition, trading 
one tensor multiplet for $29$ hypermultiplets. (iii) There are $\mathcal{P}(n)$ 
zero-dimensional discrete gauging transitions. Taking all of these into account 
leads to a description of the Higgs branch at the origin of the tensor branch. 

Returning to the single \Mf\ case, the geometry of the Higgs branches as a 
symplectic singularity has been studied in Section \ref{sec:Hasse}. Assuming 
minimal transitions only, the previously computed Hasse diagram 
\cite{Bourget:2019aer} has been rederived using (i) brane configurations with 
\Os\ orientifold planes as well as (ii) quiver subtraction for magnetic quivers 
with orthosymplectic gauge nodes. This results provide a crucial consistency 
check for the proposal of this paper.

\paragraph{Outlook.}
An interesting subject is the understanding the Higgs branches of $6$d 
$\Ncal=(1,0)$ theories from multiple \Mf\ branes near an \Mn\ plane on a 
$D$-type ALE space. For the $A$-type case, this has been answered in 
\cite{Cabrera:2019izd}. In order to derive magnetic quivers for these 
systems, there are two necessary ingredients: (i) the rules 
established in Conjecture \ref{conj:magnetic_quiver}, and (ii) the embedding of 
$\D_{k-2} \hookrightarrow E_8$. In contrast to the $A$-type case, the latter is 
not straightforward and progress \cite{Frey:2018vpw} has only been achieved 
recently.

From the experience gained with magnetic quivers, the changes of Higgs branches 
over the tensor branch can be compared to known F-theory descriptions. In 
particular, a singularity on the tensor branch corresponds to the collapse of 
some $-n$ curve. Recently, the following transitions in $6$d have been 
understood:
\begin{itemize}
 \item collapse of a single $-1$ curve $\leftrightarrow$ small $E_8$ instanton 
transition
\begin{compactitem}
 \item $\surm(N)$ gauge group with $N_f=N+8$ fundamental flavours and one 2nd 
rank antisymmetric hypermultiplet \cite{Mekareeya:2017jgc,Cabrera:2019izd}
\item $\sprm(N)$ gauge group with $N_f =N+8$ flavours,  
\cite{Mekareeya:2017jgc,Hanany:2018uhm,Cabrera:2019izd} and Section 
\ref{sec:1M5_generic_k}
\end{compactitem}
 \item collapse of a single $-2$ curve $\leftrightarrow$ discrete gauging 
transition
    \begin{compactitem}
     \item $\surm(N)$ gauge group with $N_f=2N$ flavours 
\cite{Hanany:2018vph,Cabrera:2019izd}
    \end{compactitem}
\end{itemize}
While this paper provides evidence for a new entry in the list, namely 
following the $1$d transition \eqref{eq:1d_transition}:
\begin{itemize}
 \item collapse of $-4$ curve $\leftrightarrow$ partial Higgsing 
$\sorm(2k)\to\sorm(8)$ transition, i.e.\ the $D_4$ transition.
\end{itemize}
The simplest set-up, to test this further, corresponds to one full \NS\ brane 
fractionating on 
a stack of $k$ full \Ds\ branes on top of an \Osp\ orientifold in Type IIA, 
such that the $6$d $\Ncal=(1,0)$ becomes
\begin{align}
 \raisebox{-.5\height}{
 	\begin{tikzpicture}
	\tikzstyle{gauge} = [circle, draw];
	\tikzstyle{flavour} = [regular polygon,regular polygon sides=4, draw];
	\node (g1) [gauge,label=below:{\SO{2k+8}}] {};
	\node (f1) [flavour,above of=g1,label=above:{\Sp{2k}}] {};
	\draw (g1)--(f1);
	\end{tikzpicture}
	} \,.
	\label{eq:Sp-theory}
\end{align}
Conjecture \ref{conj:magnetic_quiver} provides \emph{candidate magnetic 
quivers} for the Higgs branch at finite and infinite coupling, i.e.\
\begin{subequations}
\label{eq:magQuiver_Sp}
\begin{align}
 \raisebox{-.5\height}{
 	\begin{tikzpicture}
 	\tikzset{node distance = 0.5cm};
	\tikzstyle{gauge} = [circle, draw,inner sep=2.5pt];
	\tikzstyle{flavour} = [regular polygon,regular polygon sides=4,inner 
sep=2.5pt, draw];
	\node (g1) [gauge,label={[rotate=-45]below right:{\bb{0}}}] {};
	\node (g2) [gauge,right of=g1,label={[rotate=-45]below right:{\cc{1}}}] 
{};
	\node (g3) [gauge,right of=g2,label={[rotate=-45]below right:{\bb{1}}}] {};
\	\node (g4) [gauge,right of=g3,label={[rotate=-45]below 
right:{\cc{2}}}] {};
	\node (g5) [gauge,right of=g4,label={[rotate=-45]below right:{\bb{2}}}] 
{};
	\node (g6) [right of=g5] {$\scriptstyle{\ldots}$};
    \node (g7) [gauge,right of=g6,label={[rotate=-45]below 
right:{\bb{k{-}1}}}] {};
	\node (g8) [gauge,right of=g7,label={[rotate=-45]below right:{\cc{k}}}] 
{};
    \node (g9) [gauge,right of=g8,label={[rotate=-45]below 
right:{\bb{k}}}] {};
	\node (g10) [gauge,right of=g9,label={[rotate=-45]below 
right:{\cc{k}}}] 
{};
    \node (g11) [gauge,right of=g10,label={[rotate=-45]below 
right:{\bb{k{-}1}}}] {};
    \node (g12) [right of=g11] {$\scriptstyle{\ldots}$};
	\node (g13) [gauge,right of=g12,label={[rotate=-45]below right:{\bb{2}}}] 
{};
    \node (g14) [gauge,right of=g13,label={[rotate=-45]below right:{\cc{2}}}] 
{};
	\node (g15) [gauge,right of=g14,label={[rotate=-45]below right:{\bb{1}}}] 
{};
	\node (g16) [gauge,right of=g15,label={[rotate=-45]below right:{\cc{1}}}] 
{};
	\node (g17) [gauge,right of=g16,label={[rotate=-45]below right:{\bb{0}}}] 
{};
	\node (b1) [flavour,above of=g9,label=above:{\cc{1}}] {};
	\draw  (g1)--(g2) (g2)--(g3) (g3)--(g4) (g4)--(g5) (g5)--(g6) (g6)--(g7) 
(g7)--(g8) (g8)--(g9) (g9)--(g10) (g10)--(g11) (g11)--(g12) (g12)--(g13)
(g13)--(g14) (g14)--(g15) (g15)--(g16) (g16)--(g17) (g9)--(b1);
	\end{tikzpicture}
	} 
	\,,
		\label{eq:magQuiver_Sp_finite} \\
 \raisebox{-.5\height}{
 	\begin{tikzpicture}
 	\tikzset{node distance = 0.5cm};
	\tikzstyle{gauge} = [circle, draw,inner sep=2.5pt];
	\tikzstyle{flavour} = [regular polygon,regular polygon sides=4,inner 
sep=2.5pt, draw];
	\node (g1) [gauge,label={[rotate=-45]below right:{\bb{0}}}] {};
	\node (g2) [gauge,right of=g1,label={[rotate=-45]below right:{\cc{1}}}] 
{};
	\node (g3) [gauge,right of=g2,label={[rotate=-45]below right:{\bb{1}}}] {};
\	\node (g4) [gauge,right of=g3,label={[rotate=-45]below 
right:{\cc{2}}}] {};
	\node (g5) [gauge,right of=g4,label={[rotate=-45]below right:{\bb{2}}}] 
{};
	\node (g6) [right of=g5] {$\scriptstyle{\ldots}$};
    \node (g7) [gauge,right of=g6,label={[rotate=-45]below 
right:{\bb{k{-}1}}}] {};
	\node (g8) [gauge,right of=g7,label={[rotate=-45]below right:{\cc{k}}}] 
{};
    \node (g9) [gauge,right of=g8,label={[rotate=-45]below 
right:{\bb{k}}}] {};
	\node (g10) [gauge,right of=g9,label={[rotate=-45]below 
right:{\cc{k}}}] 
{};
    \node (g11) [gauge,right of=g10,label={[rotate=-45]below 
right:{\bb{k{-}1}}}] {};
    \node (g12) [right of=g11] {$\scriptstyle{\ldots}$};
	\node (g13) [gauge,right of=g12,label={[rotate=-45]below right:{\bb{2}}}] 
{};
    \node (g14) [gauge,right of=g13,label={[rotate=-45]below right:{\cc{2}}}] 
{};
	\node (g15) [gauge,right of=g14,label={[rotate=-45]below right:{\bb{1}}}] 
{};
	\node (g16) [gauge,right of=g15,label={[rotate=-45]below right:{\cc{1}}}] 
{};
	\node (g17) [gauge,right of=g16,label={[rotate=-45]below right:{\bb{0}}}] 
{};
	\node (b1) [gauge,above of=g9,label=above:{\cc{1}}] {};
	\draw  (g1)--(g2) (g2)--(g3) (g3)--(g4) (g4)--(g5) (g5)--(g6) (g6)--(g7) 
(g7)--(g8) (g8)--(g9) (g9)--(g10) (g10)--(g11) (g11)--(g12) (g12)--(g13)
(g13)--(g14) (g14)--(g15) (g15)--(g16) (g16)--(g17) (g9)--(b1);
	\end{tikzpicture}
	} 
	\,,
		\label{eq:magQuiver_Sp_infinite}
\end{align}
\end{subequations}
such that 
\begin{align}
 \Higgs(\eqref{eq:Sp-theory})_\fin = \Coulomb(\eqref{eq:magQuiver_Sp_finite})
 \qquad \text{and} \qquad
 \Higgs(\eqref{eq:Sp-theory})_\infty = 
\Coulomb(\eqref{eq:magQuiver_Sp_infinite}) \,.
\end{align}
However, the nature of the transition needs to be analysed more carefully; for 
instance, what is the  geometry of the transverse slice? In addition, it is 
imperative to study the Hasse diagram of \eqref{eq:Sp-theory} using 
\eqref{eq:magQuiver_Sp} and compare to \cite[Fig.\ 3]{DelZotto:2018tcj} for a 
single $-4$ curve. This is left for future research.
%
%
\paragraph{Further predictions.}
With Conjecture \ref{conj:magnetic_quiver} at hand, one can derive predictions 
for the Higgs branches of a single \NS\ brane on either a \Osmt\ or \Ospt\  
plane with $k$ \Ds\ branes. In contrast to configurations on a \Osm\ plane or a 
\Osp\ plane, discussed above, there exists no gauge theory description and the 
\NS\ brane cannot split along the orientifold. In addition, only the 
configuration \eqref{eq:1M5_electric_brane_system} with \Osm\ admits an M-theory 
dual, while all other three \Os\ planes only exist as Type IIA systems. 
Following the prescription outlined in this paper, one finds:
\begin{align}
 &\text{1 \NS\ on \Osmt\ with $k$ \Ds :}
 \qquad \qquad
 \raisebox{-.5\height}{
 	\begin{tikzpicture}
 	\tikzset{node distance = 0.5cm};
	\tikzstyle{gauge} = [circle, draw,inner sep=2.5pt];
	\tikzstyle{flavour} = [regular polygon,regular polygon sides=4,inner 
sep=2.5pt, draw];
	\node (g2) [gauge,label={[rotate=-45]below right:{\dd{1}}}] 
{};
	\node (g3) [gauge,right of=g2,label={[rotate=-45]below right:{\cc{1}}}] {};
\	\node (g4) [gauge,right of=g3,label={[rotate=-45]below 
right:{\dd{2}}}] {};
	\node (g5) [gauge,right of=g4,label={[rotate=-45]below right:{\cc{2}}}] 
{};
	\node (g6) [right of=g5] {$\scriptstyle{\ldots}$};
    \node (g7) [gauge,right of=g6,label={[rotate=-45]below 
right:{\cc{k{-}1}}}] {};
	\node (g8) [gauge,right of=g7,label={[rotate=-45]below right:{\dd{k}}}] 
{};
    \node (g9) [gauge,right of=g8,label={[rotate=-45]below 
right:{\cc{k}}}] {};
	\node (g10) [gauge,right of=g9,label={[rotate=-45]below 
right:{\dd{k}}}] 
{};
    \node (g11) [gauge,right of=g10,label={[rotate=-45]below 
right:{\cc{k{-}1}}}] {};
    \node (g12) [right of=g11] {$\scriptstyle{\ldots}$};
	\node (g13) [gauge,right of=g12,label={[rotate=-45]below right:{\cc{2}}}] 
{};
    \node (g14) [gauge,right of=g13,label={[rotate=-45]below right:{\dd{2}}}] 
{};
	\node (g15) [gauge,right of=g14,label={[rotate=-45]below right:{\cc{1}}}] 
{};
	\node (g16) [gauge,right of=g15,label={[rotate=-45]below right:{\dd{1}}}] 
{};
	\node (b1) [gauge,above of=g9,label=above:{\dd{1}}] {};
	\draw  (g2)--(g3) (g3)--(g4) (g4)--(g5) (g5)--(g6) (g6)--(g7) 
(g7)--(g8) (g8)--(g9) (g9)--(g10) (g10)--(g11) (g11)--(g12) (g12)--(g13)
(g13)--(g14) (g14)--(g15) (g15)--(g16)  (g9)--(b1);
	\end{tikzpicture}
	} 
	\,,
	\label{eq:BB-matter}
 \\
&\text{1 \NS\ on \Ospt\ with $k$ \Ds :}
\qquad \qquad 
\raisebox{-.5\height}{
 	\begin{tikzpicture}
 	\tikzset{node distance = 0.5cm};
	\tikzstyle{gauge} = [circle, draw,inner sep=2.5pt];
	\tikzstyle{flavour} = [regular polygon,regular polygon sides=4,inner 
sep=2.5pt, draw];
	\node (g2) [gauge,label={[rotate=-45]below right:{\bb{1}}}] 
{};
	\node (g3) [gauge,right of=g2,label={[rotate=-45]below right:{\cc{1}}}] {};
\	\node (g4) [gauge,right of=g3,label={[rotate=-45]below 
right:{\bb{2}}}] {};
	\node (g5) [gauge,right of=g4,label={[rotate=-45]below right:{\cc{2}}}] 
{};
	\node (g6) [right of=g5] {$\scriptstyle{\ldots}$};
    \node (g7) [gauge,right of=g6,label={[rotate=-45]below 
right:{\cc{k{-}1}}}] {};
	\node (g8) [gauge,right of=g7,label={[rotate=-45]below right:{\bb{k}}}] 
{};
    \node (g9) [gauge,right of=g8,label={[rotate=-45]below 
right:{\cc{k}}}] {};
	\node (g10) [gauge,right of=g9,label={[rotate=-45]below 
right:{\bb{k}}}] 
{};
    \node (g11) [gauge,right of=g10,label={[rotate=-45]below 
right:{\cc{k{-}1}}}] {};
    \node (g12) [right of=g11] {$\scriptstyle{\ldots}$};
	\node (g13) [gauge,right of=g12,label={[rotate=-45]below right:{\cc{2}}}] 
{};
    \node (g14) [gauge,right of=g13,label={[rotate=-45]below right:{\bb{2}}}] 
{};
	\node (g15) [gauge,right of=g14,label={[rotate=-45]below right:{\cc{1}}}] 
{};
	\node (g16) [gauge,right of=g15,label={[rotate=-45]below right:{\bb{1}}}] 
{};
	\node (b1) [gauge,above of=g9,label=above:{\bb{1}}] {};
	\draw  (g2)--(g3) (g3)--(g4) (g4)--(g5) (g5)--(g6) (g6)--(g7) 
(g7)--(g8) (g8)--(g9) (g9)--(g10) (g10)--(g11) (g11)--(g12) (g12)--(g13)
(g13)--(g14) (g14)--(g15) (g15)--(g16)  (g9)--(b1);
	\end{tikzpicture}
	} 
	\,.
	\label{eq:CC-matter}
\end{align}
By the rules of Appendix \ref{app:global_sym}, one would conclude that the 
magnetic quiver \eqref{eq:BB-matter} is \emph{good}, with all nodes except the 
central $\dalg_{1}$ being \emph{balanced}. Similarly,  all nodes in 
\eqref{eq:CC-matter} are \emph{good}. 
In view of these predictions and the results of 
\cite{Cabrera:2019izd}, one can summarise the magnetic quiver 
for a single \NS\ brane on $k$ \Ds\ branes with or without an \Os\ orientifold 
as in Table \ref{tab:GG-matter}.
\begin{table}[t]
 \centering
 \begin{tabular}{c|c}
 \toprule
  Type IIA system & magnetic quiver \\ \midrule
1 \NS\  with $k$ \Ds & $\left(T_{(1^k)}[\surm(k)] \times T_{(k-1,1)}[\surm(k)] 
\times T_{(1^k)}[\surm(k)]\right)  \slash \slash \slash \surm(k) $ \\
1 \NS\ on \Osm\ with $k$ \Ds &  $\left(T_{(1^{2k})}[\sorm(2k)] \times 
T_{(2k-3,3)}[\sorm(2k)] 
\times T_{(1^{2k})}[\sorm(2k)]\right)  \slash \slash \slash \sorm(2k) $\\
1 \NS\ on \Osp\ with $k$ \Ds & $\left(T_{(1^{2k})}[\sorm(2k{+}1)] \times 
T_{(2k,2)}[\sorm(2k{+}1)] 
\times T_{(1^{2k})}[\sorm(2k{+}1)]\right)  \slash \slash \slash \sorm(2k{+}1) 
$\\
1 \NS\ on \Osmt\ with $k$ \Ds & $\left(T_{(1^{2k+1})}[\usprm(2k)] \times 
T_{(2k-1,1^2)}[\usprm(2k)] 
\times T_{(1^{2k+1})}[\usprm(2k)]\right)  \slash \slash \slash 
\usprm(2k) 
$ \\
1 \NS\ on \Ospt\ with $k$ \Ds & $\left(T_{(1^{2k})}[\usprm'(2k)] \times 
T_{(2k-2,2)}[\usprm'(2k)] 
\times T_{(1^{2k})}[\usprm'(2k)]\right)  \slash \slash \slash 
\usprm'(2k) 
$\\
\bottomrule
 \end{tabular}
\caption{The Higgs branch at the origin of the tensor branch can be described 
by a magnetic quiver obtained from three $T_\rho[G]$ theories 
\cite{Cremonesi:2014uva} glued
along the common $G$ flavour node, which is denoted by 
$\slash \slash \slash G$.}
\label{tab:GG-matter}
\end{table}
\paragraph{Acknowledgements.}
We are indebted to 
Antoine Bourget,
Julius Grimminger,
Rudolph Kalveks,
Noppadol Mekareeya,
Tom Rudelius, and
Zhenghao Zhong
for useful discussions.
We thank the Simons Center for Geometry and Physics, Stony Brook University for 
the hospitality and the partial support during the initial
stage of this work at the Simons Summer workshop 2018. 
A.H. and M.S. gratefully acknowledge support from the Simons Center for 
Geometry and Physics, Stony
Brook University during the Simons Summer workshop 2019, where part of the 
research for this paper was performed.
S.C. was supported by an EPSRC DTP studentship EP/M507878/1. 
A.H. is supported by STFC grant ST/P000762/1. 
M.S. had been supported by Austrian Science Fund (FWF) grant P28590. 
M.S. thanks the Faculty of Physics of the University of Vienna for travel 
support via the “Jungwissenschaftsförderung”. 
The work of M.S. was supported by the National Thousand-Young-Talents Program 
of China and the China Postdoctoral Science Foundation (grant no. 2019M650616).
M.S. thanks the Theoretical Physics Group of Imperial College London for 
hospitality.

\appendix
    \section{Background material}
\label{app:background}
\subsection{Brane creation and annihilation}
\label{app:orientifolds}
Following \cite{Hanany:1996ie}, in a system of D$p$-D$(p{+}2)$-\NS\ branes, 
D$p$ brane 
creation or annihilation happens whenever a \NS\ passes through an D$(p{+}2)$. 
In 
the presence of O$p$ planes, which carry non-trivial brane charge, a \NS\ brane 
can pass through an D$(p{+}2)$ with or without creation of an additional D$p$ 
brane. 
To begin with, recall \cite{Evans:1997hk,Hanany:1999sj,Hanany:2000fq}
\begin{compactitem}
 \item An O$p^\pm$ becomes an O$p^\mp$ when passing through a half \NS ; 
likewise, $\widetilde{\text{O}p}^\pm$ turns into $\widetilde{\text{O}p}^\mp$.
\item An O$p^\pm$ becomes an $\widetilde{\text{O}p}^\pm$ when passing through 
a half D$(p{+}2)$, and vice versa.
\end{compactitem}
According to \cite{Hanany:1999sj,Feng:2000eq}, the charges of the O$p$ planes 
(in unites of the physical D$p$ branes) are given by 
\begin{align}
 \text{charge}(\text{O}p^{\pm}) = \pm 2^{p-5} 
 \; , \quad
 \text{charge}(\widetilde{\text{O}p}^{-}) = \frac{1}{2}- 2^{p-5}
 \; , \quad
 \text{charge}(\widetilde{\text{O}p}^{+}) =  2^{p-5} \; .
 \label{eq:charges_orientifold}
\end{align}
Following the conventions of \cite{Gaiotto:2008ak}, the different 
orientifolds are denoted by:
\begin{alignat}{2}
\Osm \; \&\; 2n\cdot \frac{1}{2}\Ds &: \quad 
  \raisebox{-.5\height}{
 	\begin{tikzpicture}
\DsixOMinus{n}{0} 	 
 	 	\end{tikzpicture}
	} 
\quad ,& \qquad \qquad 
	\Osmt\; \&\; 2n\cdot \frac{1}{2}\Ds &: \quad 
  \raisebox{-.5\height}{
 	\begin{tikzpicture}
\DsixOMinusTilde{n}{0} 	 
 	 	\end{tikzpicture}
	} 
	\quad , \\
\Osp\; \&\; 2n\cdot \frac{1}{2}\Ds &: \quad 
  \raisebox{-.5\height}{
 	\begin{tikzpicture}
\DsixOPlus{n}{0} 	 
 	 	\end{tikzpicture}
	} 
\quad ,& \qquad \qquad 
	\Ospt\; \&\; 2n\cdot \frac{1}{2}\Ds &: \quad 
  \raisebox{-.5\height}{
 	\begin{tikzpicture}
\DsixOPlusTilde{n}{0} 	 
 	 	\end{tikzpicture}
	} 
	\quad ,
\end{alignat}
i.e.\ 
O$6^-$ empty line, $\widetilde{\text{O}6}^-$ solid line, O$6^+$ dotted line, 
$\widetilde{\text{O}6}^+$ dashed line.

Next, there a four scenarios for brane creation and annihilation. These follow 
from preservation of the linking number before and after the transition.
The linking numbers $l_{\NS\ }$ for half \NS\ or $l_{\text{D}(p+2) }$ for half 
D$(p{+}2)$ are 
defined as \cite{Hanany:1996ie}
\begin{subequations}
\label{eq:linking_numbers}
\begin{align}
l_{\NS\ } &= \frac{1}{2} \left( R_{\text{D}(p{+}2)} - L_{\text{D}(p{+}2)} 
\right) + \left( L_{\text{D}p} - R_{\text{D}p} \right) \,,\\
l_{\text{D}(p+2) } &= \frac{1}{2} \left( R_{\NS\ } - L_{\NS\ } \right) + \left( 
L_{\text{D}p} - R_{\text{D}p} \right) \,,
\end{align}
\end{subequations}
where $L_X$, $R_X$ denote the total number of branes of type $X$ to the left 
or right, respectively.
Note that the O$p$ planes contribute to $L_{Dp}$ and $R_{Dp}$ according to 
\eqref{eq:charges_orientifold}; naturally, 
half \NS\ or half D$(p{+}2)$ branes contribute with charge 
$\frac{1}{2}$ to the numbers $L$ and $R$, respectively.
It then follows that
\begin{subequations}
\begin{align}
\raisebox{-.5\height}{
\begin{tikzpicture}
        \draw[white] (2,0)--(3,0);
        \OPlusTilde{0,0}{1,0}
        \Dbrane{1,0}{2,0}
		\ns{1,0}
        \DeightMany{1}{2}{1}
        \draw (0.5,0.5) node {$\widetilde{+}$};
        \draw (1.5,0.5) node {$\widetilde{-}$};
        \draw (2.5,0.5) node {$-$};
	\end{tikzpicture}
	}
	\qquad
	&\leftrightarrow
	\qquad
	\raisebox{-.5\height}{
	\begin{tikzpicture}
        \OPlusTilde{0,0}{1,0}
        \OPlus{1,0}{2,0}
		\ns{2,0}
        \DeightMany{1}{1}{1}
        \draw (0.5,0.5) node {$\widetilde{+}$};
        \draw (1.5,0.5) node {$+$};
        \draw (2.5,0.5) node {$-$};
	\end{tikzpicture}
	} 
	\label{eq:brane_creation_1}
	\\
\raisebox{-.5\height}{
\begin{tikzpicture}
        \OPlus{0,0}{1,0}
        \Dbrane{2,0}{3,0}
		\ns{1,0}
        \DeightMany{1}{2}{1}
        \draw (0.5,0.5) node {$+$};
        \draw (1.5,0.5) node {$-$};
        \draw (2.5,0.5) node {$\widetilde{-}$};
	\end{tikzpicture}
	}
	\qquad
	&\leftrightarrow
	\qquad
	\raisebox{-.5\height}{
	\begin{tikzpicture}
        \OPlus{0,0}{1,0}
        \OPlusTilde{1,0}{2,0}
        \Dbrane{2,0}{3,0}
        \DsixEvenFree{}{1}
		\ns{2,0}
        \DeightMany{1}{1}{1}
        \draw (0.5,0.5) node {$+$};
        \draw (1.5,0.5) node {$\widetilde{+}$};
        \draw (2.5,0.5) node {$\widetilde{-}$};
	\end{tikzpicture}
	} 
		\label{eq:brane_creation_2}
	\\
\raisebox{-.5\height}{
\begin{tikzpicture}
        \draw[white] (0,0)--(1,0);
        \OPlus{1,0}{2,0}
        \OPlusTilde{2,0}{3,0}
		\ns{1,0}
        \DeightMany{1}{2}{1}
        \draw (0.5,0.5) node {$-$};
        \draw (1.5,0.5) node {$+$};
        \draw (2.5,0.5) node {$\widetilde{+}$};
	\end{tikzpicture}
	}
	\qquad
	&\leftrightarrow
	\qquad
	\raisebox{-.5\height}{
	\begin{tikzpicture}
        \draw[white] (0,0)--(1,0);
        \Dbrane{1,0}{2,0}
        \OPlusTilde{2,0}{3,0}
		\ns{2,0}
        \DeightMany{1}{1}{1}
        \draw (0.5,0.5) node {$-$};
        \draw (1.5,0.5) node {$\widetilde{-}$};
        \draw (2.5,0.5) node {$\widetilde{+}$};
	\end{tikzpicture}
	} 
		\label{eq:brane_creation_3}
	\\
\raisebox{-.5\height}{
\begin{tikzpicture}
        \Dbrane{0,0}{1,0}
        \OPlusTilde{1,0}{2,0}
        \OPlus{2,0}{3,0}
		\ns{1,0}
        \DeightMany{1}{2}{1}
        \draw (0.5,0.5) node {$\widetilde{-}$};
        \draw (1.5,0.5) node {$\widetilde{+}$};
        \draw (2.5,0.5) node {$+$};
	\end{tikzpicture}
	}
	\qquad
	&\leftrightarrow
	\qquad
	\raisebox{-.5\height}{
	\begin{tikzpicture}
        \Dbrane{0,0}{1,0}
        \OPlus{2,0}{3,0}
        \DsixEvenFree{}{1}
		\ns{2,0}
        \DeightMany{1}{1}{1}
        \draw (0.5,0.5) node {$\widetilde{-}$};
        \draw (1.5,0.5) node {$-$};
        \draw (2.5,0.5) node {$+$};
	\end{tikzpicture}
	} 
		\label{eq:brane_creation_4}
\end{align}
\label{eq:brane_creation_all}
\end{subequations}
by requiring that all linking numbers \eqref{eq:linking_numbers} remain 
constant.
%
%
\subsection{Global symmetry for orthosymplectic quiver}
\label{app:global_sym}
Following \cite[Sec.\ 5.1-5.2]{Gaiotto:2008ak}, there are conditions upon which 
orthogonal and symplectic gauge nodes in a $3$d $\Ncal=4$ gauge theory are 
called \emph{good}, \emph{bad}, or \emph{ugly}. A subset of good gauge nodes 
are \emph{balanced} gauge nodes, for which monopole operators of spin 1 under 
the R-charge are expected to lead to symmetry enhancement. 

An $\sorm(k)$ (or $\orm(k)$) gauge theory coupled to fundamental 
hypermultiplets with $\usprm(2n)$ flavour symmetry is called
 \begin{equation}
 \text{\emph{good} if } \quad   n \geq k-1\, , \qquad \text{and \emph{balanced} 
if}\quad   n =k-1 \,.
\label{eq:balanced_SO}
\end{equation}

Analogously, an $\usprm(2l)=\sprm(l)$ gauge theory coupled to fundamental 
hypermultiplets with $\orm(2n)$ flavour symmetry is called
\begin{equation}
 \text{\emph{good} if } \quad   n \geq 2l+1 \,, \qquad 
 \text{ and \emph{balanced} if } \quad   n =2l+1 \,.
\end{equation}
Considering an orthosymplectic quiver, i.e.\ a linear quiver with alternating 
orthogonal and symplectic gauge nodes, a chain of $p$ balanced nodes gives rise 
to the following enhanced Coulomb branch symmetry:
\begin{compactitem}
 \item An $\sorm(p+1)$ symmetry, if there are no $\sorm(2)$ (or $\orm(2)$) 
gauge nodes at the ends.
\item An $\sorm(p+2)$ symmetry, if there is an $\sorm(2)$ (or $\orm(2)$) 
gauge node at one of the two ends.
\item An $\sorm(p+3)$ symmetry, if there is an $\sorm(2)$ (or $\orm(2)$) 
gauge node at each end.
\end{compactitem}

%
%
 \bibliographystyle{JHEP}     
 {\footnotesize{\bibliography{references}}}

\providecommand{\href}[2]{#2}\begingroup\raggedright\begin{thebibliography}{10}

\bibitem{Witten:1995zh}
E.~Witten, {\it {Some comments on string dynamics}},  in {\em {Future
  perspectives in string theory. Proceedings, Conference, Strings'95, Los
  Angeles, USA, March 13-18, 1995}}, pp.~501--523, 1995.
\newblock \href{http://arxiv.org/abs/hep-th/9507121}{{\tt hep-th/9507121}}.

\bibitem{Strominger:1995ac}
A.~Strominger, {\it {Open p-branes}},  {\em Phys. Lett.} {\bf B383} (1996)
  44--47, [\href{http://arxiv.org/abs/hep-th/9512059}{{\tt hep-th/9512059}}].
  [,116(1995)].

\bibitem{Hanany:1997gh}
A.~Hanany and A.~Zaffaroni, {\it {Branes and six-dimensional supersymmetric
  theories}},  {\em Nucl. Phys.} {\bf B529} (1998) 180--206,
  [\href{http://arxiv.org/abs/hep-th/9712145}{{\tt hep-th/9712145}}].

\bibitem{Brunner:1997gk}
I.~Brunner and A.~Karch, {\it {Branes and six-dimensional fixed points}},  {\em
  Phys. Lett.} {\bf B409} (1997) 109--116,
  [\href{http://arxiv.org/abs/hep-th/9705022}{{\tt hep-th/9705022}}].

\bibitem{Brunner:1997gf}
I.~Brunner and A.~Karch, {\it {Branes at orbifolds versus Hanany Witten in
  six-dimensions}},  {\em JHEP} {\bf 03} (1998) 003,
  [\href{http://arxiv.org/abs/hep-th/9712143}{{\tt hep-th/9712143}}].

\bibitem{Hanany:1999sj}
A.~Hanany and A.~Zaffaroni, {\it {Issues on orientifolds: On the brane
  construction of gauge theories with SO(2n) global symmetry}},  {\em JHEP}
  {\bf 07} (1999) 009, [\href{http://arxiv.org/abs/hep-th/9903242}{{\tt
  hep-th/9903242}}].

\bibitem{Heckman:2013pva}
J.~J. Heckman, D.~R. Morrison, and C.~Vafa, {\it {On the Classification of 6D
  SCFTs and Generalized ADE Orbifolds}},  {\em JHEP} {\bf 05} (2014) 028,
  [\href{http://arxiv.org/abs/1312.5746}{{\tt arXiv:1312.5746}}]. [Erratum:
  JHEP06,017(2015)].

\bibitem{Heckman:2015bfa}
J.~J. Heckman, D.~R. Morrison, T.~Rudelius, and C.~Vafa, {\it {Atomic
  Classification of 6D SCFTs}},  {\em Fortsch. Phys.} {\bf 63} (2015) 468--530,
  [\href{http://arxiv.org/abs/1502.05405}{{\tt arXiv:1502.05405}}].

\bibitem{Green:1984bx}
M.~B. Green, J.~H. Schwarz, and P.~C. West, {\it {Anomaly Free Chiral Theories
  in Six-Dimensions}},  {\em Nucl. Phys.} {\bf B254} (1985) 327--348.

\bibitem{RandjbarDaemi:1985wc}
S.~Randjbar-Daemi, A.~Salam, E.~Sezgin, and J.~A. Strathdee, {\it {An Anomaly
  Free Model in Six-Dimensions}},  {\em Phys. Lett.} {\bf 151B} (1985)
  351--356.

\bibitem{Dabholkar:1996zi}
A.~Dabholkar and J.~Park, {\it {An Orientifold of type IIB theory on K3}},
  {\em Nucl. Phys.} {\bf B472} (1996) 207--220,
  [\href{http://arxiv.org/abs/hep-th/9602030}{{\tt hep-th/9602030}}].

\bibitem{Sagnotti:1992qw}
A.~Sagnotti, {\it {A Note on the Green-Schwarz mechanism in open string
  theories}},  {\em Phys. Lett.} {\bf B294} (1992) 196--203,
  [\href{http://arxiv.org/abs/hep-th/9210127}{{\tt hep-th/9210127}}].

\bibitem{Danielsson:1997kt}
U.~H. Danielsson, G.~Ferretti, J.~Kalkkinen, and P.~Stjernberg, {\it {Notes on
  supersymmetric gauge theories in five-dimensions and six-dimensions}},  {\em
  Phys. Lett.} {\bf B405} (1997) 265--270,
  [\href{http://arxiv.org/abs/hep-th/9703098}{{\tt hep-th/9703098}}].

\bibitem{Bershadsky:1997sb}
M.~Bershadsky and C.~Vafa, {\it {Global anomalies and geometric engineering of
  critical theories in six-dimensions}},
  \href{http://arxiv.org/abs/hep-th/9703167}{{\tt hep-th/9703167}}.

\bibitem{Hitchin:1986ea}
N.~J. Hitchin, A.~Karlhede, U.~Lindstrom, and M.~Rocek, {\it {Hyperkahler
  Metrics and Supersymmetry}},  {\em Commun. Math. Phys.} {\bf 108} (1987) 535.

\bibitem{Mekareeya:2017sqh}
N.~Mekareeya, K.~Ohmori, H.~Shimizu, and A.~Tomasiello, {\it {Small instanton
  transitions for M5 fractions}},  {\em JHEP} {\bf 10} (2017) 055,
  [\href{http://arxiv.org/abs/1707.05785}{{\tt arXiv:1707.05785}}].

\bibitem{Cabrera:2019izd}
S.~Cabrera, A.~Hanany, and M.~Sperling, {\it {Magnetic quivers, Higgs branches,
  and 6d $N$=(1,0) theories}},  {\em JHEP} {\bf 06} (2019) 071,
  [\href{http://arxiv.org/abs/1904.12293}{{\tt arXiv:1904.12293}}]. [erratum:
  JHEP 07 (2019) 137].

\bibitem{DelZotto:2014kka}
M.~Del~Zotto and A.~Hanany, {\it {Complete Graphs, Hilbert Series, and the
  Higgs branch of the 4d $\mathcal{N} =$ 2 $(A_n,A_m)$ SCFTs}},  {\em Nucl.
  Phys.} {\bf B894} (2015) 439--455,
  [\href{http://arxiv.org/abs/1403.6523}{{\tt arXiv:1403.6523}}].

\bibitem{Cremonesi:2015lsa}
S.~Cremonesi, G.~Ferlito, A.~Hanany, and N.~Mekareeya, {\it {Instanton
  Operators and the Higgs Branch at Infinite Coupling}},  {\em JHEP} {\bf 04}
  (2017) 042, [\href{http://arxiv.org/abs/1505.06302}{{\tt arXiv:1505.06302}}].

\bibitem{Ferlito:2017xdq}
G.~Ferlito, A.~Hanany, N.~Mekareeya, and G.~Zafrir, {\it {3d Coulomb branch and
  5d Higgs branch at infinite coupling}},  {\em JHEP} {\bf 07} (2018) 061,
  [\href{http://arxiv.org/abs/1712.06604}{{\tt arXiv:1712.06604}}].

\bibitem{Cabrera:2018jxt}
S.~Cabrera, A.~Hanany, and F.~Yagi, {\it {Tropical Geometry and Five
  Dimensional Higgs Branches at Infinite Coupling}},  {\em JHEP} {\bf 01}
  (2019) 068, [\href{http://arxiv.org/abs/1810.01379}{{\tt arXiv:1810.01379}}].

\bibitem{Mekareeya:2017jgc}
N.~Mekareeya, K.~Ohmori, Y.~Tachikawa, and G.~Zafrir, {\it {E$_{8}$ instantons
  on type-A ALE spaces and supersymmetric field theories}},  {\em JHEP} {\bf
  09} (2017) 144, [\href{http://arxiv.org/abs/1707.04370}{{\tt
  arXiv:1707.04370}}].

\bibitem{Hanany:2018uhm}
A.~Hanany and N.~Mekareeya, {\it {The small E$_{8}$ instanton and the Kraft
  Procesi transition}},  {\em JHEP} {\bf 07} (2018) 098,
  [\href{http://arxiv.org/abs/1801.01129}{{\tt arXiv:1801.01129}}].

\bibitem{Hanany:2018vph}
A.~Hanany and G.~Zafrir, {\it {Discrete Gauging in Six Dimensions}},  {\em
  JHEP} {\bf 07} (2018) 168, [\href{http://arxiv.org/abs/1804.08857}{{\tt
  arXiv:1804.08857}}].

\bibitem{beauville2000symplectic}
A.~Beauville, {\it Symplectic singularities},  {\em Invent. Math.} {\bf 139}
  (2000), no.~3 541--549, [\href{http://arxiv.org/abs/math/9903070}{{\tt
  math/9903070}}].

\bibitem{Bourget:2019aer}
A.~Bourget, S.~Cabrera, J.~F. Grimminger, A.~Hanany, M.~Sperling, A.~Zajac, and
  Z.~Zhong, {\it {The Higgs mechanism — Hasse diagrams for symplectic
  singularities}},  {\em JHEP} {\bf 01} (2020) 157,
  [\href{http://arxiv.org/abs/1908.04245}{{\tt arXiv:1908.04245}}].

\bibitem{DelZotto:2014hpa}
M.~Del~Zotto, J.~J. Heckman, A.~Tomasiello, and C.~Vafa, {\it {6d Conformal
  Matter}},  {\em JHEP} {\bf 02} (2015) 054,
  [\href{http://arxiv.org/abs/1407.6359}{{\tt arXiv:1407.6359}}].

\bibitem{Evans:1997hk}
N.~J. Evans, C.~V. Johnson, and A.~D. Shapere, {\it {Orientifolds, branes, and
  duality of 4-D gauge theories}},  {\em Nucl. Phys.} {\bf B505} (1997)
  251--271, [\href{http://arxiv.org/abs/hep-th/9703210}{{\tt hep-th/9703210}}].

\bibitem{Intriligator:1997kq}
K.~A. Intriligator, {\it {RG fixed points in six-dimensions via branes at
  orbifold singularities}},  {\em Nucl. Phys.} {\bf B496} (1997) 177--190,
  [\href{http://arxiv.org/abs/hep-th/9702038}{{\tt hep-th/9702038}}].

\bibitem{Blum:1997mm}
J.~D. Blum and K.~A. Intriligator, {\it {New phases of string theory and 6-D RG
  fixed points via branes at orbifold singularities}},  {\em Nucl. Phys.} {\bf
  B506} (1997) 199--222, [\href{http://arxiv.org/abs/hep-th/9705044}{{\tt
  hep-th/9705044}}].

\bibitem{Intriligator:1997dh}
K.~A. Intriligator, {\it {New string theories in six-dimensions via branes at
  orbifold singularities}},  {\em Adv. Theor. Math. Phys.} {\bf 1} (1998)
  271--282, [\href{http://arxiv.org/abs/hep-th/9708117}{{\tt hep-th/9708117}}].

\bibitem{Ferrara:1998vf}
S.~Ferrara, A.~Kehagias, H.~Partouche, and A.~Zaffaroni, {\it {Membranes and
  five-branes with lower supersymmetry and their AdS supergravity duals}},
  {\em Phys. Lett.} {\bf B431} (1998) 42--48,
  [\href{http://arxiv.org/abs/hep-th/9803109}{{\tt hep-th/9803109}}].

\bibitem{Mekareeya:2016yal}
N.~Mekareeya, T.~Rudelius, and A.~Tomasiello, {\it {T-branes, Anomalies and
  Moduli Spaces in 6D SCFTs}},  {\em JHEP} {\bf 10} (2017) 158,
  [\href{http://arxiv.org/abs/1612.06399}{{\tt arXiv:1612.06399}}].

\bibitem{Ganor:1996mu}
O.~J. Ganor and A.~Hanany, {\it {Small E(8) instantons and tensionless
  noncritical strings}},  {\em Nucl. Phys.} {\bf B474} (1996) 122--140,
  [\href{http://arxiv.org/abs/hep-th/9602120}{{\tt hep-th/9602120}}].

\bibitem{Seiberg:1996vs}
N.~Seiberg and E.~Witten, {\it {Comments on string dynamics in
  six-dimensions}},  {\em Nucl. Phys.} {\bf B471} (1996) 121--134,
  [\href{http://arxiv.org/abs/hep-th/9603003}{{\tt hep-th/9603003}}].

\bibitem{Hanany:1996ie}
A.~Hanany and E.~Witten, {\it {Type IIB superstrings, BPS monopoles, and
  three-dimensional gauge dynamics}},  {\em Nucl. Phys.} {\bf B492} (1997)
  152--190, [\href{http://arxiv.org/abs/hep-th/9611230}{{\tt hep-th/9611230}}].

\bibitem{Feng:2000eq}
B.~Feng and A.~Hanany, {\it {Mirror symmetry by O3 planes}},  {\em JHEP} {\bf
  11} (2000) 033, [\href{http://arxiv.org/abs/hep-th/0004092}{{\tt
  hep-th/0004092}}].

\bibitem{Cremonesi:2013lqa}
S.~Cremonesi, A.~Hanany, and A.~Zaffaroni, {\it {Monopole operators and Hilbert
  series of Coulomb branches of $3d$ $\mathcal{N} = 4$ gauge theories}},  {\em
  JHEP} {\bf 01} (2014) 005, [\href{http://arxiv.org/abs/1309.2657}{{\tt
  arXiv:1309.2657}}].

\bibitem{Cabrera:2016vvv}
S.~Cabrera and A.~Hanany, {\it {Branes and the Kraft-Procesi Transition}},
  {\em JHEP} {\bf 11} (2016) 175, [\href{http://arxiv.org/abs/1609.07798}{{\tt
  arXiv:1609.07798}}].

\bibitem{Cabrera:2017njm}
S.~Cabrera and A.~Hanany, {\it {Branes and the Kraft-Procesi transition:
  classical case}},  {\em JHEP} {\bf 04} (2018) 127,
  [\href{http://arxiv.org/abs/1711.02378}{{\tt arXiv:1711.02378}}].

\bibitem{Cabrera:2018ann}
S.~Cabrera and A.~Hanany, {\it {Quiver Subtractions}},  {\em JHEP} {\bf 09}
  (2018) 008, [\href{http://arxiv.org/abs/1803.11205}{{\tt arXiv:1803.11205}}].

\bibitem{Hanany:2018cgo}
A.~Hanany and M.~Sperling, {\it {Discrete quotients of 3-dimensional $
  \mathcal{N}=4 $ Coulomb branches via the cycle index}},  {\em JHEP} {\bf 08}
  (2018) 157, [\href{http://arxiv.org/abs/1807.02784}{{\tt arXiv:1807.02784}}].

\bibitem{Hanany:2018dvd}
A.~Hanany and A.~Zajac, {\it {Discrete Gauging in Coulomb branches of Three
  Dimensional $\mathcal{N}=4$ Supersymmetric Gauge Theories}},  {\em JHEP} {\bf
  08} (2018) 158, [\href{http://arxiv.org/abs/1807.03221}{{\tt
  arXiv:1807.03221}}].

\bibitem{Intriligator:1996ex}
K.~A. Intriligator and N.~Seiberg, {\it {Mirror symmetry in three-dimensional
  gauge theories}},  {\em Phys. Lett.} {\bf B387} (1996) 513--519,
  [\href{http://arxiv.org/abs/hep-th/9607207}{{\tt hep-th/9607207}}].

\bibitem{Goddard:1976qe}
P.~Goddard, J.~Nuyts, and D.~I. Olive, {\it {Gauge Theories and Magnetic
  Charge}},  {\em Nucl. Phys.} {\bf B125} (1977) 1--28.

\bibitem{Gaiotto:2008ak}
D.~Gaiotto and E.~Witten, {\it {S-Duality of Boundary Conditions In N=4 Super
  Yang-Mills Theory}},  {\em Adv. Theor. Math. Phys.} {\bf 13} (2009), no.~3
  721--896, [\href{http://arxiv.org/abs/0807.3720}{{\tt arXiv:0807.3720}}].

\bibitem{Benini:2010uu}
F.~Benini, Y.~Tachikawa, and D.~Xie, {\it {Mirrors of 3d Sicilian theories}},
  {\em JHEP} {\bf 09} (2010) 063, [\href{http://arxiv.org/abs/1007.0992}{{\tt
  arXiv:1007.0992}}].

\bibitem{Cremonesi:2014uva}
S.~Cremonesi, A.~Hanany, N.~Mekareeya, and A.~Zaffaroni, {\it
  {T$_{\rho}^{\sigma}$ (G) theories and their Hilbert series}},  {\em JHEP}
  {\bf 01} (2015) 150, [\href{http://arxiv.org/abs/1410.1548}{{\tt
  arXiv:1410.1548}}].

\bibitem{Cabrera:2017ucb}
S.~Cabrera, A.~Hanany, and Z.~Zhong, {\it {Nilpotent orbits and the Coulomb
  branch of $T^\sigma (G)$ theories: special orthogonal vs orthogonal gauge
  group factors}},  {\em JHEP} {\bf 11} (2017) 079,
  [\href{http://arxiv.org/abs/1707.06941}{{\tt arXiv:1707.06941}}].

\bibitem{Chacaltana:2011ze}
O.~Chacaltana and J.~Distler, {\it {Tinkertoys for the $D_N$ series}},  {\em
  JHEP} {\bf 02} (2013) 110, [\href{http://arxiv.org/abs/1106.5410}{{\tt
  arXiv:1106.5410}}].

\bibitem{Zhenghao}
{Zhong, Zhenghao}, {\it {Quiver gauge theories in 3d, 5d and 6d}},  Master's
  thesis, {Imperial College, London}, 2018.

\bibitem{Frey:2018vpw}
D.~D. Frey and T.~Rudelius, {\it {6D SCFTs and the Classification of
  Homomorphisms $\Gamma_{ADE} \rightarrow E_8$}},
  \href{http://arxiv.org/abs/1811.04921}{{\tt arXiv:1811.04921}}.

\bibitem{DelZotto:2018tcj}
M.~Del~Zotto and G.~Lockhart, {\it {Universal Features of BPS Strings in
  Six-dimensional SCFTs}},  {\em JHEP} {\bf 08} (2018) 173,
  [\href{http://arxiv.org/abs/1804.09694}{{\tt arXiv:1804.09694}}].

\bibitem{Hanany:2000fq}
A.~Hanany and B.~Kol, {\it {On orientifolds, discrete torsion, branes and M
  theory}},  {\em JHEP} {\bf 06} (2000) 013,
  [\href{http://arxiv.org/abs/hep-th/0003025}{{\tt hep-th/0003025}}].

\end{thebibliography}\endgroup

\end{document}